\documentclass{article}
\usepackage[english]{babel}
\usepackage{cite}
\usepackage{jheppub} 

\usepackage{gensymb}

\usepackage{placeins}

\usepackage{epsf}
\usepackage{amssymb}
\usepackage{amsmath}
\usepackage{amsfonts}
\usepackage{psfrag,epsfig,graphicx,graphics}

\usepackage{bm,slashed,multirow,
	soul,mathtools,xspace,array,comment}                
\allowdisplaybreaks
\usepackage{bbold}
\usepackage{color}
\usepackage{xcolor}
\usepackage{subfigure}
\usepackage[printonlyused]{acronym}
\usepackage{hyperref}

\usepackage{float}

\def\slashchar#1{\setbox0=\hbox{$#1$}
   \dimen0=\wd0
   \setbox1=\hbox{/} \dimen1=\wd1
   \ifdim\dimen0>\dimen1
      \rlap{\hbox to \dimen0{\hfil/\hfil}}
      #1
   \else
      \rlap{\hbox to \dimen1{\hfil$#1$\hfil}}
      /
   \fi}

\newcommand{\APP}{App.~}

\newcommand{\FIG}{Fig.~}
\newcommand{\SEC}{Sec.~}

\newcommand{\EQ}{Eq.~}

\newcommand{\GeV}{\,\mbox{GeV}}

\def\bei{\begin{itemize}}
\def\ei{\end{itemize}}

\def\beeq{\begin{eqnarray}} 
\def\beqa{\begin{eqnarray}}
\def\bea{\begin{eqnarray}}

\def\eea{\end{eqnarray}}
\def\eqa{\end{eqnarray}}
\def\eeeq{\end{eqnarray}}

\def\eqar{\end{array}}
\def\beqar{\begin{array}}

\def\beas{\begin{eqnarray*}}
\def\beqas{\begin{eqnarray*}}

\def\eqas{\end{eqnarray*}}
\def\eeas{\end{eqnarray*}}

\def\beq{\begin{equation}} 
\def\be{\begin{equation}}

\def\ee{\end{equation}}
\def\eq{\end{equation}}
\def\eeq{\end{equation}}

\def\beqd{\begin{displaymath}}
\def\eeqd{\end{displaymath}}
\def\eqd{\end{displaymath}}

\def\beeq{\begin{eqnarray}} \def\eeeq{\end{eqnarray}}

\newcommand{\fin}{\end{document}}

\def\pv{\vec{p}_t}
\def\dv{\vec{\Delta}_t}

\def\meson{ \pi }

\def\fin{\end{document}}

\newcommand{\alb}{\bar{\alpha}}

\newcommand{\pt}{ \vec{p}_{t} }
\newcommand{\SgN}{ S_{\gamma N} }
\newcommand{\Msq}{ M_{\gamma \meson}^2 }

\def\zb{\bar{z}}
\def\phiAS{\phi_{\rm as}}
\def\phiLC{\phi_{\rm hol}}

\DeclareMathOperator{\sgn}{sgn}

\title{Accessing chiral-even quark generalised parton distributions in the exclusive photoproduction of a $ \gamma  \pi ^{\pm} $ pair with large invariant mass in both fixed-target and collider experiments}

\author[1]{Goran Duplan\v{c}i\'{c},}
\author[2]{Saad Nabeebaccus,}
\author[1]{Kornelija Passek-Kumeri\v{c}ki,}
\author[3]{Bernard Pire,}
\author[4]{Lech Szymanowski,}
\author[2]{Samuel Wallon}

\affiliation[1]{Theoretical Physics Division, Rudjer Bo{\v s}kovi{\'c} Institute,
	HR-10002 Zagreb, Croatia}
\affiliation[2]{Universit\'e Paris-Saclay, CNRS/IN2P3, IJCLab, 91405 Orsay, France}
\affiliation[3]{CPHT, CNRS, Ecole polytechnique, Institut Polytechnique de Paris, 91128 Palaiseau, France}
\affiliation[4]{National Center for Nuclear Research (NCBJ), Warsaw, Poland}

\emailAdd{gorand@thphys.irb.hr}
\emailAdd{passek@irb.hr}
\emailAdd{saad.nabeebaccus@ijclab.in2p3.fr}
\emailAdd{bernard.pire@polytechnique.edu}
\emailAdd{Lech.Szymanowski@ncbj.gov.pl}
\emailAdd{samuel.wallon@ijclab.in2p3.fr}

\abstract{
We compute the exclusive photoproduction of a $\gamma\,\pi^\pm$ pair using the collinear
factorisation framework, in the
kinematic regime where the pair has a large invariant mass. This exclusive channel presents a new avenue for the investigation of GPDs. It is particularly interesting as the high centre of mass energies available at future experiments
will allow the study of GPDs at small skewness $  \xi  $. We compute the scattering amplitude of the process, at leading twist and leading order in
$\alpha_s$, which is used to estimate its cross-section and linear polarisation asymmetries with respect to the incoming photon, for JLab~12-GeV, COMPASS, future EIC and LHC (in ultra-peripheral collisions) kinematics. We find that the order of magnitude of estimates are sufficiently large for a dedicated experimental analysis to be performed, especially at JLab. We also compare the results from an asymptotic distribution amplitude (DA) to those using a recently proposed \textit{holographic} DA.
}

\date{\today}
\begin{document}
	
	\maketitle

\preprint{}

\pagestyle{empty}
\newpage

\mbox{}

\pagestyle{plain}

\section{Introduction}

\label{Sec:Introduction}

We pursue our aim to access generalised parton distributions (GPDs) through a new family of $ 2 \to 3$ exclusive processes \cite{Ivanov:2002jj,Enberg:2006he, ElBeiyad:2010pji,Pedrak:2017cpp,Pire:2019hos,Pedrak:2020mfm,Cosyn:2021dyr}, in addition to the well-known $2 \to 2$ channels such as deeply-virtual Compton scattering (DVCS), deeply-virtual meson production (DVMP) and timelike Compton scattering (TCS), see e.g \cite{Goeke:2001tz,Diehl:2003ny,Belitsky:2005qn,Boffi:2007yc,Burkert:2007zz,Guidal:2008zza,CLAS:2021lky} and references therein. In the present work, we focus on the exclusive photoproduction of a $\gamma \pi^{\pm} $ pair with a large invariant mass which provides the hard scale for justifying QCD collinear factorisation,
\begin{align}
	\label{eq:process}
	\gamma (q,\epsilon _{q})+N(p_1, \lambda _{1}) \longrightarrow \gamma (k,\epsilon _{k})+N'(p_2, \lambda _{2})+{\meson}^{\pm}(p_{\meson})\,.
\end{align}
In this way, one is able to probe the leading twist chiral even quark GPDs. The work presented here builds up on our previous publication \cite{Duplancic:2018bum} in the following aspects:
 \begin{itemize}
 	\item the kinematics is extended from JLab to COMPASS, EIC and LHC (in ultraperipheral collisions),
 	\item results for a new distribution amplitude (DA), the so-called  `holographic' model are worked out and presented,
 	\item in addition to the unpolarised cross-sections, linear polarisation asymmetries are computed.
 \end{itemize}

\begin{figure}[h]
	
	\psfrag{TH}{$\Large T_H$}
	\psfrag{Pi}{$m'$}
	\psfrag{P1}{$\,\phi$}
	\psfrag{P2}{$\,\phi$}
	\psfrag{Phi}{$\,\phi$}
	\psfrag{Rho}{$ m $}
	\psfrag{tp}{$t'$}
	\psfrag{s}{$s$}
	\psfrag{x1}{$\!\!\!\!\!\!x+\xi$}
	\psfrag{x2}{$\!\!x-\xi$}
	\psfrag{RhoT}{$M_T$}
	\psfrag{t}{$t$}
	\psfrag{N}{$N$}
	\psfrag{Np}{$N'$}
	\psfrag{M}{$M^2_{\gamma m }$}
	\psfrag{GPD}{$\!GPD$}

	\centerline{
		\raisebox{1.6cm}{\includegraphics[width=12pc]{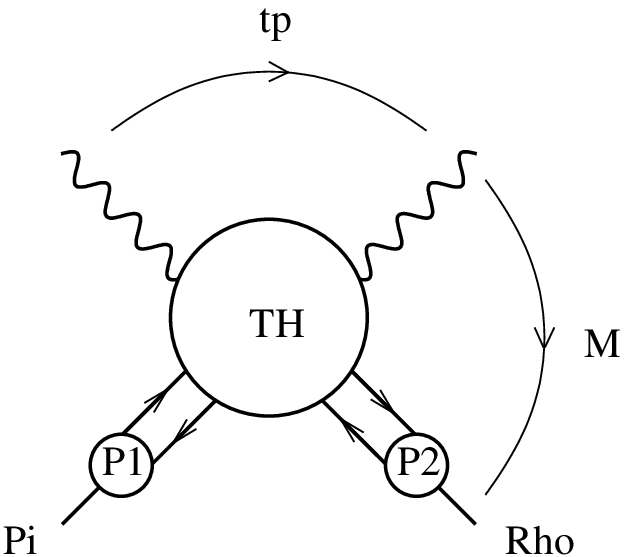}}~~~~~~~~~~~~~~
		\psfrag{TH}{$\,\Large T_H$}
		\psfrag{Rho}{$ \meson $}
		\psfrag{M}{$M^2_{\gamma \meson }$}
		\includegraphics[width=12pc]{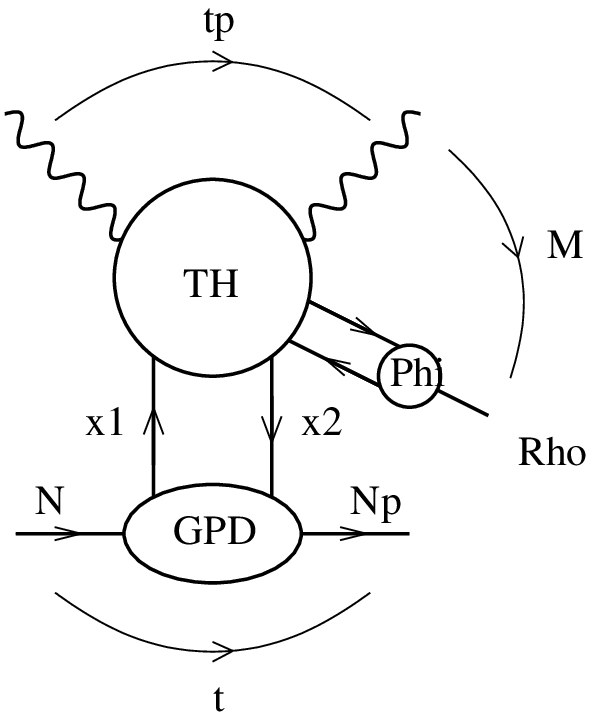}}
	
	\caption{\small Left: factorisation of the amplitude for the process $\gamma + m' \rightarrow \gamma + m  $ at large $s$ and fixed angle (i.e. fixed ratio $t'/s$). Right:  Replacing the incoming meson distribution amplitude by a nucleon generalised parton distribution  leads to the factorisation of the amplitude  for $\gamma + N \rightarrow \gamma + \meson  +N'$ at large $M_{\gamma \meson }^2$ and small $ t $.}
	\label{Fig:feyndiag}
\end{figure}

Originally, the motivation for using QCD collinear factorisation for the process we consider was obtained by making a comparison with the photon meson scattering process, $\gamma + m' \rightarrow \gamma + m  $, at large $s$ and fixed angle (i.e. fixed ratio $t'/s$), see \FIG\ref{Fig:feyndiag}. By replacing the incoming meson distribution amplitude with the generalised parton distribution of the nucleon, one can infer that in order to benefit from the factorisation properties of the former process, the invariant mass of the photon-meson pair, $ M_{\gamma \meson} $, should be large, while $ t= \left( p_{N}-p_{N'} \right)^2  $ should be small.

Recently, QCD collinear factorisation has been proven in a very similar process, the exclusive production of a photon pair in pion-nucleon collisions  \cite{Qiu:2022bpq, Qiu:2022pla}. The proof relies on the transverse momentum of each photon in the final state to be large, while in the process under consideration the invariant mass of the photon-meson pair $ M_{\gamma \meson}^2 $ is taken to be large. In fact, the latter condition is \textit{sufficient}, since it is more constraining than the former condition (see \EQ\eqref{skewness2}). We note that such a proof of factorisation at leading twist is applicable for the family of exclusive $2 \to 3$ processes, and in particular for our process as well. We point out that a recent 1-loop computation for the similar process of exclusive photoproduction of a photon pair \cite{Grocholski:2021man,Grocholski:2022rqj} shows indeed that collinear QCD holds at the NLO level.
 
 The paper is organised as follows: In \SEC\ref{sec:kinematics}, we discuss the kinematics of our process. Then, in \SEC\ref{sec:non-pert-inputs}, we present the non-perturbative inputs to our calculation, namely the GPDs and the DAs. The decomposition of the amplitude in terms of tensorial structures and basic building block integrals, as well as the calculation of the fully-differential cross-section, are the subject of \SEC\ref{sec:computation}. Our results, for the fully-differential, single-differential and integrated cross-sections and linear polarisation asymmetry with respect to the incoming photon are shown in \SEC\ref{sec:results}. Both cross-sections and linear polarisation asymmetries are shown, for JLab, COMPASS, EIC and ultraperipheral collisions (UPCs) at LHC kinematics. This section finishes with an estimation of counting rates at various experiments in order to assess the feasibility of measuring our process. We end with conclusions in \SEC\ref{sec:conclusion}. In \APP\ref{app:holographic-DA}, the new building block integrals involving the holographic DA are presented, 
and relevant diagrams are expressed in terms of these building block integrals. Details regarding the determination of the photon flux are found in \APP\ref{app:photon-flux}. The phase space integration is briefly discussed in \APP\ref{app:phase-space}. Finally, in \APP\ref{app:polarisation-asymmetries}, the linear polarisation asymmetry is discussed, and their expressions are explicitly given.

\section{Kinematics}

\label{sec:kinematics}

We work with the following useful momenta, see \eqref{eq:process},
\begin{align}
	P^{ \mu }=\frac{p_1^{ \mu }+p_{2}^{ \mu }}{2}\,,\quad  \Delta ^{ \mu }=p_2^{ \mu }-p_{1}^{ \mu }\,.
\end{align}
We decompose all momenta in a Sudakov basis, such that a generic vector $ v $ can be written as
\begin{align}
v^{\mu}=a\,n^{ \mu }+b\,p^{ \mu }+v^{ \mu }_{\perp}\,.
\end{align}
The two light-cone vectors $p$ and $n$ are chosen such that
\begin{align}
\label{sudakov2}
p^\mu = \frac{\sqrt{s}}{2}(1,0,0,1)\,,\qquad n^\mu = 
\frac{\sqrt{s}}{2}(1,0,0,-1) \,,\qquad p\cdot n = \frac{s}{2}\,.
\end{align}
For the transverse vectors, we use the following convention,
\begin{equation}
	\label{sudakov3}
	v_\bot^\mu = (0,v^x,v^y,0) \,, \qquad v_\bot^2 = -\vec{v}_t^2\,.
\end{equation}
Thus, the particle momenta for the process we consider can be written as
\begin{align}
	\label{eqn:impini}
	p_1^\mu &= (1+\xi)\,p^\mu + \frac{M^2}{s(1+\xi)}\,n^\mu\,,\\[5pt]
	p_2^\mu &= (1-\xi)\,p^\mu + \frac{M^2+\vec{\Delta}^2_t}{s(1-\xi)}n^\mu + \Delta^\mu_\bot\,,\\[5pt]
	q^\mu &= n^\mu\,,\\[5pt]
	\label{eq:momentum-outgoing-photon}
	k^\mu &= \alpha \, n^\mu + \frac{(\vec{p}_t-\vec\Delta_t/2)^2}{\alpha s}\,p^\mu + p_\bot^\mu -\frac{\Delta^\mu_\bot}{2}\,,\\[5pt]
	\label{eq:p-pi}
	p_\meson^\mu &= \alpha_\meson \, n^\mu + \frac{(\vec{p}_t+\vec\Delta_t/2)^2+M^2_\meson}{\alpha_\meson s}\,p^\mu - p_\bot^\mu-\frac{\Delta^\mu_\bot}{2}\,,
\end{align}
where $M$ and $M_\meson$ are the masses of the nucleon and the pion respectively. The  square of the centre of mass energy of the $\gamma$-N system is
then
\begin{align}
S_{\gamma N} = (q + p_1)^2 = (1+\xi)s + M^2\,,
\end{align}
while the squared transferred momentum is
\begin{align}
	t = (p_2 - p_1)^2 = -\frac{1+\xi}{1-\xi}\vec{\Delta}_t^2 -\frac{4\xi^2M^2}{1-\xi^2}\,.
\end{align}
The hard scale $M^2_{\gamma\meson}$ is the invariant mass squared of the $\gamma\meson^{\pm}$ system. This hardness is guaranteed by having a large \textit{relative} transverse momentum $  \vec{p}_{t}  $ between the outgoing photon and meson.

Collinear QCD factorisation implies that
\begin{equation}
 -u'= \left( p_{\meson}-q \right)^2\,,\qquad -t'= \left( k-q \right)^2\,,\qquad   M_{\gamma \meson}^2 =  \left(  p_{\meson}+k\right)^2\,,
\end{equation}
 are large, while 
 \begin{equation}
-t =  \left( p_2-p_1 \right)^2  \,,
 \end{equation}
   needs to be small. For this, we employ the cuts
 \begin{align}
 -u',-t'&>1 \GeV ^2\,,\\[5pt]
   -t &< 0.5  \GeV ^2\,.
 \end{align}
 
 We note that these cuts are sufficient to ensure that $ M^2_{\gamma\meson} > 1 $ GeV$ ^2 $. In fact, as pointed out in the introduction, these cuts are \textit{sufficient} for collinear factorisation to hold, since the necessary condition is that $  \vec{p}_{t}^{\,2}  $ should be large. The above kinematical cuts ensure that the $\pi N'$ invariant mass is out of the resonance region.

In the generalised Bjorken limit, neglecting  $\dv$ in front of $\pv$, as well as hadronic masses, we have that the approximate kinematics is
\begin{eqnarray}
	\label{skewness2}
	M^2_{\gamma\meson} \approx  \frac{\vec{p}_t^{\,2}}{\alpha\bar{\alpha}} ~, \qquad
	\alpha _{\meson} \approx 1-\alpha \equiv \bar{\alpha} ~,\qquad
	\xi =  \frac{\tau}{2-\tau} ~,
\end{eqnarray}
\begin{eqnarray}
	\tau \approx 
	\frac{M^2_{\gamma\meson}}{S_{\gamma N}-M^2}~,\qquad
	~-t'  \approx  \bar\alpha\, M_{\gamma\meson}^2  ~,\qquad -u'  \approx  \alpha\, M_{\gamma\meson}^2 \,.\quad \,\nonumber
\end{eqnarray}

We choose as independent variables $(-t)$, $(-u')$ and $M_{\gamma\meson}^2$. More details on the kinematics can be found in the previous two papers on the subject \cite{Boussarie:2016qop,Duplancic:2018bum}.

\section{Non-perturbative inputs}

\label{sec:non-pert-inputs}

\subsection{Generalised Parton Distributions}

\label{sec:GPDs}

In our studies, both the $p \to n$ and $n \to p$ quark chiral even transition GPDs are needed. By 
isospin symmetry, they are identical and are related to the proton GPD by the relation \cite{Mankiewicz:1997aa}
\beqa
\label{TransitionGPD}
\langle n | \bar{d} \, \Gamma \, u | p \rangle = \langle p | \bar{u} \, \Gamma 
\, d | n \rangle =
\langle p | \bar{u} \, \Gamma \, u | p \rangle  - \langle p |
\bar{d} \, \Gamma \, d | p \rangle\,.
\eqa
Therefore, we only use the proton GPDs in practice. The chiral-even GPDs of a parton $q$ (where $q = u,\ d$) in the nucleon target   are  defined by~\cite{Diehl:2003ny}:
\beqa
\label{defGPDEvenV}
&&\langle p(p_2,\lambda_2)|\, \bar{q}\left(-\frac{y}{2}\right)\,\gamma^+q \left(\frac{y}{2}\right)|p(p_1,\lambda_1) \rangle \\ \nonumber 
&&= \int_{-1}^1dx\ e^{-\frac{i}{2}x(p_1^++p_2^+)y^-}\bar{u}(p_2,\lambda_2)\, \left[ \gamma^+ H^{q}(x,\xi,t)   +\frac{i}{2m}\sigma^{+ \,\alpha}\Delta_\alpha  \,E^{q}(x,\xi,t) \right]
u(p_1,\lambda_1)\,,
\eqa
for the chiral-even vector GPDs, and
\beqa
\label{defGPDEvenA}
&&\langle p(p_2,\lambda_2)|\, \bar{q}\left(-\frac{y}{2}\right)\,\gamma^+ \gamma^5 q\left(\frac{y}{2}\right)|p(p_1,\lambda_1)\rangle \\ \nonumber
&&= \int_{-1}^1dx\ e^{-\frac{i}{2}x(p_1^++p_2^+)y^-}\bar{u}(p_2,\lambda_2)\, \left[ \gamma^+ \gamma^5 \tilde H^{q}(x,\xi,t)   +\frac{1}{2m}\gamma^5 \Delta^+  \,\tilde E^{q}(x,\xi,t) \right]
u(p_1,\lambda_1)\,.
\eqa
for chiral-even axial GPDs. In the above, $\lambda_1$ and $\lambda_2$ are the light-cone helicities of the nucleons with momenta $p_1$ and $p_2$.

In our analysis, 
the contributions from $ E^{q} $ and $  \tilde{E}^{q}  $  are neglected, since they are suppressed by kinematical factors at the cross-section level, see \eqref{squareCEresult}. The GPDs are parametrised in terms of double distributions \cite{Radyushkin:1998es}. The details can be found in \cite{Boussarie:2016qop,Duplancic:2018bum}, and we do not repeat them here. The $t$-dependence of the GPDs is modelled by a simplistic dipole ansatz, discussed in \APP\ref{app:phase-space}.

We note that in the current leading order in $\alpha_s$ study, we neglect any evolution of the GPDs/PDFs, and take a fixed factorisation scale of $  \mu _{F}^2=10 \GeV^2 $. As in \cite{Boussarie:2016qop,Duplancic:2018bum}, the PDF datasets that we use to construct the GPDs are
\begin{itemize}
	\item 
	For $x q(x)$,  the GRV-98 parameterisation~\cite{Gluck:1998xa}, as 
	made available from the Durham database. 
		
	\item
	For $x \Delta q(x)\,,$  the  GRSV-2000 
	parameterisation~\cite{Gluck:2000dy}, also available from the Durham 
	database. Two scenarios are proposed within this parameterisation:
\begin{itemize}
	\item The \textit{standard} scenario, for which the light sea quark and anti-quark distributions are \textit{flavour-symmetric},
	\item The \textit{valence} scenario, which corresponds to \textit{flavour-asymmetric} light sea quark densities.
\end{itemize}
		We use both of them in order to estimate the order of magnitude of
	theoretical uncertainties.
	\end{itemize}
We note that using more recent tables for the PDFs leads to variations that are smaller than the above-mentioned theoretical uncertainties. This effect was studied in \cite{Boussarie:2016qop} (see e.g. Fig.~8).

\subsection{Distribution Amplitudes}

\label{sec:DAs}

The chiral-even light-cone DA for the $\pi^{+}$ meson is defined, at the leading 
twist 2, by the matrix element~\cite{Ball:1998je},
\begin{equation}
	\langle \pi^{+}(p_\pi)|\bar{u}(y)\gamma^5 \gamma^\mu  d(-y)|0 \rangle = i f_{\pi} 
	p_\pi^\mu \int_0^1dz\ e^{-i(z - \bar{z}) p_\pi \cdot y}\ \phi_{\pi}(z),
	\label{defDApi}
\end{equation}
and analogously for the $ \pi^{-} $ meson, with $f_{\pi}=131\,\mbox{MeV}$. 

For the computation, we use the asymptotic form of the distribution amplitude, $ \phi^{\rm as} $, as well as an alternative form, which is often  called `holographic' DA, $ \phi^{\rm hol} $. They are given by
\begin{align}
\label{DA-asymp}
\phi^{\rm as}(z)&= 6 z (1-z)\,,\\[5pt]
\label{DA-hol}
 \phi^{\rm hol}(z)&= \frac{8}{ \pi } \sqrt{z (1-z)}\,,
\end{align}
where both are normalised to 1. The alternative form, first proposed in \cite{Mikhailov:1986be}, has been suggested in the literature in the context of AdS-QCD holographic correspondence \cite{Brodsky:2006uqa} (hence the name `holographic' DA) and dynamical chiral symmetry breaking on the light-front \cite{Shi:2015esa}. In fact, recent lattice results indicate an even further departure from the asymptotic form, with $  \phi (z) \propto z^{ \alpha } \left( 1-z \right)^{ \alpha }  $ and $  \alpha \approx 0.2-0.32 $ \cite{Gao:2022vyh}. In the present work, we restrict ourselves to the asymptotic and holographic DAs, as this allows us to perform the integral over $z$ analytically, see \APP\ref{app:holographic-DA}.

\section{The Computation}

\label{sec:computation}

\subsection{Amplitude}

\subsubsection{Gauge invariant decomposition of the hard amplitude}
\label{SubSec:gauge-decomposition}
We now deal with the amplitude at the partonic level, and focus on the twist 2 coefficient function.
The  $\pi^+$ meson is described by $u\bar{d}$, and $  \pi ^{-} $ by $ d\bar{u} $.

\def\diagici{2.65cm}
\begin{figure}[h]
	\begin{center}
		\psfrag{z}{\begin{small} $z$ \end{small}}
		\psfrag{zb}{\raisebox{0cm}{ \begin{small}$\bar{z}$\end{small}} }
		\psfrag{gamma}{\raisebox{+.1cm}{ $\,\gamma$} }
		\psfrag{pi}{$\,\pi$}
		\psfrag{rho}{$\,\pi$}
		\psfrag{TH}{\hspace{-0.2cm} $T_H$}
		\psfrag{tp}{\raisebox{.5cm}{\begin{small}     $t'$       \end{small}}}
		\psfrag{s}{\hspace{.6cm}\begin{small}$s$ \end{small}}
		\psfrag{Phi}{ \hspace{-0.3cm} $\phi$}
		\hspace{-.4cm}
		\begin{picture}(430,170)
			\put(0,20){\includegraphics[width=15.2cm]{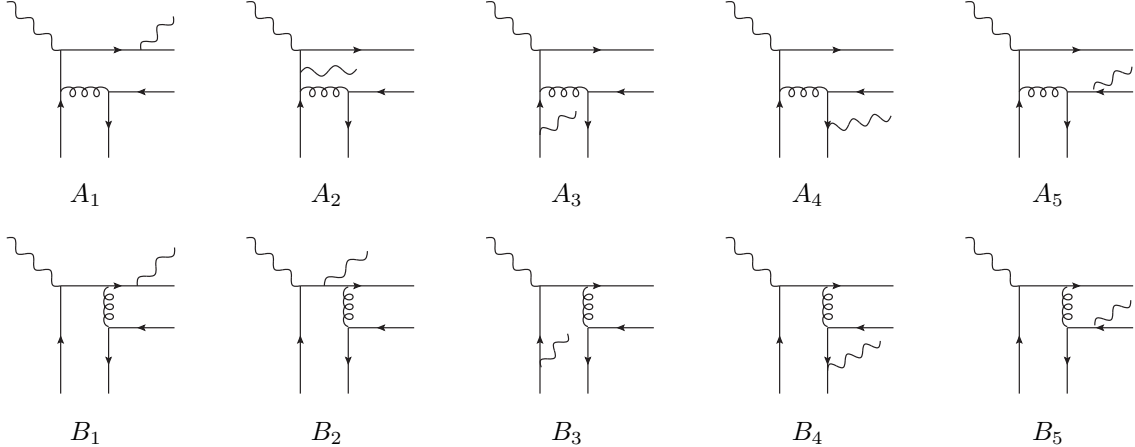}}
			\put(28,95){$A_1$}
			\put(119,95){$A_2$}
			\put(210,95){$A_3$}
			\put(301,95){$A_4$}
			\put(392,95){$A_5$}
			\put(28,5){$B_1$}
			\put(119,5){$B_2$}
			\put(210,5){$B_3$}
			\put(301,5){$B_4$}
			\put(392,5){$B_5$}
		\end{picture}
		\caption{Half of the Feynman diagrams contributing to the hard part of the amplitude.}
		\label{Fig:diagrams}
	\end{center}
\end{figure}

For the sake of completeness, we remind the reader of the properties of the diagrams contributing to the coefficient function, which significantly simplify the calculation \cite{Duplancic:2018bum}. This hard part is described at lowest order in $\alpha_s$ by 20 
Feynman diagrams.
Half of these diagrams, denoted $A$ and $B$, are drawn in Figure~\ref{Fig:diagrams}. The $ A $ and $ B $ diagrams are distinguished by the order in which the incoming photon and virtual gluon join one of the quark lines. The numbers (1 to 5) denote the five different ways of attaching the outgoing photon to the quark lines. The remaining set of diagrams, $C$ and $D$, is  obtained by exchanging the role of the two quarks in the $t-$channel. This $C-$parity transformation\footnote{Strictly speaking, this corresponds to a $C-$parity transformation \textit{after} the electric charges have been factored out, such that effectively, $q$ and $\bar{q}$ have a charge of 1.} corresponds to $z \leftrightarrow 1-z$ and $x \leftrightarrow -x$.

The sets of diagrams (without including charge factors) are denoted as $(\cdots  
)$.
We denote $(AB)_{123}$ the contribution of the sum of diagrams  
$A_1+A_2+A_3+B_1+B_2+B_3$, and $(AB)_{45}$ the contribution of the sum of 
diagrams  $A_4+A_5+B_4+B_5,$ 
and similarly for $(CD)_{12}$ and $(CD)_{345}$. They are separately QED gauge invariant. Indeed,
the colour factor factorises, and the discussion reduces to a pure QED one.
In the block $(AB)_{123}$, the three bosons are  connected to a single 
quark line in all possible ways.
In the block $(AB)_{45}$, a photon and a gluon are  connected to each quark line 
in all possible ways. The same reasoning applies to  $(CD)_{12}$ and $(CD)_{345}$ 
after exchanging the role of the initial and final state photons.

Using the notation $e_q=Q_q |e|$, by QED gauge invariance, one can write any 
amplitude for photon meson production 
as the sum of three separate gauge invariant terms, in the form
\beqa
\label{generic-decomposition}
{\cal M}=
(Q_1^2 + Q_2^2) {\cal M}_{\rm sum} + (Q_1^2 - Q_2^2) {\cal M}_{\rm diff} + 2 
Q_1 \, Q_2 {\cal M}_{\rm prod} \,,
\eqa
where $Q_1$ is the charge of the quark entering the DA and $Q_2$ is
the charge of the  quark leaving the DA, in each diagram.

Considering the parity properties of the $q \bar{q}$ correlators appearing
in the DA and in the GPDs,
we separate the contributions for parity $(+)$, denoted as $S$ and parity $(-)$, 
denoted as $P$.
Only two structures occur in the hard part, namely $PP$ (two $\gamma^5$ 
matrices) and $SP$ (one $\gamma^5$).

A close inspection of the $C-$parity transformation which relates the two 
sets of 10 diagrams  gives the following results. In the present case, for the vector contribution, 
the sum of 
diagrams reads
\beqa
\label{sumVpi}
&&{\cal M}_\pi^V \\
&&=
Q_1^2 [(AB)_{123}]_{SP} \otimes f + Q_1 Q_2 [(AB)_{45}]_{SP} \otimes f
- Q_2^2 [(AB)_{123}]_{SP}^{(C)}  \otimes f- Q_1 Q_2 [(AB)_{45}]_{SP}^{(C)} 
\otimes f\, , \nonumber
\eqa
while for the axial contribution one gets
\beqa
\label{sumApi}
&&{\cal M}_\pi^A \\
&&=
Q_1^2 [(AB)_{123}]_{PP}  \otimes \tilde{f} + Q_1 Q_2 [(AB)_{45}]_{PP} \otimes 
\tilde{f}
+ Q_2^2 [(AB)_{123}]_{PP}^{(C)}  \otimes \tilde{f} + Q_1 Q_2 
[(AB)_{45}]_{PP}^{(C)} \otimes \tilde{f}\,.\nonumber
\eqa
Here, $f$ denotes a GPD of the set  $H, E$ appearing in the decomposition of the 
vector correlator  (\ref{defGPDEvenV}), while 
$\tilde{f}$ denotes a GPD of the set  $\tilde{H}, \tilde{E}$ appearing in the 
decomposition of the axial correlator  (\ref{defGPDEvenA}). The symbol $\otimes$ represents the integration over $x$. The integration over $z$ 
for the pion DA is implicit, since the DA is symmetric 
over $z \leftrightarrow 1-z$. The above decomposition is convenient since the integration over $z$ is performed \textit{analytically}, while the integration over $x$ is performed \textit{numerically}. This allows us to evaluate the amplitude in blocks which can be used for computing various observables. 
 Equations \eqref{sumVpi} and \eqref{sumApi} are obtained by making the identification
\begin{align}
	 \left[  \left( CD \right)_{345}  \right] _{SP}&=	- \left[  \left( AB \right)_{123}  \right] _{SP}^{(C)}\,,\\[5pt]
	 \left[  \left( CD \right)_{12}  \right] _{SP}&=	- \left[  \left( AB \right)_{45}  \right] _{SP}^{(C)}\,,\\[5pt]
	 \left[  \left( CD \right)_{345}  \right] _{PP}&=	 \left[  \left( AB \right)_{123}  \right] _{PP}^{(C)}\,,\\[5pt]
	 \left[  \left( CD \right)_{12}  \right] _{PP}&=	 \left[  \left( AB \right)_{45}  \right] _{PP}^{(C)}\,.
\end{align}

We introduce a few convenient notations.
A superscript $s$ (resp. $a$) refers to the symmetric (resp. 
antisymmetric) structures of the hard amplitude and of the GPD wrt $ x $, i.e.
\beqa
\label{Def:a-s}
f(x) = \frac{1}{2} (f(x) + f(-x)) +  \frac{1}{2} (f(x) - f(-x))  = f^s(x) + 
f^a(x)\,.
\eqa
This thus leads to
\beqa
\label{generic-decomposition-pion-vector}
{\cal M}^V_\pi&=&
(Q_1^2 + Q_2^2) \,[(AB)_{123}]^a_{SP} \otimes f^a 
+ (Q_1^2 - Q_2^2) \,[(AB)_{123}]^s_{SP} \otimes f^s \nonumber \\
&&+ 2 Q_1 \, Q_2 \,
[(AB)_{45}]^a_{SP} \otimes f^a\,,
\eqa
and for the axial GPD contribution, i.e. $PP$:
\beqa
\label{generic-decomposition-pion-axial}
{\cal M}^A_\pi &=&
(Q_1^2 + Q_2^2)\, [(AB)_{123}]^s_{PP} \otimes \tilde{f}^s 
+ (Q_1^2 - Q_2^2) \,[(AB)_{123}]^a_{PP} \otimes \tilde{f}^a \nonumber \\
&& + 2 Q_1 \, Q_2 \,
[(AB)_{45}]^s_{PP} \otimes \tilde{f}^s\,,
\eqa
with $Q_1=Q_u$ and $Q_2=Q_d$ for a $\pi^+,$ and  $Q_1=Q_d$ and $Q_2=Q_u$ for a 
$\pi^-.$

In the case of $\rho^0$ meson 
production~\cite{Boussarie:2016qop}, which is $C (-)$, the exchange in the $t$-channel is fixed to be $C(-)$, while for the $\pi^0$ meson, which is $C (+)$, the exchange in the $t$-channel is fixed to be $C(+)$. On the other hand, $\pi^+$ production (and similarly for $\pi^-$) 
involves
both $C$-parity exchanges in $t-$channel, which explains why both symmetrical and 
antisymmetrical parts of  the GPDs are involved in 
equations \eqref{generic-decomposition-pion-vector} and \eqref{generic-decomposition-pion-axial}. 

The detailed evaluation of one diagram was already illustrated in \cite{Duplancic:2018bum}, and therefore, we do not repeat it here.

\subsubsection{Tensor structure}

For convenience, we introduce the common normalization coefficient
\begin{equation}
	C_\pi =  i \frac{4}{9}\,f_\pi \, \alpha_{em}\,\alpha_s\,\pi^2 \,.
	\label{coefpi}
\end{equation}   
Note that we include the charge factors $Q_u$ and $Q_d$  inside
the hard matrix element, using the decompositions obtained in 
equations \eqref{generic-decomposition-pion-vector} and
\eqref{generic-decomposition-pion-axial}.

For the $PP$ sector, 
two tensor structures appear, namely
\beqa
\label{def:TA-TB}
T_A &=& (\varepsilon_{q\perp} \cdot \varepsilon_{k\perp}^*)\,, \nonumber \\                                                  
T_B &=& (\varepsilon_{q\perp} \cdot p_\perp) (p_\perp \cdot                      
\varepsilon_{k\perp}^*)\,,
\eqa
while for the $SP$ sector, the two following structures appear
\beqa
\label{def:TA5-TB5}
T_{A_5} &=& (p_\perp \cdot                                      
\varepsilon_{k\perp}^*) \,  \epsilon^{n \,p \,\varepsilon_{q\perp}\, p_\perp}\,, 
\nonumber \\
T_{B_5} &=& -(p_\perp \cdot \varepsilon_{q\perp})\, \epsilon^{n \,p 
	\varepsilon_{k\perp}^*\, p_\perp}\,.
\eqa

\subsubsection{Organisation of the amplitude}

\label{sec:organising-amplitude}
The scattering amplitude of the process \eqref{eq:process}, in the factorised 
form,
is expressed in terms of  form factors ${\cal H}_\pi$, ${\cal E}_\pi$, $\tilde 
{\cal H}_\pi,$ $\tilde {\cal E}_\pi$, analogous to Compton form factors in DVCS, 
and reads
\begin{eqnarray}
	\mathcal{M}_\pi \equiv 
	\frac{1}{n\cdot p}\bar{u}(p_2,\lambda_2) \!\! \left[   \slashed {n}  {\cal 
		H}_\pi(\xi,t) +\frac{i\,\sigma^{n\,\alpha}\Delta_\alpha}{2m}  {\cal 
		E}_\pi(\xi,t) +    \slashed{n} \gamma^5  \tilde {\cal H}_\pi(\xi,t)
	+ \frac{n\cdot \Delta}{2m} \,\gamma^5\, \tilde {\cal E}_\pi(\xi,t)
	\right] \!\! u(p_1,\lambda_1). \!\!\!\!\!\nonumber
	\\
	\label{CEGPD}
\end{eqnarray}

We isolate the tensor structures of the form factors as
\begin{eqnarray}
	\label{dec-tensors-quarks}
	\mathcal{H}_\pi(\xi , t) &=&  \mathcal{H}_{\pi A_5} (\xi , t) T_{A_5} + 
	\mathcal{H}_{\pi B_5} (\xi , t) T_{B_5} \,,\nonumber\\
	\mathcal{\tilde{H}}_\pi(\xi , t) &=& \mathcal{\tilde{H}}_{\pi A} (\xi , t) T_A + 
	\mathcal{\tilde{H}}_{\pi B} (\xi , t) T_B \,.
\end{eqnarray}

These coefficients can be expressed in terms  of the sum over diagrams of the 
integral of the product of their traces, of GPDs and DAs, as defined and given 
explicitly in  
appendix~\ref{app:holographic-DA} for the case of the holographic DA, and appendix D in \cite{Duplancic:2018bum}.
We introduce dimensionless coefficients $N$ and $\tilde{N}$ as follows:
\beqa
\label{form-factors-NA5}
{\mathcal{H}}_{\pi A_5} = \frac{1}{s^3}C_\pi N_{\pi A_5} \,, \\
\label{form-factors-NB5}
{\mathcal{H}}_{\pi B_5} = \frac{1}{s^3}C_\pi N_{\pi B_5}\,,
\eqa
and
\beqa
\label{form-factors-TildeNA}
\tilde{\mathcal{H}}_{\pi A} = \frac{1}s C_\pi \tilde{N}_{\pi A} \,, \\
\label{form-factors-TildeNB}
\tilde{\mathcal{H}}_{\pi B} = \frac{1}{s^2} C_\pi \tilde{N}_{\pi B}\,.
\eqa
In order to emphasise the gauge invariant structure and to organise the 
numerical study, we factorise out the charge coefficients, and put an explicit
index $q$ for the flavour of the quark GPDs $f^q$ and $\tilde{f}^q$. In 
accordance
with the decompositions (\ref{generic-decomposition-pion-vector})
and (\ref{generic-decomposition-pion-axial}) we thus introduce\footnote{Typos in our previous publication \cite{Duplancic:2018bum} have been corrected here, as well as in \eqref{list-N-to-be-computed}.}
\beqa
\label{gauge-NA5}
&&\hspace{-1cm}N^q_{\pi A_5}(Q_1,Q_2) \\ 
&=& \!
(Q_1^2 + Q_2^2) 
N^q_{A_5}[(AB)_{123}]^a
+
(Q_1^2 - Q_2^2) N^q_{A_5}[(AB)_{123}]^s
+ 2 Q_1 \, Q_2 \,N^q_{A_5}[(AB)_{45}]^a\,, \nonumber
\\
\label{gauge-NB5}
&&\hspace{-1cm}N^q_{\pi B_5}(Q_1,Q_2) \\
&=& \!
(Q_1^2 + Q_2^2) 
N^q_{B_5}[(AB)_{123}]^a
+
(Q_1^2 - Q_2^2) N^q_{B_5}[(AB)_{123}]^s
+ 2 Q_1 \, Q_2 \,N^q_{B_5}[(AB)_{45}]^a\,, \nonumber
\eqa
and
\beqa
\label{gauge-TildeNA}
&&\hspace{-1cm}\tilde{N}^q_{\pi A}(Q_1,Q_2)\\ 
&=&\!
(Q_1^2 + Q_2^2) \tilde{N}^q_A[(AB)_{123}]^s
+ (Q_1^2 - Q_2^2) \tilde{N}^q_A[(AB)_{123}]^a
+ 2 Q_1 \, Q_2 \,
\tilde{N}^q_A[(AB)_{45}]^s\,, \nonumber 
\\
\label{gauge-TildeNB}
&&\hspace{-1cm}\tilde{N}^q_{\pi B}(Q_1,Q_2) \\
&=&\!
(Q_1^2 + Q_2^2) \tilde{N}^q_B[(AB)_{123}]^s
+ (Q_1^2 - Q_2^2) \tilde{N}^q_B[(AB)_{123}]^a
+ 2 Q_1 \, Q_2 \,
\tilde{N}^q_B[(AB)_{45}]^s\,. \nonumber 
\eqa
For the specific case of our two processes, namely $\gamma \pi^+$ production on 
a proton and $\gamma \pi^-$ production on a neutron, taking into account the 
structure (\ref{TransitionGPD}) of the transition GPDs structure
we thus need to compute the coefficients
\beqa
\label{Npi+A5}
N_{\pi^+ A_5}&=&N^u_{\pi A_5}(Q_u,Q_d) - N^d_{\pi A_5}(Q_u,Q_d) \,, \\
\label{Npi+B5}
N_{\pi^+ B_5}&=&N^u_{\pi B_5}(Q_u,Q_d) - N^d_{\pi B_5}(Q_u,Q_d)\,, 
\eqa
and
\beqa
\label{Npi-A5}
N_{\pi^- A_5}&=&N^u_{\pi A_5}(Q_d,Q_u) - N^d_{\pi A_5}(Q_d,Q_u)\,, \\
\label{Npi-B5}
N_{\pi^- B_5}&=&N^u_{\pi B_5}(Q_d,Q_u) - N^d_{\pi B_5}(Q_d,Q_u)\,,
\eqa
as well as
\beqa
\label{Npi+A}
\tilde{N}_{\pi^+ A}&=&\tilde{N}^u_{\pi A}(Q_u,Q_d) - \tilde{N}^d_{\pi 
	A}(Q_u,Q_d)\,, \\
\label{Npi+B}
\tilde{N}_{\pi^+ B}&=&\tilde{N}^u_{\pi B}(Q_u,Q_d) - \tilde{N}^d_{\pi 
	B}(Q_u,Q_d)\,, 
\eqa
and
\beqa
\label{Npi-A}
\tilde{N}_{\pi^- A}&=&\tilde{N}^u_{\pi A}(Q_d,Q_u) - \tilde{N}^d_{\pi 
	A}(Q_d,Q_u)\,, \\
\label{Npi-B}
\tilde{N}_{\pi^- B}&=&\tilde{N}^u_{\pi B}(Q_d,Q_u) - \tilde{N}^d_{\pi 
	B}(Q_d,Q_u)\,.
\eqa
Therefore, for each flavour $u$ and $d$, 
knowing the 12 numerical coefficients 
\beqa
\label{list-N-to-be-computed}
&& N^q_{A_5}[(AB)_{123}]^s, \ N^q_{A_5}[(AB)_{123}]^a, \ N^q_{A_5}[(AB)_{45}]^a, 
\nonumber \\
&& N^q_{B_5}[(AB)_{123}]^s, \ N^q_{B_5}[(AB)_{123}]^a, \ N^q_{B_5}[(AB)_{45}]^a, 
\nonumber \\ 
&& \tilde{N}^q_A[(AB)_{123}]^s, \ \tilde{N}^q_A[(AB)_{123}]^a, \ 
\tilde{N}^q_A[(AB)_{45}]^s, \nonumber \\
&& \tilde{N}^q_B[(AB)_{123}]^s, \ \tilde{N}^q_B[(AB)_{123}]^a, \  
\tilde{N}^q_B[(AB)_{45}]^s,
\eqa
for two given GPDs $f$ and $\tilde{f}$ (in practice $H$ and 
$\tilde{H}$, see next subsection), one can reconstruct the scattering amplitudes of the two processes.
These 12 coefficients can be expanded in terms of 5 building block integrals which we label as $I_b$, $I_c$, $I_h$, $I_i$ and $I_e$ for the asymptotic DA case, and 2 extra building blocks labelled as $\chi_b$, $\chi_c$
for the case of the holographic DA. The building block integrals can be found in appendix D of ref.~\cite{Duplancic:2018bum}, and in appendix~\ref{app:hol-DA-explicit}.

\subsection{Cross-section}
\label{sec:cross-section}

In the forward limit $\Delta_{\bot} = 0 = P_{\bot}$, one can show that the 
square of $\mathcal{M}_\pi$  reads, after summing over nucleon helicities\footnote{We note that this equation corrects a mistake from previous publications, cf. (4.19) in \cite{Boussarie:2016qop} and (5.23) in \cite{Duplancic:2018bum}.}
\begin{eqnarray}
	\label{squareCEresult}
	\mathcal{M}_{\pi} \mathcal{M}_{\pi}^{*} &\equiv  &
	\sum_{\lambda_2,\, \lambda_1}
	\mathcal{M}_{\pi} (\lambda_1,\lambda_2)\,
	\mathcal{M}_{\pi}^{*}(\lambda_1,\lambda_2) \\  
	&=&   8(1-\xi^2) 
	\left(  {\cal H}(\xi,t)  {\cal H}^{*}_\pi(\xi,t)    +  \tilde {\cal 
		H}_\pi(\xi,t) \tilde {\cal H}^{*}_\pi(\xi,t)  \right) \nonumber \\
	&&+8\,\frac{\xi^4}{1-\xi^2}   
	\left(  {\cal E}_\pi(\xi,t)
	{\cal E}^{*}_\pi(\xi,t)
	+  \tilde {\cal E}_\pi(\xi,t)
	\tilde {\cal E}^{*}_\pi(\xi,t)
	\right)\nonumber
	\\ 
	&&-8\, \xi^2   \left(  {\cal H}_\pi(\xi,t) {\cal E}^{ *}_\pi(\xi,t) + 
	{\cal H}^{*}_\pi(\xi,t) {\cal E}_\pi(\xi,t)
	+
	\tilde {\cal H}_\pi(\xi,t)\tilde {\cal E}^{*}_\pi(\xi,t) 
	+
	\tilde {\cal H}^{*}_\pi(\xi,t)\tilde {\cal E}_\pi(\xi,t)
	\right) .\nonumber
\end{eqnarray}
For moderately small values of $\xi$, this becomes
\begin{eqnarray}
	\label{squareCEresultsmallxi}
	\mathcal{M}_\pi \mathcal{M}^{*}_\pi &\simeq&   8
	\left(  {\cal H}_\pi(\xi,t) \, {\cal H}^{*}_\pi(\xi,t)    +  \tilde {\cal 
		H}_\pi(\xi,t)\, \tilde {\cal H}^{*}_\pi(\xi,t)  \right).
\end{eqnarray}
Hence we will restrict ourselves to the GPDs $H$, $\tilde{H}$ to perform our 
estimates of the cross-section\footnote{In practice, we keep the first line in 
	the r.h.s. of eq.~(\ref{squareCEresult}).}. We note that this approximation remains valide for the linear polarisation asymmetry wrt the incoming photon, as the above equation still contains the helicities of the incoming and outgoing photons.

We now perform the sum/averaging over the polarisations of the incoming and outgoing photons,
\begin{eqnarray}
	\label{FF-squared-H}
	|\tilde{\mathcal{H}}_\pi(\xi , t)|^2 & \equiv & \sum_{\lambda_k \lambda_q} 
	\tilde{\mathcal{H}}_\pi(\xi , t, \lambda_k, \lambda_q) \, 
	\tilde{\mathcal{H}}(\xi , t, \lambda_k, \lambda_q) \\ \nonumber
	&=& 2|\tilde{\mathcal{H}}_A (\xi , t)|^2 + p_\bot^4 | \tilde{\mathcal{H}}_B 
	(\xi , t)|^2 + p_\bot^2 \left[ \tilde{\mathcal{H}}_A (\xi , 
	t)\tilde{\mathcal{H}}^{\ast}_B (\xi , t) + \tilde{\mathcal{H}}^{\ast}_A (\xi , 
	t)\tilde{\mathcal{H}}_B (\xi , t) \right], \\ \nonumber \\
	|\mathcal{H}_\pi(\xi , t)|^2 &\equiv & \sum_{\lambda_k \lambda_q} 
	\mathcal{H}(\xi , t, \lambda_k, \lambda_q) \, \mathcal{H}^*(\xi , t, \lambda_k, 
	\lambda_q)
	\\ \nonumber
	\label{FF-squared-HTilde}
	&=& \frac{s^2 p_\bot^4}{4} \left(| \mathcal{H}_{A_5} (\xi , t)|^2 + | 
	\mathcal{H}_{B_5} (\xi , t) |^2\right). \nonumber 
\end{eqnarray}
Finally, we define the averaged amplitude squared $|\mathcal{\overline{M}}_\pi|^2,$ 
which includes 
the factor 1/4 coming from the averaging over the 
polarizations of the initial particles. Collecting all prefactors, which read 
\beq
\label{coefficients}
\frac{1}{s^2}   8 (1-\xi^2) |C_\pi|^2 \frac{1}{2^2}\,,
\eq
we have that
\beqa
\label{all-pi}
&&|\mathcal{\overline{M}}_{\pi}|^2 = \frac{2}{s^2}   (1-\xi^2)  |C_\pi|^2 
\left\{ 
2 \left|\tilde{N}_{\pi A} \right|^2 
+ \frac{p_\perp^4}{s^2} \left|\tilde{N}_{\pi B} \right|^2 
\right.\\
&&
\left.
+ \frac{p_\perp^2}s \left(\tilde{N}_{\pi A} 
\tilde{N}_{\pi B}^* + c.c. \right)
+ \frac{p_\perp^4}{4 s^2} 
\left|N_{\pi A_5} \right|^2 
+ \frac{p_\perp^4}{4 s^2} \left|N_{\pi B_5} \right|^2
\right\}.\nonumber
\eqa
Here $\pi$ is either a $\pi^+$ or a $\pi^-$, and the corresponding coefficients
$\tilde{N}_{\pi^+ A}$, $\tilde{N}_{\pi^+ B}$, $N_{\pi^+ A_5}$, $N_{\pi^+ B_5}$, 
and $\tilde{N}_{\pi^- A}$, $\tilde{N}_{\pi^- B}$, $N_{\pi^- A_5}$, $N_{\pi^- 
	B_5}$ are given by eqs.~(\ref{Npi+A}, \ref{Npi+B}, \ref{Npi+A5}, \ref{Npi+B5}) 
and eqs.~(\ref{Npi-A}, \ref{Npi-B}, \ref{Npi-A5}, \ref{Npi-B5}) respectively.

The differential cross-section as a function of $t$, $M^2_{\gamma\pi},$ $-u'$ 
then reads
\begin{equation}
	\label{difcrosec}
	\left.\frac{d\sigma}{dt \,du' \, dM^2_{\gamma\pi}}\right|_{\ -t=(-t)_{ \mathrm{min} }} = 
	\frac{|\mathcal{\overline{M}}_\pi|^2}{32S_{\gamma 
			N}^2M^2_{\gamma\pi}(2\pi)^3}\,.
\end{equation}

\section{Results}

\label{sec:results}

\subsection{Conventions for plots}

\label{sec:conventions-plots}

When showing the results, we will typically include 4 cases, considering 2 models for the DA (asymptotic or holographic), and 2 GPD models (valence or standard scenario). For consistency, the conventions used throughout this section are:
\begin{itemize}
	\item Solid line: asymptotic DA, valence scenario
	\item Dashed line: Holographic DA, valence scenario
	\item Dotted line: asymptotic DA, standard scenario
	\item Dot-dashed line: Holographic DA, standard scenario
\end{itemize}
In other words, being dashed implies the use of the holographic DA, while being dotted implies the use of the standard scenario for the GPD.

\subsection{Description of the numerics}

The GPDs are computed as tables in $ x $, for different $  \xi  $. For the amplitudes, we compute tables at different $ (-u') $ and $ \Msq $, at a particular value of $ \SgN $. We remind the reader that to compute the fully differential cross-section (and hence amplitudes), $ (-t) $ is fixed to its minimum value $ (-t)_{ \mathrm{min} } $, see \eqref{difcrosec}.

In practice, we want to compute the cross-section covering the \textit{full} phase space in the region $20 \GeV^2 <\SgN< 20000 \GeV^2 $, since this covers the full kinematical range of JLab, COMPASS, EIC, and most of the relevant kinematical range for UPCs at LHC, see \SEC\ref{sec:UPC-LHC}. We compute 7 sets of amplitude tables in total:
\begin{itemize}
	\item $ \SgN = 20 \GeV^2  $, $ 1.6 \leq \Msq \leq 10 \GeV^2$ with a uniform step of 0.1~GeV$^2$
	\item $ \SgN = 200 \GeV^2  $, $ 1.6 \leq \Msq \leq 51.4 \GeV^2$ with a uniform step of 0.2~GeV$^2$
	\item $ \SgN = 200 \GeV^2  $, $ 1.6 \leq \Msq \leq 110.5 \GeV^2$ with a uniform step of 1.1~GeV$^2$
	\item $ \SgN = 2000 \GeV^2  $, $ 1.6 \leq \Msq \leq 51.4 \GeV^2$ with a uniform step of 0.2~GeV$^2$
	\item $ \SgN = 2000 \GeV^2  $, $ 1.6 \leq \Msq \leq 1041.1 \GeV^2$ with a uniform step of 10.5~GeV$^2$
	\item $ \SgN = 20000 \GeV^2  $, $ 1.6 \leq \Msq \leq 51.4 \GeV^2$ with a uniform step of 0.2~GeV$^2$
	\item $ \SgN = 20000 \GeV^2  $, $ 1.6 \leq \Msq \leq 10396.6 \GeV^2$ with a uniform step of 105~GeV$^2$
\end{itemize}
The first, third, fifth and seventh sets cover the full range of the phase space, while the second, fourth and sixth sets are needed to resolve the peak in $M_{\gamma \meson}^2$ (importance sampling). For each amplitude table, the whole range of $ (-u') $ is covered. More details regarding the boundaries of the kinematic variables can be found in appendix \ref{app:phase-space}, and in Appendix E of \cite{Duplancic:2018bum}.
At each value of $ \SgN = 200,\,2000,\,20000 \GeV^2 $, two separate datasets were needed, one to cover the whole range of the phase space, and the other to ensure that peaks in the distribution of $ \Msq $ were well-resolved. This is not needed for the $ \SgN=20 \GeV^2 $ case, as the peak is moderate in that case. More details on the importance sampling procedure can be found in \SEC\ref{sec:importance-sampling}.

To obtain the amplitude tables in $ (-u') $ for each of value of $ \Msq $,
\begin{itemize}
	\item 
	we calculate, for each of the above types of GPDs (in the present paper $H$ and 
	$\tilde{H}$), sets of $u$ and $d$ quarks GPDs indexed by $M^2_{\gamma\pi}$, {\it 
		i.e.} ultimately by $\xi$ given by 
	\beqa
	\label{rel-xi-M2-S}
	\xi = \frac{M^2_{\gamma \pi}}{2(S_{\gamma N}-M^2)-M^2_{\gamma \pi}}\,.
	\eqa
	The GPDs are computed as tables of 1000 values for $x$ ranging from $-1$ to 
	$1$, unless importance sampling is needed, in which case 1000 more values around the peak is added, see \SEC\ref{sec:importance-sampling}.
	
	\item we compute the building block integrals which do not depend on 
	$-u'$. In the asymptotic DA case, this corresponds to $ I_e $ (see appendix D in  \cite{Duplancic:2018bum} for the notation), while in the holographic DA case, this corresponds to both $ I_e $ and  $  \chi _{c} $, see appendix \ref{app:holographic-DA}.
	
	\item
	we choose 100 values of  $(-u')$, linearly varying from $(-u')_{ \mathrm{min} }=1~{\rm GeV}^2$ 
	up to its maximum possible value
	$(-u')_{ \mathrm{maxMax} }$ (see Appendix E in \cite{Duplancic:2018bum} for how this is computed). Again, if importance sampling is needed (when the cross-section varies rapidly at the boundaries), an extra set of 100 values of $ (-u') $ is added at each boundary.
	
	\item 
	at each value of $(-u')$, we compute, for each GPD and each flavour $u$ and $d$, the remaining building 
	block integrals, which are $I_b$, $I_c$, $I_h$, $I_i$ in the asymptotic DA case, and only $  \chi _{b} $ in the holographic DA case.
	
	\item 
	this gives, for each of these couples of values of ($M^2_{\gamma \pi}, -u')$
	and each flavour, a set of 12 coefficients listed in 
	equation \eqref{list-N-to-be-computed}.
	
	\item 
	one can then get the desired cross-sections using equations \eqref{all-pi} and \eqref{difcrosec}.
	
\end{itemize}

To obtain corresponding tables at other \textit{lower} values of $ \SgN $, which is needed to span the whole phase space, we use a mapping procedure, which we describe below, from the appropriate set of tables. First, note that the building block integrals only depend on $\alpha$, $\xi$ and on the GPDs (which are computed as grids indexed by $\xi$).
The crucial point to observe is that since $\alpha = -u'/\Msq\,,$
it is possible to use exactly the set of already computed amplitudes, provided one selects the same set of $(\alpha,\xi)$.

Second, one should note that a given value of $ \xi $
corresponds to an infinite set of couples of values
$(\Msq,S_{\gamma N})$, see \eqref{rel-xi-M2-S}.

On the other hand, in practice, we index our amplitude tables by 
$\Msq$ and $-u'$. Thus, by choosing a new value of $\tilde{S}_{\gamma N}$,
we obtain a new set of values of $\tilde{M}^2_{\gamma \meson}$ indexed by the original set of values of $ \Msq $, through the relation
\beqa
\label{set-M2new}
\tilde{M}^2_{\gamma \meson} = \Msq \frac{\tilde{S}_{\gamma N}-M^2}{S_{\gamma N}-M^2}\,,
\eqa
which is deduced from eq.~(\ref{rel-xi-M2-S}). For each of these $\tilde{M}^2_{\gamma \meson}$, a set of values of $-\tilde{u}'\,,$ is obtained using the relation
\beq
\label{set-u'new}
-\tilde{u}'= \frac{\tilde{M}^2_{\gamma \meson}}{M^2_{\gamma \meson}} (-u')\,.
\eq
which gives the indexation of allowed values of
$-\tilde{u}'$ as function of known values of $(-u').$

It is easy to check that this mapping procedure from a given $S_{\gamma N}$ to a lower 
$\tilde{S}_{\gamma N}$ provides a set of 
$(\tilde{M}^2_{\gamma \meson},-\tilde{u}')$ which exhaust the required domain. This has been shown explicitly in \cite{Boussarie:2016qop} already, and we do not reproduce this here.

Thus, from a single set of computation at a fixed $ \SgN $, one can obtain the complete dependence of amplitudes and thus of cross-sections for the whole range of $ \tilde{S} _{\gamma N} < \SgN$. This allows for a significant decrease in computing time, from the order of months to a few days.

\subsubsection{Importance Sampling}

\label{sec:importance-sampling}

Unfortunately, extending the kinematical range from previous papers \cite{Boussarie:2016qop,Duplancic:2018bum} is not as simple as merely changing the maximum $ \SgN $. This is due to the fact that when $ \SgN$ increases, the GPDs and cross-sections vary more rapidly over a smaller range of the variable $  x  $ for GPDs, and $ (-u') $ and $ \Msq $ for cross-sections. To get around this problem, tables had to be generated using \textit{importance sampling}. This needed to be implemented at 3 different levels:
\begin{itemize}
	\item The GPDs can vary rapidly as a function of $ x $ in a range of a few $  \xi  $ from $ x=0 $. As $\SgN$ increases up to 20000 GeV$^2$, $ \xi $ can become as small as $7.5 \times 10^{-6}$.
	\item The fully differential cross-section rises very sharply at the endpoints in $ (-u') $ as the parameters $  \alpha  $ or $\bar \alpha$ (see \eqref{skewness2}) become smaller.
	\item The single differential cross-section \textit{always} has a peak at \textit{low} values of $ \Msq $ (roughly 2-4 GeV$ ^2 $) as $ \SgN $ increases. The origin of this feature is explained in \SEC\ref{sec:sing-diff-X-section-JLab}.
\end{itemize}
Therefore, care needs to be taken during the generation of tables to ensure that there are sufficient data points to cover all the above-mentioned cases. This task is further complicated by the fact that the datasets are generated at fixed $ \SgN $, which are then mapped to lower values of $ \SgN $, as described before. One needs to ensure that the three regions described above still remain adequately covered \textit{after} the mapping. For the 7 sets of amplitude tables computed for this work, we are able to cover all such regions adequately for the whole phase space for $ \SgN \leq 20000 \GeV^2 $.

We note that importance sampling is not required when simulating tables for JLab kinematics, as the centre of mass energy $ \SgN $ is \textit{not} much larger than the imposed kinematical cuts, which are necessary for QCD collinear factorisation.

\subsection{JLab Kinematics}

At JLab, the electron beam hits a fixed target consisting of protons and neutrons, at an energy of $ 12\,\GeV $. The electron-nucleon centre-of-mass energy, $ S_{eN} $, is thus roughly 23 $ \GeV^2 $. Therefore, for most of the plots in this section, we use $ S_{\gamma N}=20\,\GeV^2 $ as a representative value for JLab kinematics. This allows us to probe GPDs for the range of skewnesses of $0.04 \leq \xi \leq 0.33$.

At this point, we would like to point out that a programming mistake, related to the sign of the interference term in the squared amplitude, c.f.~\eqref{all-pi}, was made in the previous publication~\cite{Duplancic:2018bum}. Thus, the plots that we produce here are slightly different.

\subsubsection{Fully differential cross-section}

\label{sec:jlab-fully-diff-X-section}

\begin{figure}[h!]
	\psfrag{HHH}{\hspace{-1.5cm}\raisebox{-.6cm}{\scalebox{.8}{$-u' ({\rm 
					GeV}^{2})$}}}
	\psfrag{VVV}{\raisebox{.3cm}{\scalebox{.9}{$\hspace{-.4cm}\displaystyle\left.\frac{d 
					\sigma_{\gamma\pi^+}}{d M^2_{\gamma \pi^+} d(-u') d(-t)}\right|_{(-t)_{\rm min}}({\rm pb} \cdot {\rm GeV}^{-6})$}}}
	\psfrag{TTT}{}
	\vspace{0.2cm}
	\centerline{
		{\includegraphics[width=18pc]{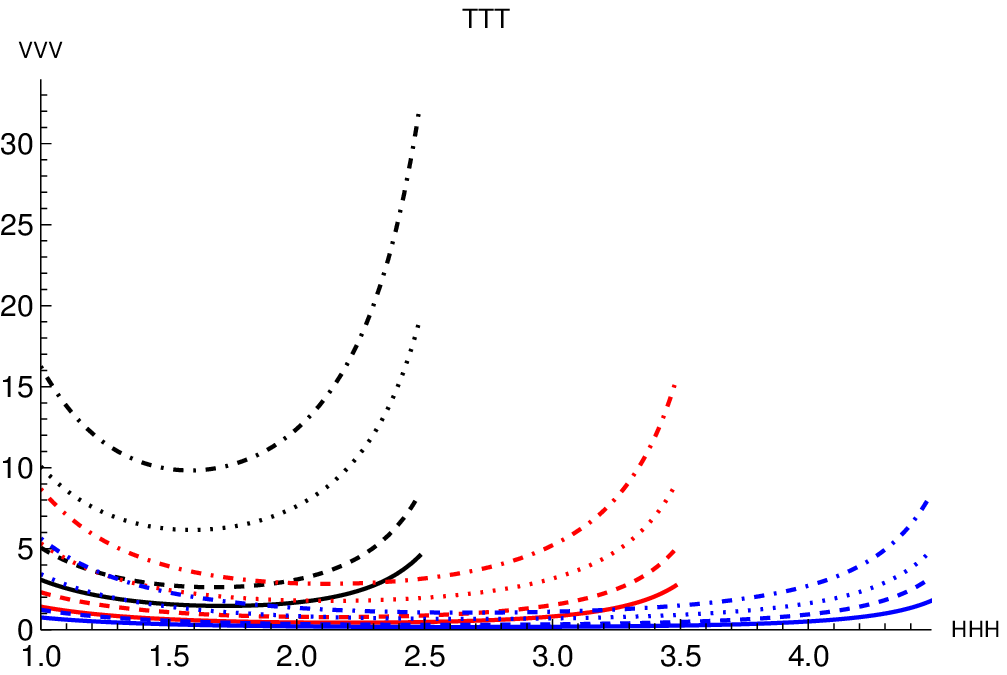}}
		\psfrag{VVV}{\raisebox{.3cm}{\scalebox{.9}{$\hspace{-.4cm}\displaystyle\left.\frac{d 
						\sigma_{\gamma\pi^-}}{d M^2_{\gamma \pi^-} d(-u') d(-t)}\right|_{(-t)_{\rm min}}({\rm pb} \cdot {\rm GeV}^{-6})$}}}
		{\includegraphics[width=18pc]{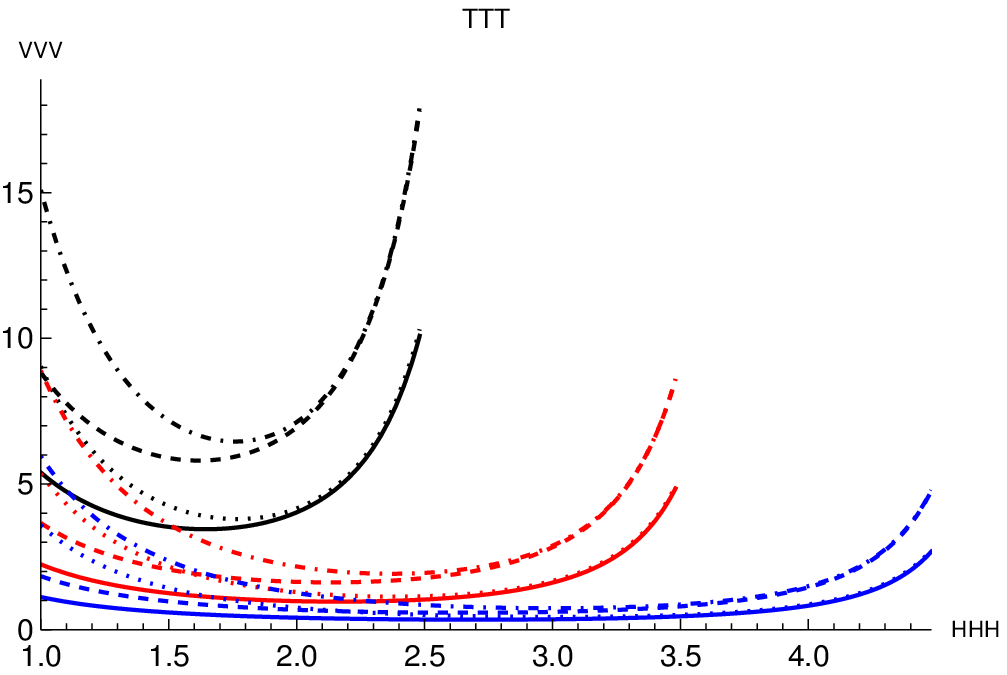}}}
	\vspace{0.2cm}
	\caption{\small The fully differential cross-section for $ \pi^{+} $ ($ \pi^{-} $) is shown as a function of $  \left( -u' \right)  $ on the left (right) for different values of $ M_{\gamma \meson}^2 $. The black, red and blue curves correspond to $ M_{\gamma \meson}^{2}=3,\,4,\,5\, $ GeV$ ^2 $ respectively. The dashed (non-dashed) lines correspond to holographic (asymptotic) DA, while the dotted (non-dotted) lines correspond to the standard (valence) scenario. $ S_{\gamma N} $ is fixed at 20 GeV$ ^2 $.}
	\label{fig:jlab-fully-diff-diff-M2}
\end{figure}

The effect of different values of $ M_{\gamma \meson}^2 $ on the cross-section  is shown in Figure \ref{fig:jlab-fully-diff-diff-M2}. The values chosen for $ M_{\gamma \meson}^2 $ are 3, 4 and 5 GeV$ ^2 $. As $ M_{\gamma \meson}^2 $ grows, the range of allowed $ (-u') $ values increases. On the other hand, the value of the cross-section itself decreases. When integrating over $(-u')$, these two competing effects will become clearer later when we show the single differential plots in \SEC\ref{sec:sing-diff-X-section-JLab} as a function of $ M_{\gamma \meson}^2 $, leading to a peak in the distribution at low values of $ M_{\gamma \meson}^2 $. In general, the GPD model corresponding to the standard scenario leads to a larger value for the cross-section. In the case of $\pi^+$, the choice of the GPD model leads to a significant difference in the value of the cross-section, whereas in the $\pi^-$ case, this happens only at low $(-u')$. When the integration over $(-u')$ is performed, this effect can be seen by the larger difference due to the choice of the GPD model in the $\pi^+$ compared to the $\pi^-$ case, see Figure \ref{fig:jlab-sing-diff}. Finally, we note that using a holographic DA gives a higher cross-section than using an asymptotic DA.

\begin{figure}[h!]
	\psfrag{HHH}{\hspace{-1.5cm}\raisebox{-.6cm}{\scalebox{.8}{$-u' ({\rm 
					GeV}^{2})$}}}
	\psfrag{VVV}{\raisebox{.3cm}{\scalebox{.9}{$\hspace{-.4cm}\displaystyle\left.\frac{d 
					\sigma_{\gamma\pi^+}}{d M^2_{\gamma \pi^+} d(-u') d(-t)}\right|_{(-t)_{\rm min}}({\rm pb} \cdot {\rm GeV}^{-6})$}}}
	\psfrag{TTT}{}
	\vspace{0.2cm}
	\centerline{
			{\includegraphics[width=18pc]{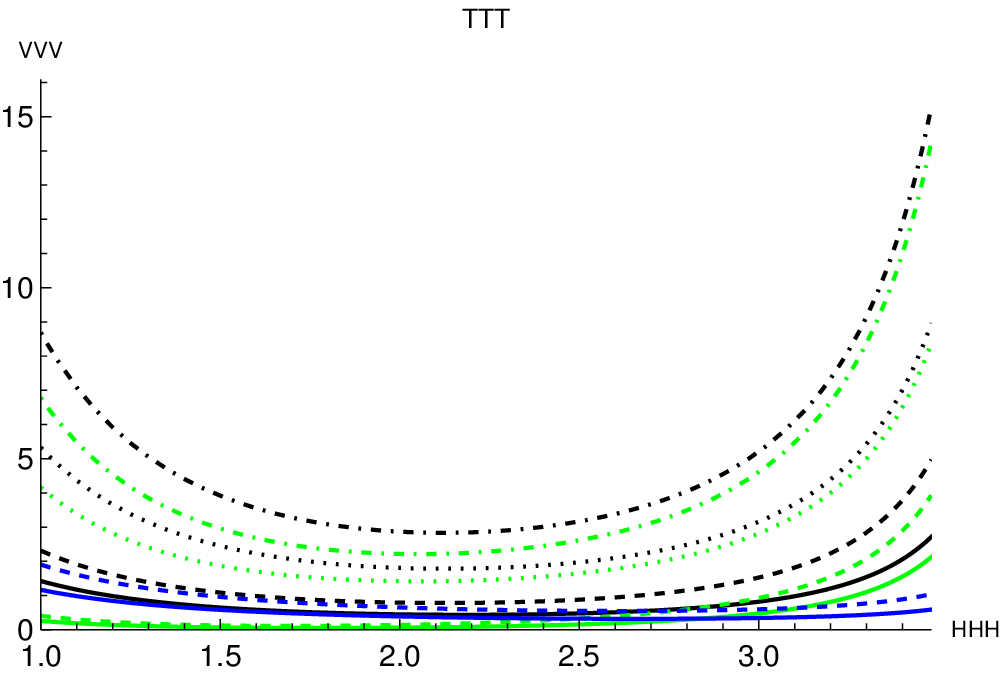}}
		\psfrag{VVV}{\raisebox{.3cm}{\scalebox{.9}{$\hspace{-.4cm}\displaystyle\left.\frac{d 
						\sigma_{\gamma\pi^-}}{d M^2_{\gamma \pi^-} d(-u') d(-t)}\right|_{(-t)_{\rm min}}({\rm pb} \cdot {\rm GeV}^{-6})$}}}
		{\includegraphics[width=18pc]{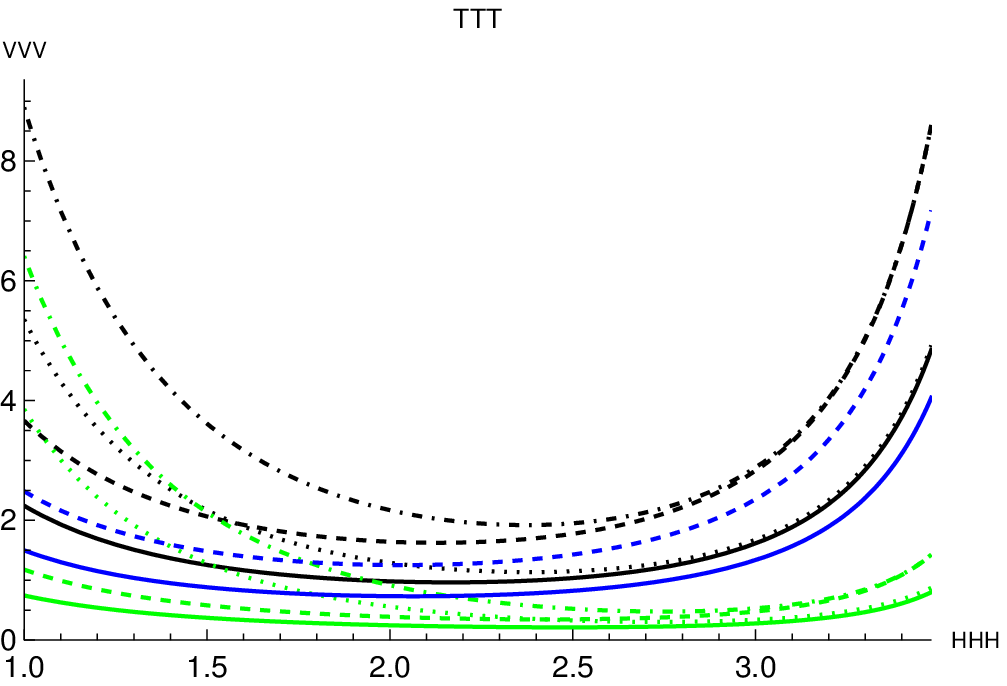}}}
	\vspace{0.2cm}
	\caption{\small The fully differential cross-section for $ \pi^{+} $ ($ \pi^{-} $) is shown as a function of $  \left( -u' \right)  $ on the left (right). The blue and green curves correspond to contributions from the vector and axial GPDs respectively. The black curves correspond to the total contribution, i.e. vector and axial GPD contributions combined. As before, the dashed (non-dashed) lines correspond to holographic (asymptotic) DA, while the dotted (non-dotted) lines correspond to the standard (valence) scenario. We fix $ S_{\gamma N}= 20\,  \mathrm{GeV}^{2}  $ and $ M_{\gamma \meson}^{2}= 4\,  \mathrm{GeV}^{2}  $. Note that the vector contributions consist of only two curves in each case, since they are insensitive to either valence or standard scenarios. Another interesting observation is that the total contribution (black) is given simply by the sum of the vector (blue) and axial (green) GPD contributions, since there is no interference between them.}
	\label{fig:jlab-fully-diff-VandA}
\end{figure}

The relative contributions of the vector and axial GPDs to the cross-section are shown in Figure \ref{fig:jlab-fully-diff-VandA}. The kinematical variables chosen for the plots are $ \SgN = 20 \GeV^2 $ and $ \Msq = 4 \GeV^2 $. The first point to note is that the vector contribution does not depend on the valence or standard scenarios, since they only enter the modelling of the axial GPDs. Hence, only two blue curves appear on each plot in the figure, corresponding to the DA model. Moreover, we note that the total contribution (black curve) corresponds simply to the sum of the vector (blue) and axial (green) contributions, since there is no interference between them, see \eqref{squareCEresult}.

\begin{figure}[h!]
	\psfrag{HHH}{\hspace{-1.5cm}\raisebox{-.6cm}{\scalebox{.8}{$-u' ({\rm 
					GeV}^{2})$}}}
	\psfrag{VVV}{\raisebox{.3cm}{\scalebox{.9}{$\hspace{-.4cm}\displaystyle\left.\frac{d 
					\sigma_{\gamma\pi^+}}{d M^2_{\gamma \pi^+} d(-u') d(-t)}\right|_{(-t)_{\rm min}}({\rm pb} \cdot {\rm GeV}^{-6})$}}}
	\psfrag{TTT}{}
	\vspace{0.2cm}
	\centerline{
		{\includegraphics[width=18pc]{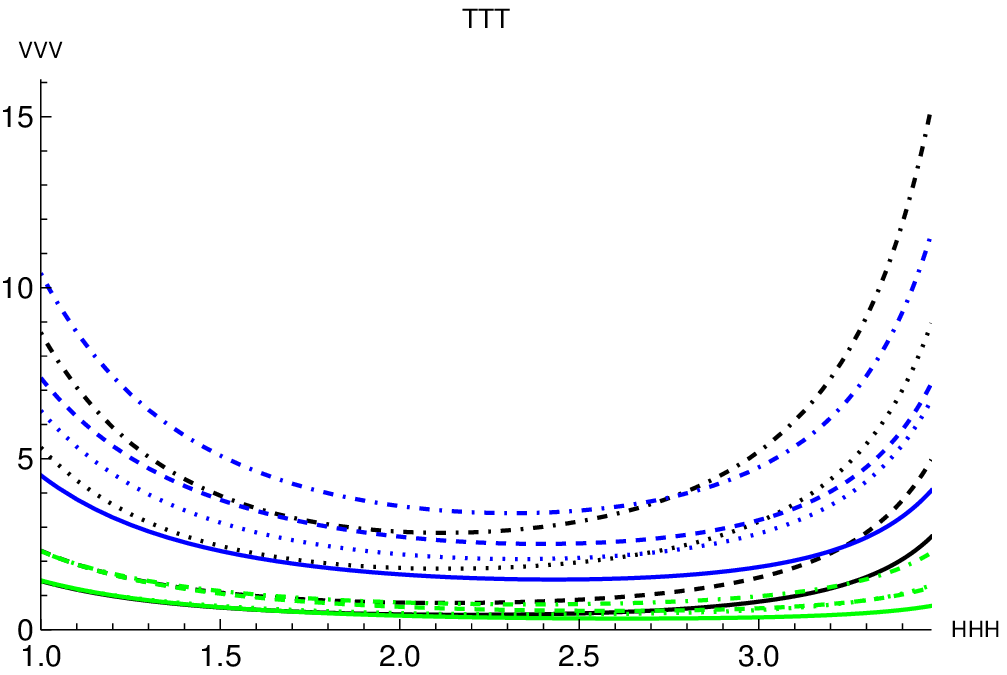}}
		\psfrag{VVV}{\raisebox{.3cm}{\scalebox{.9}{$\hspace{-.4cm}\displaystyle\left.\frac{d 
						\sigma_{\gamma\pi^-}}{d M^2_{\gamma \pi^-} d(-u') d(-t)}\right|_{(-t)_{\rm min}}({\rm pb} \cdot {\rm GeV}^{-6})$}}}
		{\includegraphics[width=18pc]{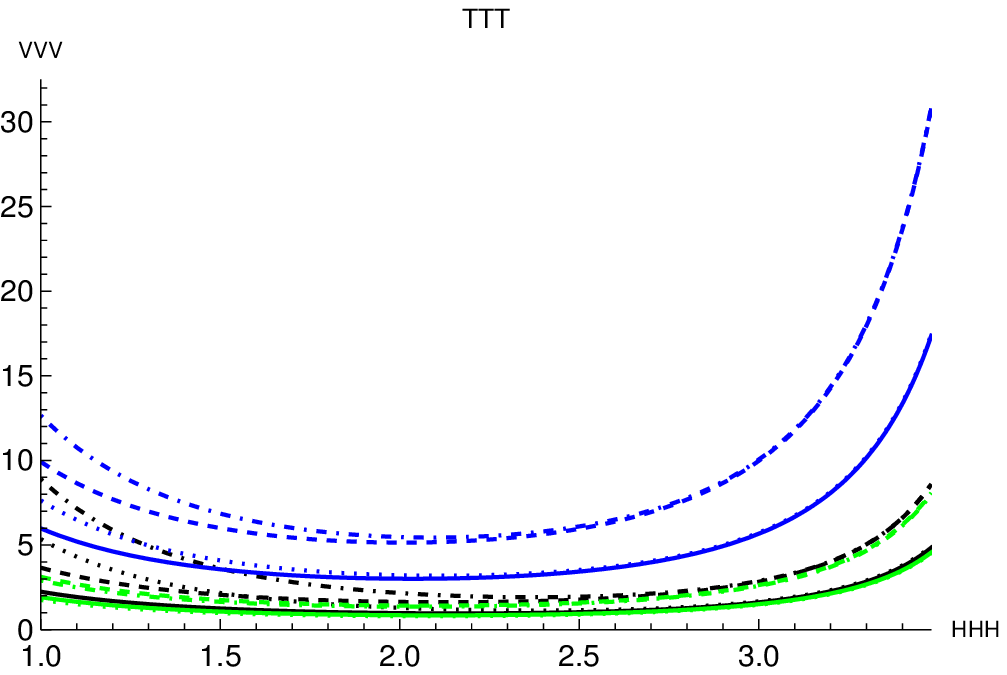}}}
	\vspace{0.2cm}
	\caption{\small The fully differential cross-section for $ \pi^{+} $ ($ \pi^{-} $) is shown as a function of $  \left( -u' \right)  $ on the left (right). The blue and green curves correspond to contributions from the u quark ($ H_{u} $ and $  \tilde{H} _{u} $) and d quark ($ H_{d} $ and $  \tilde{H} _{d} $) GPDs respectively. The black curves correspond to the total contribution. As before, the dashed (non-dashed) lines correspond to holographic (asymptotic) DA, while the dotted (non-dotted) lines correspond to the standard (valence) scenario. We fix $ S_{\gamma N}= 20\,  \mathrm{GeV}^{2}  $ and $ M_{\gamma \meson}^{2}= 4\,  \mathrm{GeV}^{2}  $. Note that the sum of u quark (blue) and d quark (green) GPD contributions to the cross-section do \textit{not} give the total cross-section (black), as there are important interference terms present, which can even be negative.}
	\label{fig:jlab-fully-diff-uandd}
\end{figure}

To conclude this subsection, the relative contributions of the u and d quark GPDs to the cross-section are shown in Figure \ref{fig:jlab-fully-diff-uandd}. To generate the plots, $ \SgN = 20 \GeV^2 $ and $ \Msq = 4 \GeV^2 $ were used. Here, unlike in Figure \ref{fig:jlab-fully-diff-VandA}, there are important interference terms between the u quark and d quark contributions, and therefore, the total contribution (black) is \textit{not} simply a sum of the individual quark GPD contributions.

\subsubsection{Single differential cross-section}

\label{sec:sing-diff-X-section-JLab}

\begin{figure}[h!]
	\psfrag{HHH}{\hspace{-1.5cm}\raisebox{-.6cm}{\scalebox{.8}{$ M_{\gamma \meson}^2 ({\rm 
					GeV}^{2})$}}}
	\psfrag{VVV}{\raisebox{.3cm}{\scalebox{.9}{$\hspace{-.4cm}\displaystyle\frac{d 
					\sigma_{\gamma\pi^+}}{d M^2_{\gamma \pi^+}}({\rm pb} \cdot {\rm GeV}^{-2})$}}}
	\psfrag{TTT}{}
	\vspace{0.2cm}
	\centerline{
		{\includegraphics[width=18pc]{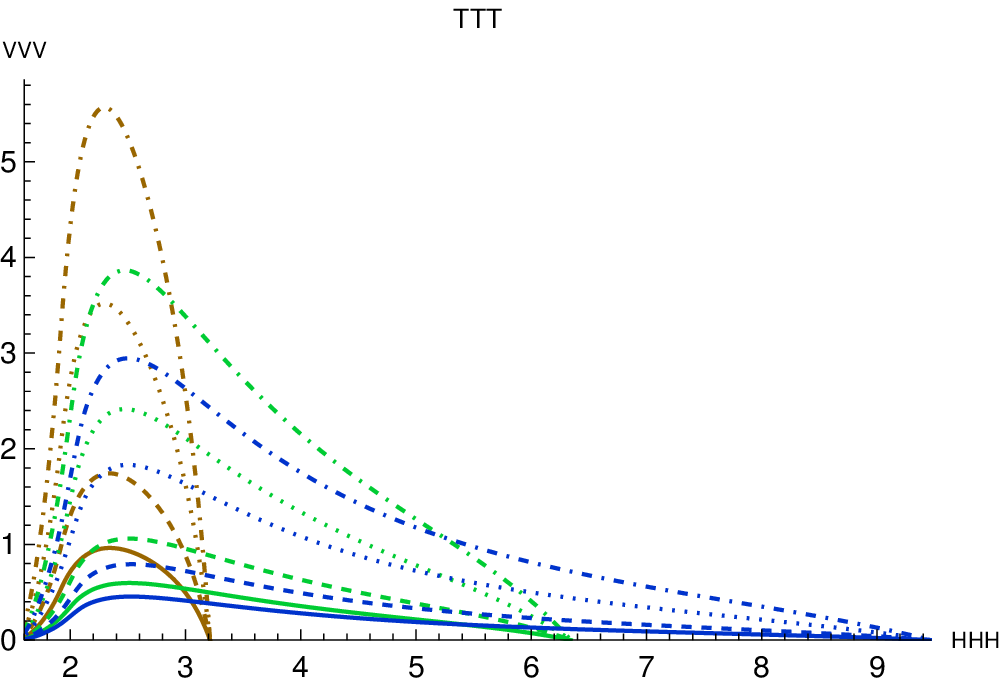}}
		\psfrag{VVV}{\raisebox{.3cm}{\scalebox{.9}{$\hspace{-.4cm}\displaystyle\frac{d 
						\sigma_{\gamma\pi^-}}{d M^2_{\gamma \pi^-}}({\rm pb} \cdot {\rm GeV}^{-2})$}}}
		{\includegraphics[width=18pc]{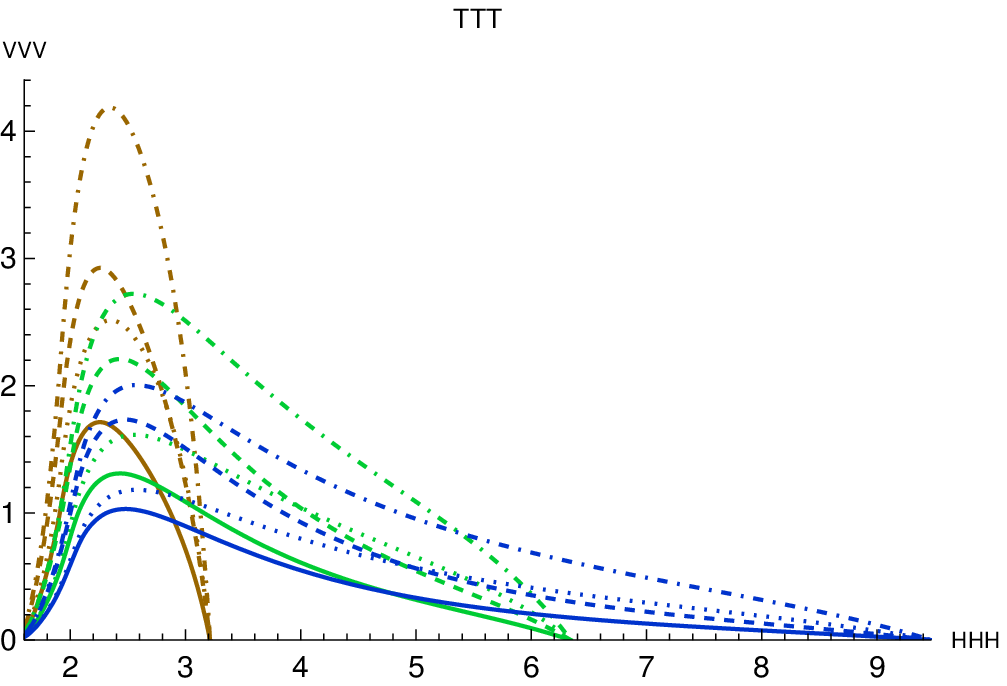}}}
	\vspace{0.2cm}
	\caption{\small The single differential cross-section for $ \pi^{+} $ ($ \pi^{-} $) is shown as a function of $  M_{\gamma \meson}^{2}  $ on the left (right) for different values of $ S_{\gamma N} $. The brown, green and blue curves correspond to $ S_{\gamma N} = 8,\,14,\,20\,\GeV^{2} $. The dashed (non-dashed) lines correspond to holographic (asymptotic) DA, while the dotted (non-dotted) lines correspond to the standard (valence) scenario.}
	\label{fig:jlab-sing-diff}
\end{figure}

Integrating over the kinematical variables $ (-u') $ and $ (-t) $ leads to single differential cross-section (as a function of $ \Msq $). The details of this integration can be found in Appendix \ref{app:phase-space}, and in Appendix E of \cite{Duplancic:2018bum}. We note that the ansatz used for the $t$-dependence of the cross-section has been modified in this work, compared to the previous papers \cite{Boussarie:2016qop,Duplancic:2018bum}, leading to slightly different values for the cross-sections.  The effect of different values of $ S_{\gamma N} $ on the single differential cross-section is shown in Figure \ref{fig:jlab-sing-diff}. The different colours, brown, green and blue, correspond to $ \SgN $ values of 8, 14 and 20 GeV$ ^2 $ respectively. As $ \SgN $ increases, the maximum value of $ \Msq $ increases (simply due to the increase in the phase space), while the value of the cross-section decreases.\footnote{A similar effect was observed in Figure \ref{fig:jlab-fully-diff-diff-M2} with increasing $M_{\gamma \meson}^2$, instead of $\SgN$.}

As previously mentioned, the peak in the plots in Figure \ref{fig:jlab-sing-diff} is the consequence of the competition between the decrease in the cross-section and the increase in the volume of the phase space as $M_{\gamma \meson}^2$ increases. An interesting point to note is that the peak of the distribution is always found at \textit{low} $ \Msq $, around $ 3 \GeV^2 $. The reason for this is that the cross-section grows rapidly as $M_{\gamma \meson}^2$ decreases, but at the same time, the kinematical cuts that we impose to use collinear QCD factorisation causes the volume of the phase space to vanish at a minimum value of $M_{\gamma \meson}^2$ of about 1.6 GeV$^2$.

\subsubsection{Integrated cross-section}

\label{sec:int-X-section-JLab}

\begin{figure}[h!]
	\psfrag{HHH}{\hspace{-1.5cm}\raisebox{-.6cm}{\scalebox{.8}{$ S_{\gamma N} ({\rm 
					GeV}^{2})$}}}
	\psfrag{VVV}{\raisebox{.3cm}{\scalebox{.9}{$\hspace{-.4cm}\displaystyle
					\sigma_{\gamma\pi^+}({\rm pb})$}}}
	\psfrag{TTT}{}
	\vspace{0.2cm}
	\centerline{
		{\includegraphics[width=18pc]{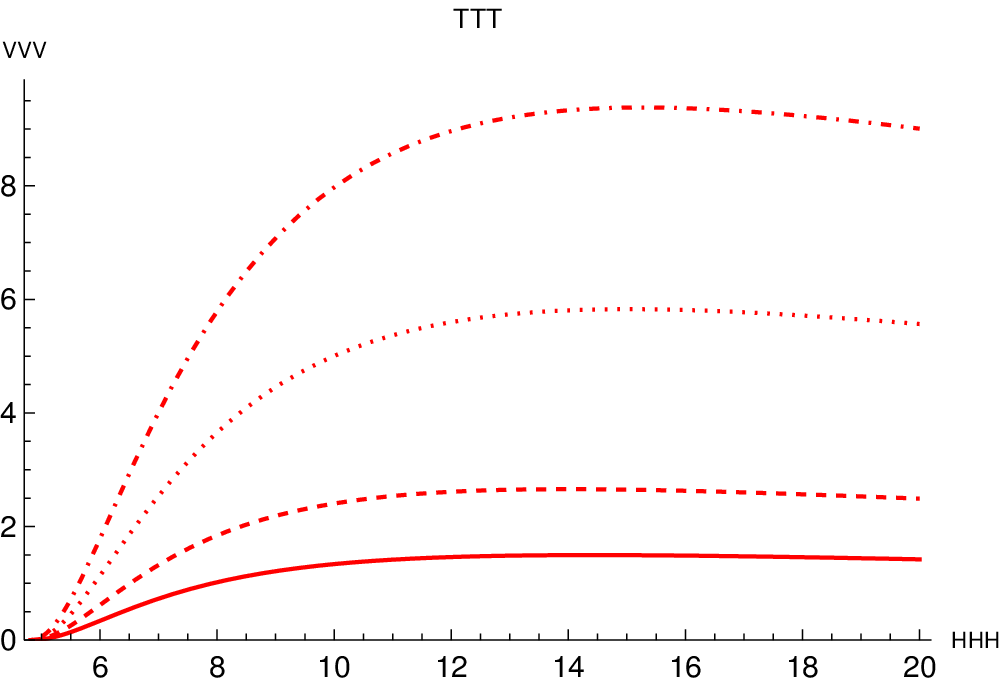}}
		\psfrag{VVV}{\raisebox{.3cm}{\scalebox{.9}{$\hspace{-.4cm}\displaystyle
					\sigma_{\gamma\pi^-}({\rm pb})$}}}
		{\includegraphics[width=18pc]{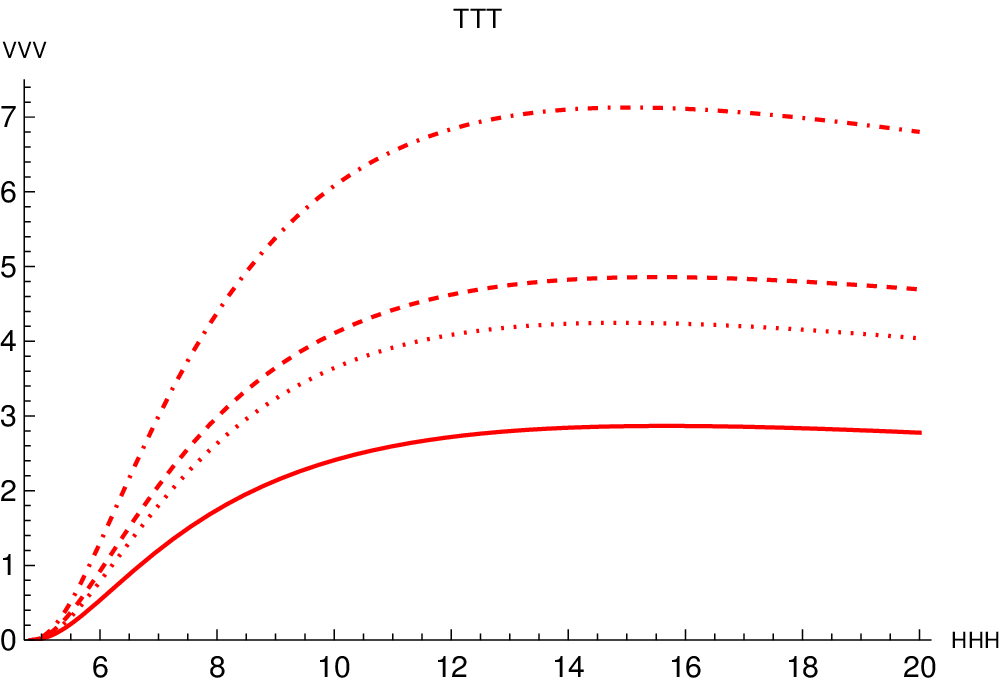}}}
	\vspace{0.2cm}
	\caption{\small The integrated cross-section for $ \pi^{+} $ ($ \pi^{-} $) is shown as a function of $   S_{\gamma N}  $ on the left (right). The dashed (non-dashed) lines correspond to holographic (asymptotic) DA, while the dotted (non-dotted) lines correspond to the standard (valence) scenario.}
	\label{fig:jlab-int-sigma}
\end{figure}

In this subsection, we discuss the variation of the cross-section as a function of $ \SgN $, after integration over $ (-u') $, $ (-t) $ and $ \Msq $. The details of the integration can be found in Appendix \ref{app:phase-space}, and in Appendix E of \cite{Duplancic:2018bum}. The variation of the cross-section as a function of $ S_{\gamma N} $ is shown in Figure \ref{fig:jlab-int-sigma}. One thus finds that using the valence or standard scenarios for modelling the axial GPDs has a much greater effect on $  \pi ^{+} $ than on $  \pi ^{-} $. On the other hand, in both cases, the holographic DA model gives a cross-section that is roughly twice that of the asymptotic DA case.

\FloatBarrier

\subsubsection{Polarisation asymmetries}

\label{sec:pol-asym-jlab}

In this section, we will show the results for the polarisation asymmetries of the incoming photon beam, in the JLab kinematics. As discussed in Appendix \ref{app:polarisation-asymmetries}, the circular polarisation vanishes as a result of conservation of parity $ P $ for an unpolarised target, which is the case we consider here\footnote{The circular double spin asymmetry does not vanish and may be an interesting observable for a polarised target experiment.}. Therefore, we compute the \textit{linear} polarisation asymmetries (LPAs) wrt the incoming photon. The basic formula for constructing the asymmetry is
\begin{align}
	\label{eq:LPA}
 \mathrm{LPA} =\frac{	\int d \sigma _{x}-\int d \sigma _{y}}{\int d \sigma _{x}+\int d \sigma _{y}}\,,
\end{align}
where $ d \sigma _{x (y)} $ corresponds to the differential cross-section with the incoming photon linearly polarised along the $ x (y) $-direction. The $ x $-direction is taken to be along $ p_{\perp} $, i.e.
\begin{align}
	\varepsilon_{x}^{ \mu }\equiv \frac{p_{\perp}^{ \mu }}{| \vec{p}_{t}| }\,.
\end{align}
In other words, the direction of the $ x $-axis is taken to be (almost) aligned to the direction of the outgoing photon in the transverse plane (since $ | \vec{p}_{t} | \gg |  \vec{\Delta}  _{t} |$), see \eqref{eq:momentum-outgoing-photon}.

The integral symbol in \EQ\eqref{eq:LPA} corresponds to phase space integration - The LPA can thus be calculated at the fully differential (by dropping the integral altogether), single differential or integrated levels. More details regarding the expressions and their derivation can be found in \APP\ref{app:polarisation-asymmetries}.

\begin{figure}[h!]
	\psfrag{HHH}{\hspace{-1.5cm}\raisebox{-.6cm}{\scalebox{.8}{$-u' ({\rm 
					GeV}^{2})$}}}
	\psfrag{VVV}{LPA}
	\psfrag{TTT}{}
	\vspace{0.2cm}
	\centerline{
		{\includegraphics[width=18pc]{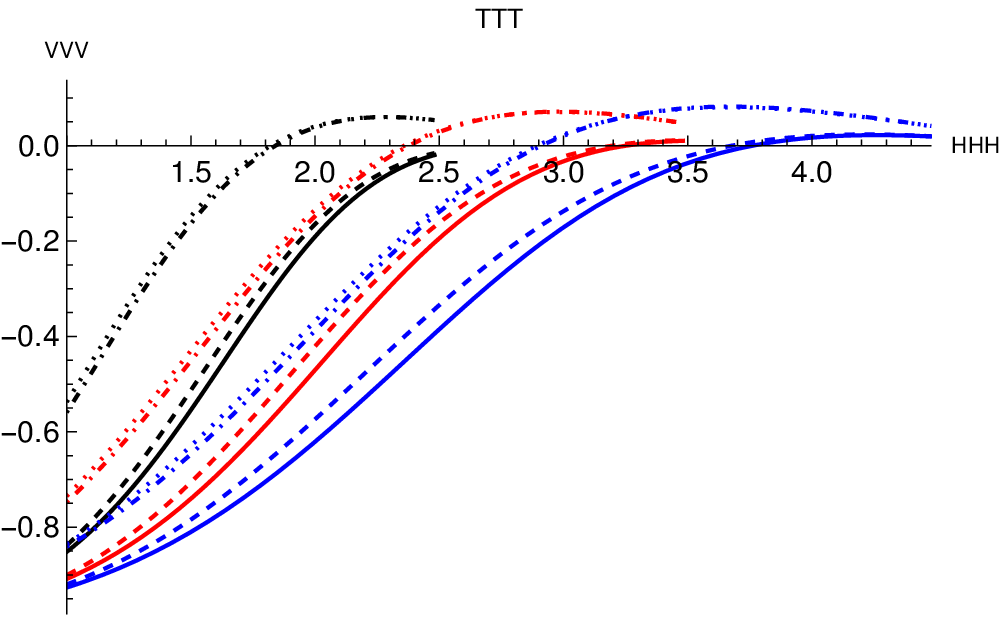}}
		\psfrag{VVV}{LPA}
			\psfrag{HHH}{\hspace{-1.5cm}\raisebox{0.4cm}{\scalebox{.8}{$-u' ({\rm 					GeV}^{2})$}}}
		{\includegraphics[width=18pc]{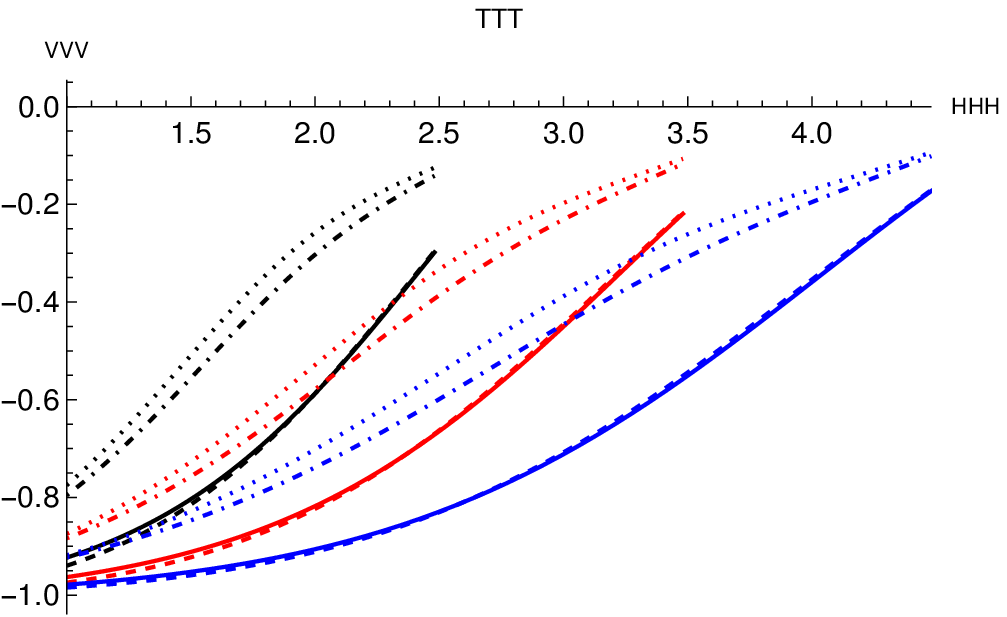}}}
	\vspace{0.2cm}
	\caption{\small The LPA at the fully-differential level for $ \pi^{+} $ ($ \pi^{-} $) is shown as a function of $  \left( -u' \right)  $ on the left (right) for different values of $ M_{\gamma \meson}^2 $. The black, red and blue curves correspond to $ M_{\gamma \meson}^{2}=3,\,4,\,5\, $ GeV$ ^2 $ respectively, and $ \SgN = 20 \GeV^{2}$. The dashed (non-dashed) lines correspond to holographic (asymptotic) DA, while the dotted (non-dotted) lines correspond to the standard (valence) scenario.}
	\label{fig:jlab-pol-asym-fully-diff-diff-M2}
\end{figure}

First, we show the effect of different $ M_{\gamma \meson}^2 $ on the LPAs at the fully differential level (i.e. differential in $ (-u') $, $ M_{\gamma \meson}^2 $ and $ (-t) $ as in \SEC\ref{sec:jlab-fully-diff-X-section}) in Figure \ref{fig:jlab-pol-asym-fully-diff-diff-M2}. As in Figure \ref{fig:jlab-fully-diff-diff-M2}, the values of $ \Msq $ used are 3, 4 and 5 GeV$ ^2 $. One thus finds that the process is dominated by incoming linearly polarised photons along the $ y $-direction, since the LPA is in general negative. Another interesting feature of the LPAs is that they are very good for distinguishing between the standard and valence scenarios for the axial GPDs, but they are relatively insensitive to the DA model being used. This is in contrast to the cross-sections themselves, for which changing the GPD model only leads to moderate differences in the values of the cross-sections.

\begin{figure}[h!]
	\psfrag{HHH}{\hspace{-1.5cm}\raisebox{0.6cm}{\scalebox{.8}{$-u' ({\rm 
					GeV}^{2})$}}}
	\psfrag{VVV}{LPA}
	\psfrag{TTT}{}
	\vspace{0.2cm}
	\centerline{
		{\includegraphics[width=18pc]{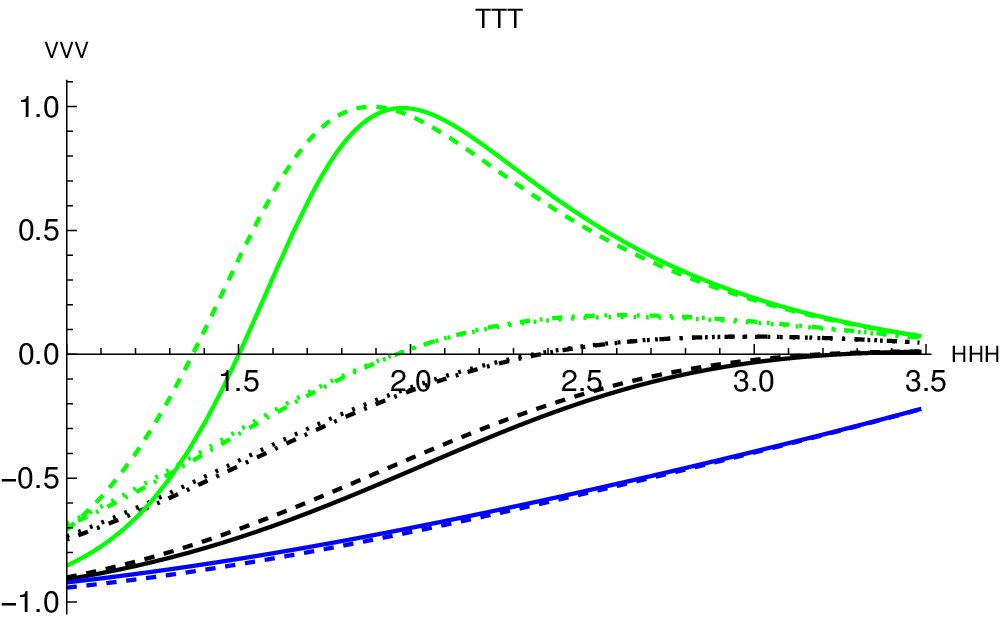}}
		\psfrag{VVV}{LPA}
			\psfrag{HHH}{\hspace{-1.5cm}\raisebox{0.4cm}{\scalebox{.8}{$-u' ({\rm 					GeV}^{2})$}}}
		{\includegraphics[width=18pc]{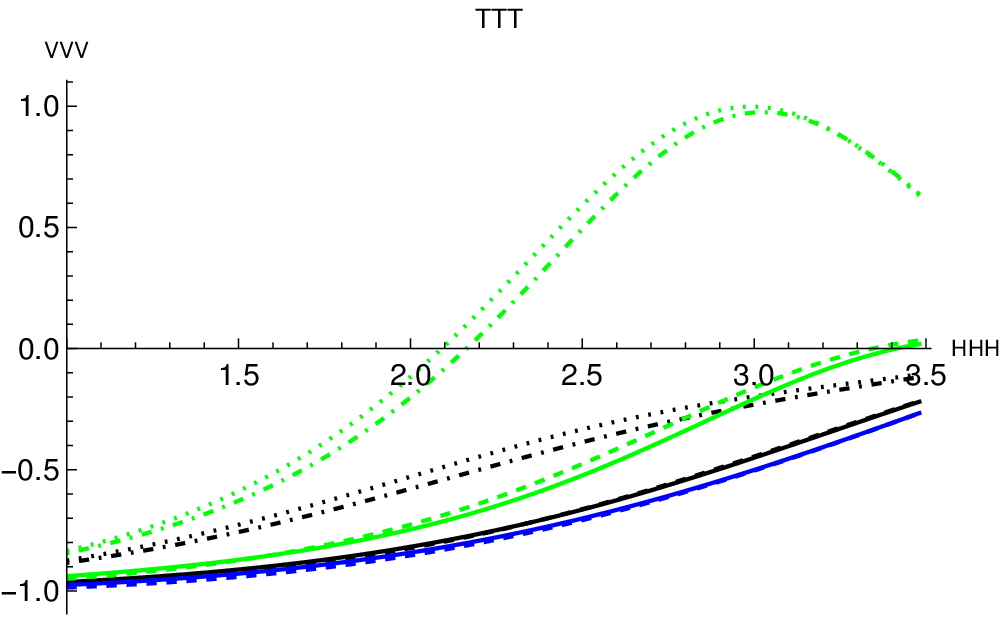}}}
	\vspace{0.2cm}
	\caption{\small The LPA at the fully-differential level for for $ \pi^{+} $ ($ \pi^{-} $) is shown as a function of $  \left( -u' \right)  $ on the left (right), using $ \Msq = 4 \GeV^2$ and $ \SgN = 20 \GeV^2 $. The blue and green curves correspond to contributions from the vector and axial GPDs respectively. The black curves correspond to the total contribution, i.e. vector and axial GPD contributions combined. As before, the dashed (non-dashed) lines correspond to holographic (asymptotic) DA, while the dotted (non-dotted) lines correspond to the standard (valence) scenario. Note that the vector contributions consist of only two curves in each case, since they are insensitive to either valence or standard scenarios.}
	\label{fig:jlab-pol-asym-fully-diff-VandA}
\end{figure}

Next, we show how the relative contributions from the vector and axial GPDs affect the LPA at the fully differential level in Figure \ref{fig:jlab-pol-asym-fully-diff-VandA}. Thus, we find that the axial contribution to the LPA (in green) changes dramatically depending on the axial GPD model used (standard vs valence scenarios), and it still leaves an imprint in the total contribution (in black). In the valence scenario, the contribution to the cross-section from the axial GPDs is very small, see \FIG\ref{fig:jlab-fully-diff-VandA}. Thus, even though the axial contribution has a very different LPA from the vector contribution, the total LPA remains closer to the one from the vector contribution only.

\begin{figure}[h!]
	\psfrag{HHH}{\hspace{-1.5cm}\raisebox{0.4cm}{\scalebox{.8}{$-u' ({\rm 
					GeV}^{2})$}}}
	\psfrag{VVV}{LPA}
	\psfrag{TTT}{}
	\vspace{0.2cm}
	\centerline{
		{\includegraphics[width=18pc]{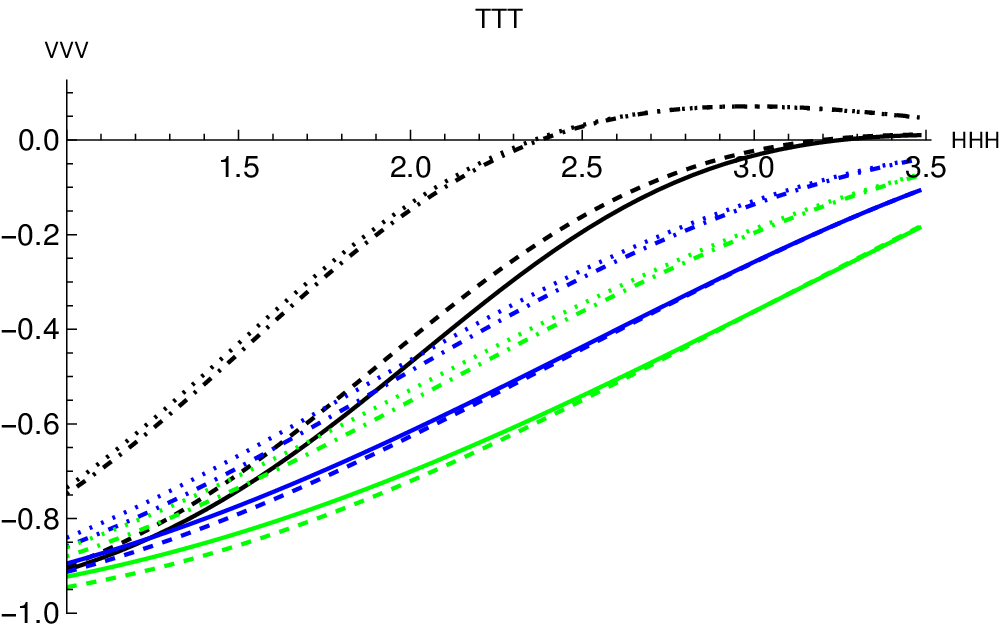}}
		\psfrag{VVV}{LPA}
		{\includegraphics[width=18pc]{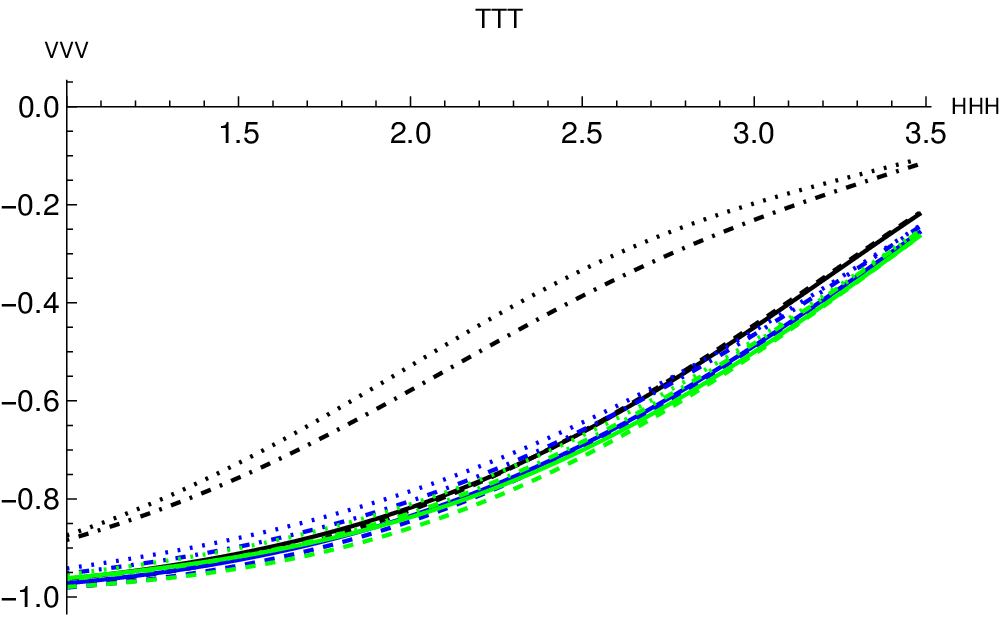}}}
	\vspace{0.2cm}
	\caption{\small The LPA at the fully-differential level for $ \pi^{+} $ ($ \pi^{-} $) is shown as a function of $  \left( -u' \right)  $ on the left (right), using $ \Msq = 4 \GeV^2$ and $ \SgN = 20 \GeV^2 $. The blue and green curves correspond to contributions from the u quark ($ H_{u} $ and $  \tilde{H} _{u} $) and d quark ($ H_{d} $ and $  \tilde{H} _{d} $) GPDs respectively. The black curves correspond to the total contribution. As before, the dashed (non-dashed) lines correspond to holographic (asymptotic) DA, while the dotted (non-dotted) lines correspond to the standard (valence) scenario.}
	\label{fig:jlab-pol-asym-fully-diff-uandd}
\end{figure}

The relative contributions from the u quark GPDs ($ H_{u} $ and $  \tilde{H} _{u} $) and d quark GPDs ($ H_{d} $ and $  \tilde{H} _{d} $) to the LPA are shown in Figure \ref{fig:jlab-pol-asym-fully-diff-uandd}.

\begin{figure}[h!]
	\psfrag{HHH}{\hspace{-1.5cm}\raisebox{-.6cm}{\scalebox{.8}{ $ M_{\gamma \meson}^{2}({\rm 
					GeV}^{2}) $}}}
	\psfrag{VVV}{LPA}
	\psfrag{TTT}{}
	\vspace{0.2cm}
	\centerline{
		{\includegraphics[width=18pc]{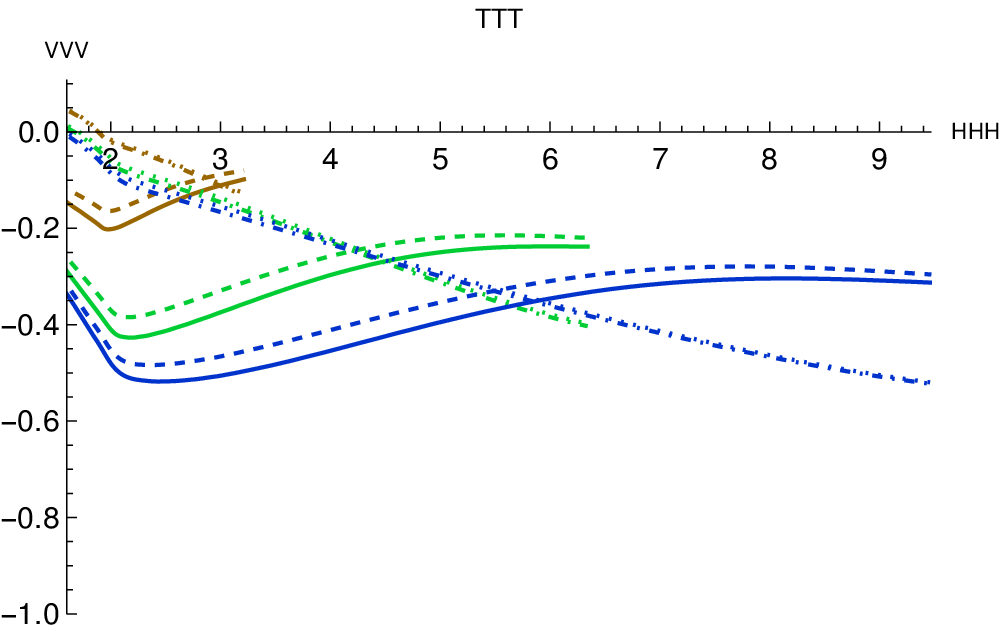}}
		\psfrag{VVV}{LPA}
		{\includegraphics[width=18pc]{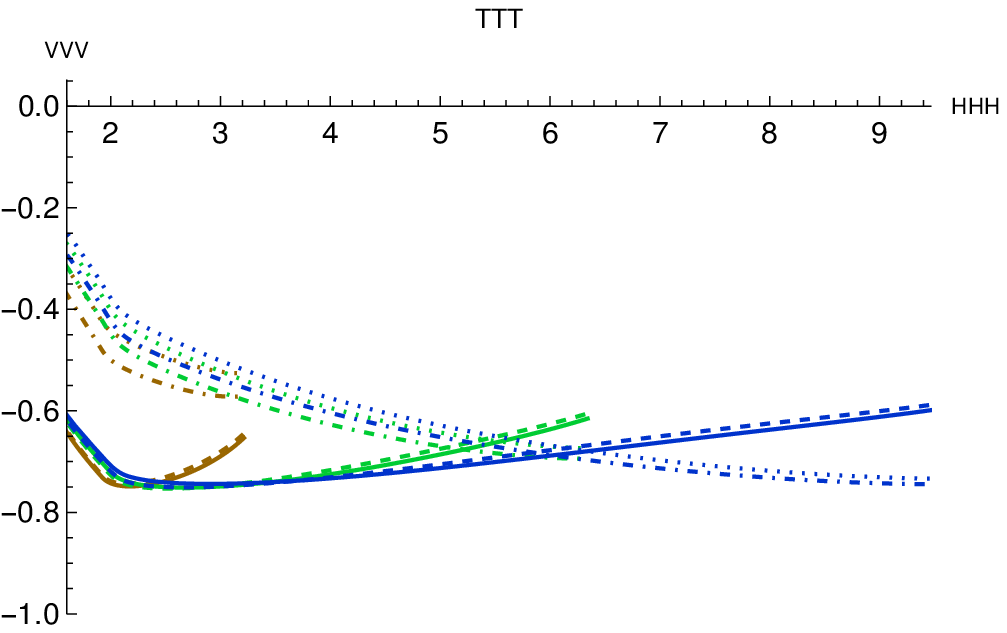}}}
	\vspace{0.2cm}
	\caption{\small The LPA at the single differential level for for $ \pi^{+} $ ($ \pi^{-} $) is shown as a function of $  M_{\gamma \meson}^{2}  $ on the left (right). The brown, green and blue curves correspond to $ S_{\gamma N} = 8,\,14,\,20\,\GeV^{2} $. The dashed (non-dashed) lines correspond to holographic (asymptotic) DA, while the dotted (non-dotted) lines correspond to the standard (valence) scenario. The same colour and line style conventions as in Figure \ref{fig:jlab-sing-diff} are used here.}
	\label{fig:jlab-pol-asym-sing-diff}
\end{figure}

Next, we show the LPA, at the single differential level, for different values of $ S_{\gamma N} $ in Figure \ref{fig:jlab-pol-asym-sing-diff}. As for the cross-section plots in \SEC\ref{sec:sing-diff-X-section-JLab}, the values of $ \SgN $ used are 8, 14 and 20 GeV$ ^2 $. We note that the type of DA used has a very small effect on the LPA. On the other hand, the models used for the axial GPDs (valence or standard) changes the shape of the curves altogether, for both $  \pi ^{+} $ and $  \pi ^{-} $ mesons. This gives hope that the LPA for $\pi^+$ and $\pi^-$ can be used for distinguishing between the 2 GPD models that we consider here.

\begin{figure}[h!]
	\psfrag{HHH}{\hspace{-1.5cm}\raisebox{-0.55cm}{\scalebox{.8}{ $ S_{\gamma N}({\rm 
					GeV}^{2}) $}}}
	\psfrag{VVV}{LPA}
	\psfrag{TTT}{}
	\vspace{0.2cm}
	\centerline{
		{\includegraphics[width=18pc]{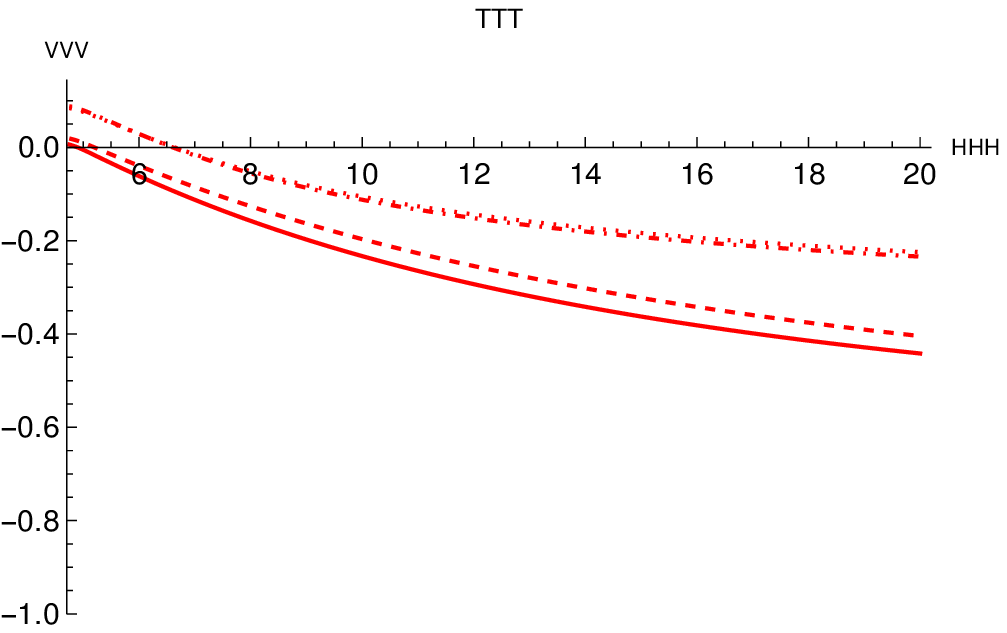}}
		\psfrag{VVV}{LPA}
		{\includegraphics[width=18pc]{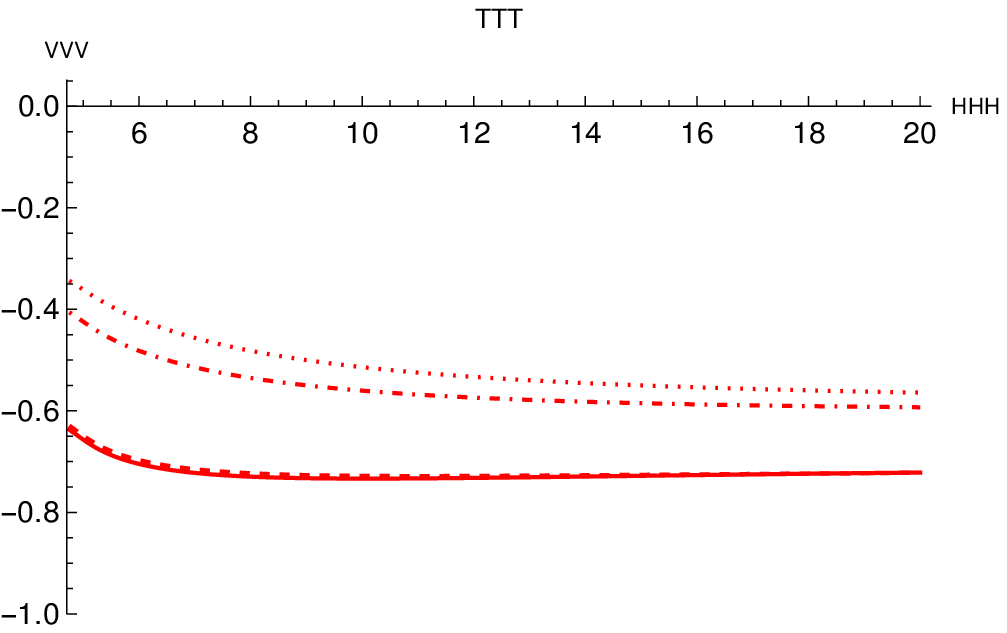}}}
	\vspace{0.2cm}
	\caption{\small The LPA integrated over all differential variables for $ \pi^{+} $ ($ \pi^{-} $) is shown on the left (right). The dashed (non-dashed) lines correspond to holographic (asymptotic) DA, while the dotted (non-dotted) lines correspond to the standard (valence) scenario.}
	\label{fig:jlab-pol-asym-int-sigma}
\end{figure}

Finally, we show the LPA, integrated over all differential variables, as a function of $ S_{\gamma N} $ in Figure \ref{fig:jlab-pol-asym-int-sigma}. In both the $ \pi^{+} $ and $ \pi ^{-} $ cases, the choice of the model for the axial GPDs (standard or valence) has a significant effect on the LPA. Moreover, in the $ \pi^{+} $ case with the standard scenario, the choice of DA has almost no effect on the LPA, while for the valence scenario, the effect of the choice of DA has a mild effect. This is the opposite of what happens in the $ \pi^{-} $ case, where it is for the valence scenario that the choice of DA has almost no effect.

In both cases, the LPA is sizeable, and goes up to 60\% in the case of $\pi^-$. This, combined with the expected counting rates found in \SEC\ref{sec:jlab-counting-rates}, makes the measurement of such an observable very promising.

\subsection{COMPASS kinematics}

Typically, COMPASS consists of colliding muons at an energy of 160 GeV onto a fixed target. This translates to a muon-nucleon centre-of-mass energy of roughly 301 GeV$ ^2 $. Since the skewness $  \xi  $ decreases with increasing $ S_{\gamma N} $ (see \EQ\eqref{skewness2}), COMPASS can in principle give us access to a kinematical region of small $  \xi  $ for GPDs ($0.0027\leq \xi \leq  0.35$), not accessible at JLab. The typical centre-of-mass energy $ S_{\gamma N} $ used for the plots that we show in this section is 200 GeV$ ^{2} $.

\subsubsection{Fully differential cross-section}

\label{sec:compass-fully-diff}

\begin{figure}[h!]
	\psfrag{HHH}{\hspace{-1.5cm}\raisebox{-.6cm}{\scalebox{.8}{$-u' ({\rm 
					GeV}^{2})$}}}
	\psfrag{VVV}{\raisebox{.3cm}{\scalebox{.9}{$\hspace{-.4cm}\displaystyle\left.\frac{d 
					\sigma_{\gamma\pi^+}}{d M^2_{\gamma \pi^+} d(-u') d(-t)}\right|_{(-t)_{\rm min}}({\rm pb} \cdot {\rm GeV}^{-6})$}}}
	\psfrag{TTT}{}
	\vspace{0.2cm}
	\centerline{
		{\includegraphics[width=18pc]{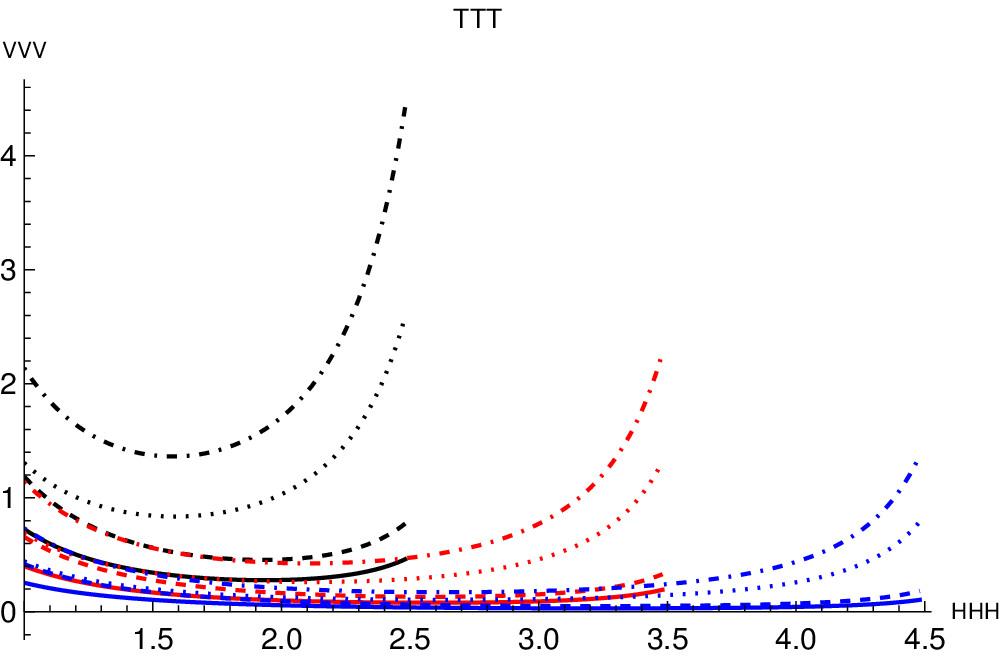}}
		\psfrag{VVV}{\raisebox{.3cm}{\scalebox{.9}{$\hspace{-.4cm}\displaystyle\left.\frac{d 
						\sigma_{\gamma\pi^-}}{d M^2_{\gamma \pi^-} d(-u') d(-t)}\right|_{(-t)_{\rm min}}({\rm pb} \cdot {\rm GeV}^{-6})$}}}
		{\includegraphics[width=18pc]{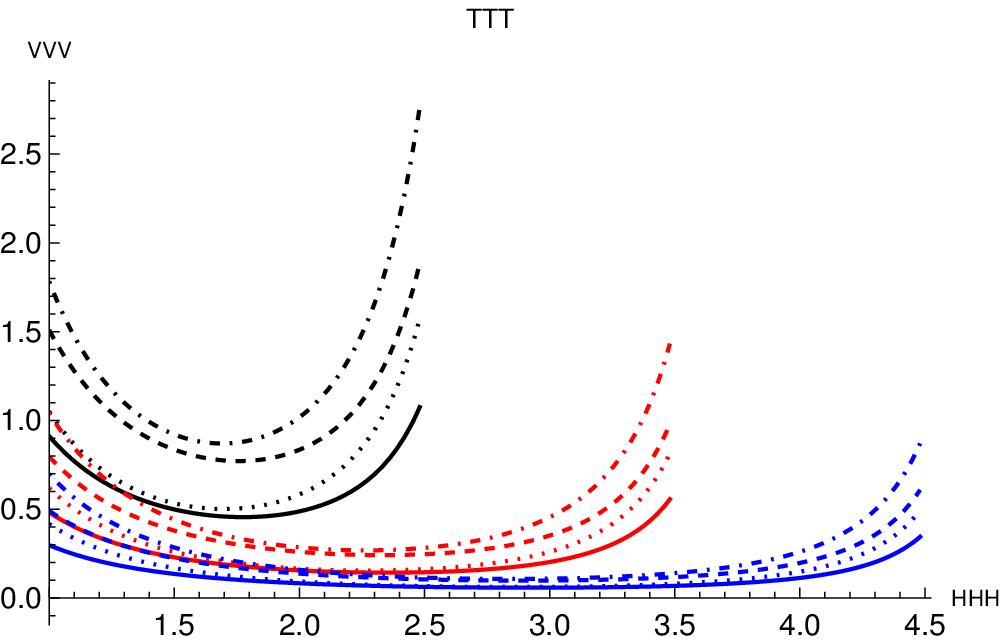}}}
	\vspace{0.2cm}
	\caption{\small The fully differential cross-section for $ \pi^{+} $ ($ \pi^{-} $) is shown as a function of $  \left( -u' \right)  $ on the left (right) for different values of $ M_{\gamma \meson}^2 $. The black, red and blue curves correspond to $ M_{\gamma \meson}^{2}=3,\,4,\,5\, $ GeV$ ^2 $ respectively. The dashed (non-dashed) lines correspond to holographic (asymptotic) DA, while the dotted (non-dotted) lines correspond to the standard (valence) scenario. As mentioned in the text, $ S_{\gamma N} $ is fixed at 200 GeV$ ^2 $ here.}
	\label{fig:compass-fully-diff-diff-M2}
\end{figure}

The effect of different values of $ M_{\gamma \meson}^2 $ on the fully differential cross-section is shown in Figure \ref{fig:compass-fully-diff-diff-M2}. The three values of $ M_{\gamma \meson}^2 $ used for the plots are 3, 4 and 5 GeV$ ^{2} $. We do not pick larger values of $\Msq$ as the values of the cross-section become too small in that case. The values of the cross-section here are suppressed by roughly a factor of 10 compared to those for the JLab kinematics, c.f. \FIG\ref{fig:jlab-fully-diff-diff-M2}.

\begin{figure}[h!]
	\psfrag{HHH}{\hspace{-1.5cm}\raisebox{-.6cm}{\scalebox{.8}{$-u' ({\rm 
					GeV}^{2})$}}}
	\psfrag{VVV}{\raisebox{.3cm}{\scalebox{.9}{$\hspace{-.4cm}\displaystyle\left.\frac{d 
					\sigma_{\gamma\pi^+}}{d M^2_{\gamma \pi^+} d(-u') d(-t)}\right|_{(-t)_{\rm min}}({\rm pb} \cdot {\rm GeV}^{-6})$}}}
	\psfrag{TTT}{}
	\vspace{0.2cm}
	\centerline{
		{\includegraphics[width=18pc]{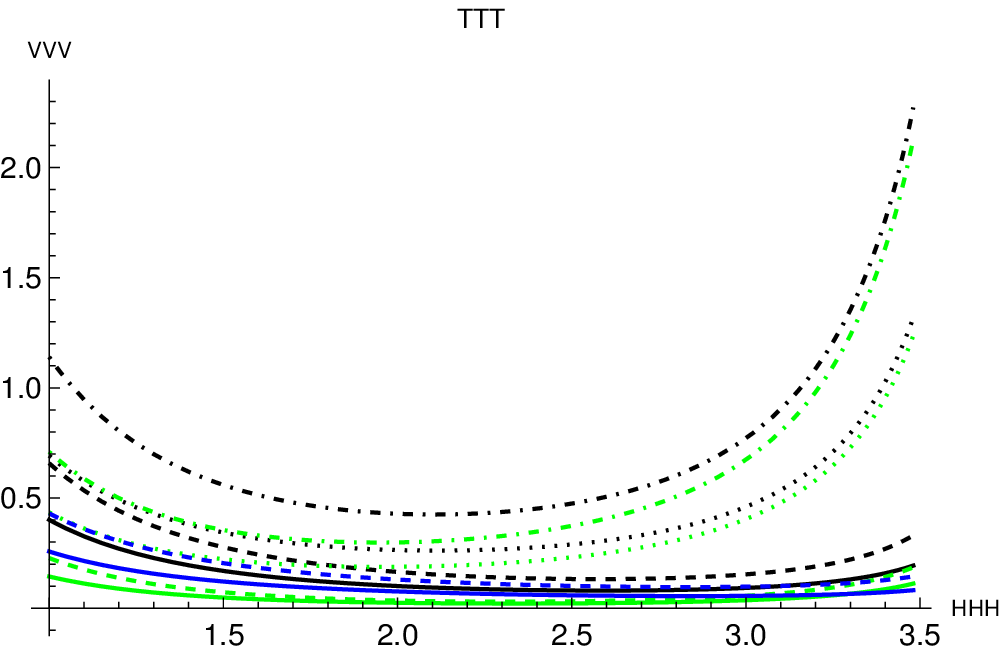}}
		\psfrag{VVV}{\raisebox{.3cm}{\scalebox{.9}{$\hspace{-.4cm}\displaystyle\left.\frac{d 
						\sigma_{\gamma\pi^-}}{d M^2_{\gamma \pi^-} d(-u') d(-t)}\right|_{(-t)_{\rm min}}({\rm pb} \cdot {\rm GeV}^{-6})$}}}
		{\includegraphics[width=18pc]{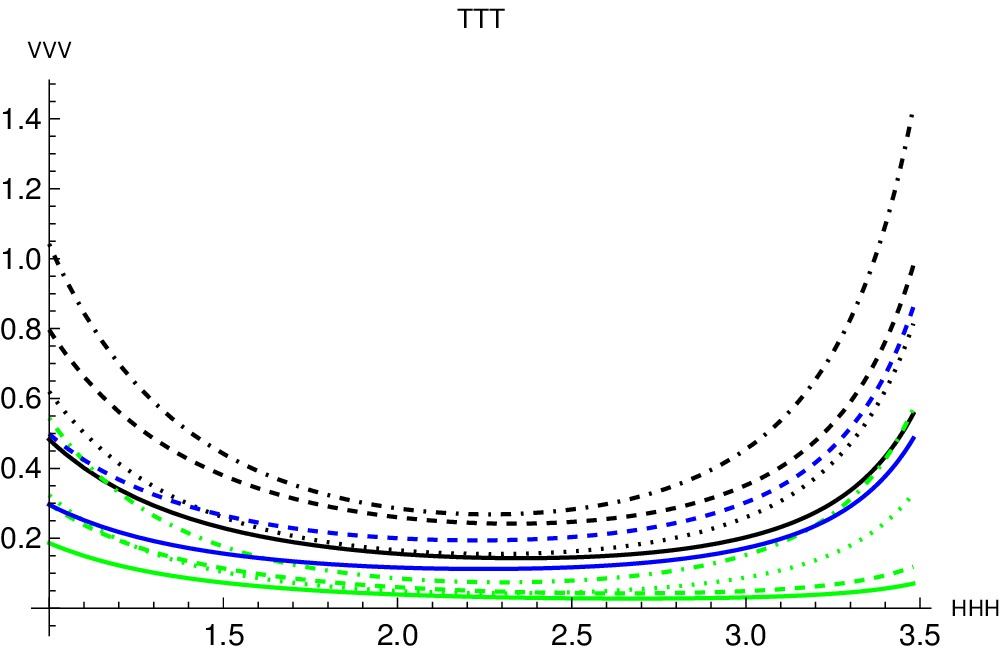}}}
	\vspace{0.2cm}
	\caption{\small The fully differential cross-section for $ \pi^{+} $ ($ \pi^{-} $) is shown as a function of $  \left( -u' \right)  $ on the left (right). The blue and green curves correspond to contributions from the vector and axial GPDs respectively. The black curves correspond to the total contribution, i.e. vector and axial GPD contributions combined. As before, the dashed (non-dashed) lines correspond to holographic (asymptotic) DA, while the dotted (non-dotted) lines correspond to the standard (valence) scenario. We fix $ S_{\gamma N}= 200\,  \mathrm{GeV}^{2}  $ and $ M_{\gamma \meson}^{2}= 4\,  \mathrm{GeV}^{2}  $. Note that the vector contributions consist of only two curves in each case, since they are insensitive to either valence or standard scenarios.}
	\label{fig:compass-fully-diff-VandA}
\end{figure}

Next,  we show the relative contributions of the vector and axial GPDs to the cross-section in Figure \ref{fig:compass-fully-diff-VandA}. Similar comments as in \SEC\ref{sec:jlab-fully-diff-X-section} apply.

\begin{figure}[h!]
	\psfrag{HHH}{\hspace{-1.5cm}\raisebox{-.6cm}{\scalebox{.8}{$-u' ({\rm 
					GeV}^{2})$}}}
	\psfrag{VVV}{\raisebox{.3cm}{\scalebox{.9}{$\hspace{-.4cm}\displaystyle\left.\frac{d 
					\sigma_{\gamma\pi^+}}{d M^2_{\gamma \pi^+} d(-u') d(-t)}\right|_{(-t)_{\rm min}}({\rm pb} \cdot {\rm GeV}^{-6})$}}}
	\psfrag{TTT}{}
	\vspace{0.2cm}
	\centerline{
		{\includegraphics[width=18pc]{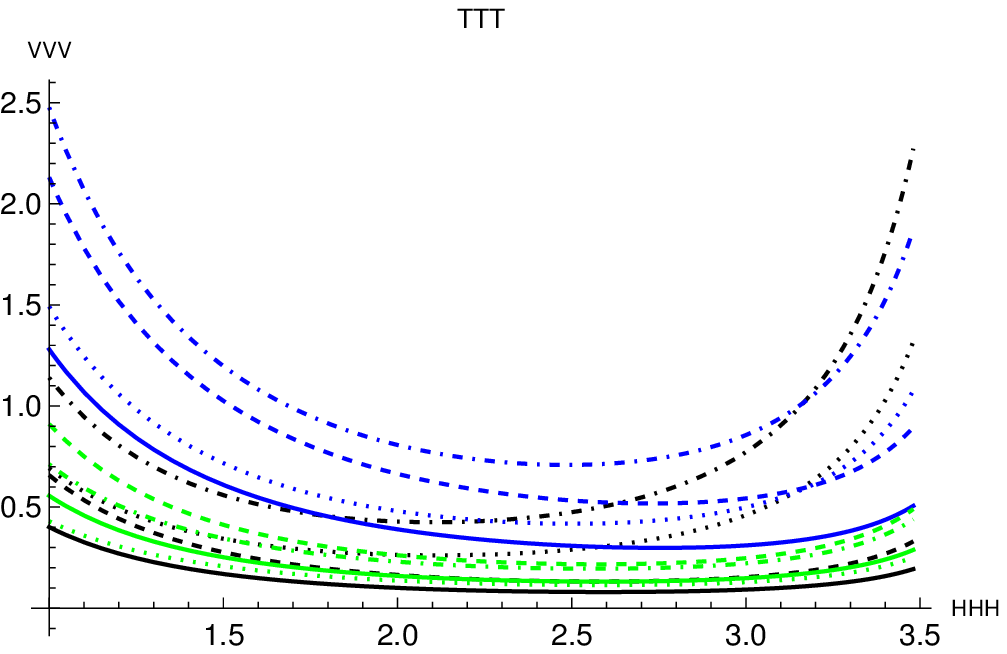}}
		\psfrag{VVV}{\raisebox{.3cm}{\scalebox{.9}{$\hspace{-.4cm}\displaystyle\left.\frac{d 
						\sigma_{\gamma\pi^-}}{d M^2_{\gamma \pi^-} d(-u') d(-t)}\right|_{(-t)_{\rm min}}({\rm pb} \cdot {\rm GeV}^{-6})$}}}
		{\includegraphics[width=18pc]{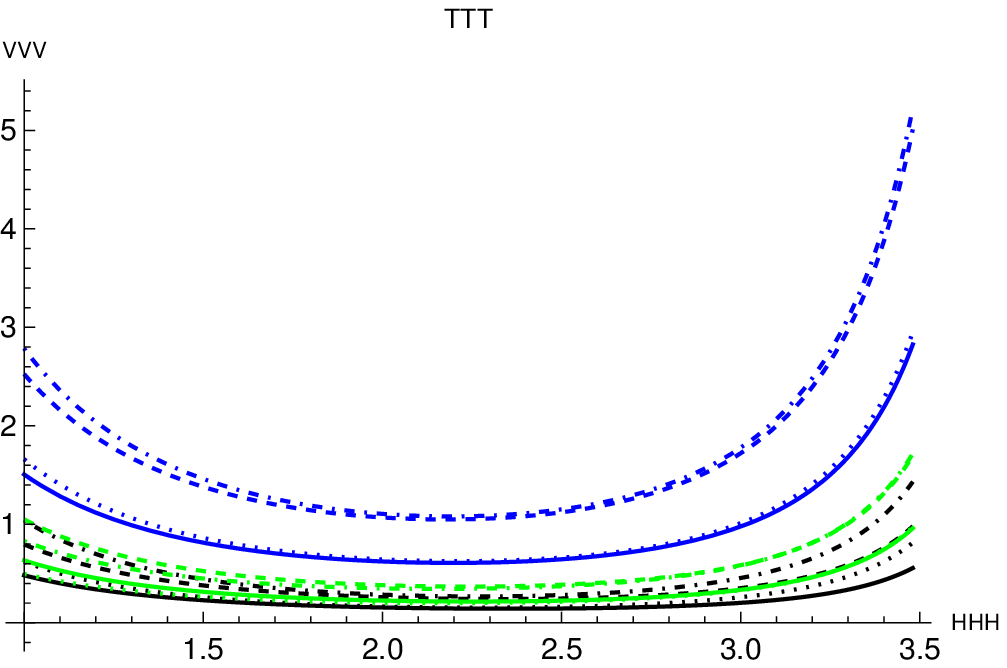}}}
	\vspace{0.2cm}
	\caption{\small The fully differential cross-section for $ \pi^{+} $ ($ \pi^{-} $) is shown as a function of $  \left( -u' \right)  $ on the left (right). The blue and green curves correspond to contributions from the u quark ($ H_{u} $ and $  \tilde{H} _{u} $) and d quark ($ H_{d} $ and $  \tilde{H} _{d} $) GPDs respectively. The black curves correspond to the total contribution. The dashed (non-dashed) lines correspond to holographic (asymptotic) DA, while the dotted (non-dotted) lines correspond to the standard (valence) scenario. Similar comments as in Figure \ref{fig:jlab-fully-diff-uandd} apply.}
	\label{fig:compass-fully-diff-uandd}
\end{figure}

Finally, to conclude this subsection, the relative contributions of the u quark and d quark GPDs to the cross-section are shown in Figure \ref{fig:compass-fully-diff-uandd}.

\subsubsection{Single differential cross-section}

\begin{figure}[h!]
	\psfrag{HHH}{\hspace{-1.5cm}\raisebox{-.6cm}{\scalebox{.8}{$M^2_{\gamma \meson} ({\rm 
					GeV}^{2})$}}}
	\psfrag{VVV}{\raisebox{.3cm}{\scalebox{.9}{$\hspace{-.4cm}\displaystyle\frac{d 
					\sigma_{\gamma\pi^+}}{d M^2_{\gamma \pi^+}}({\rm pb} \cdot {\rm GeV}^{-2})$}}}
	\psfrag{TTT}{}
	\vspace{0.2cm}
	\centerline{
		{\includegraphics[width=18pc]{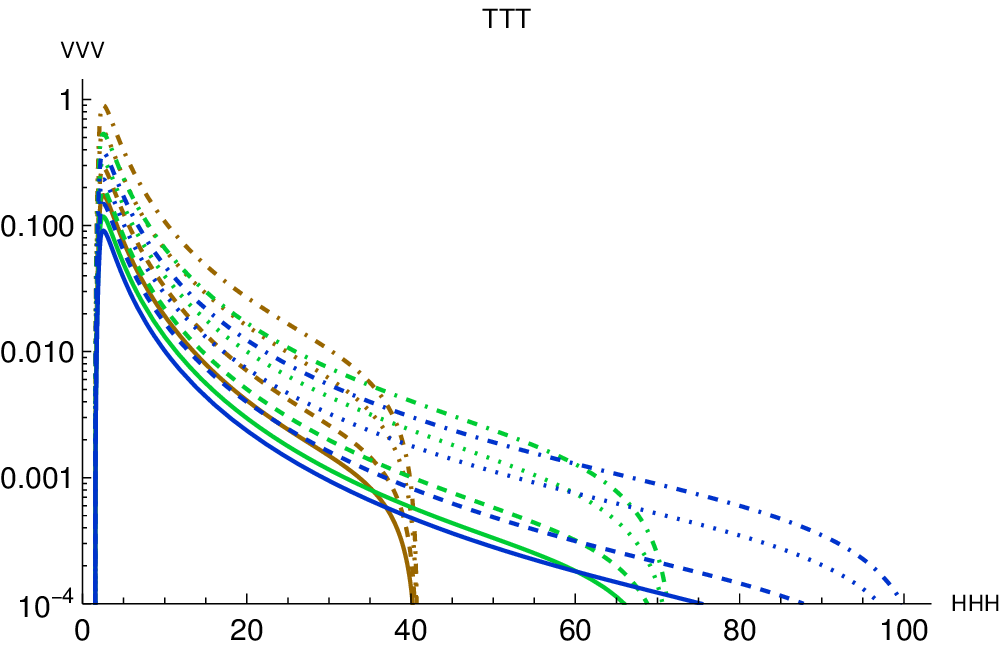}}
		\psfrag{VVV}{\raisebox{.3cm}{\scalebox{.9}{$\hspace{-.4cm}\displaystyle\frac{d 
						\sigma_{\gamma\pi^-}}{d M^2_{\gamma \pi^-}}({\rm pb} \cdot {\rm GeV}^{-2})$}}}
		{\includegraphics[width=18pc]{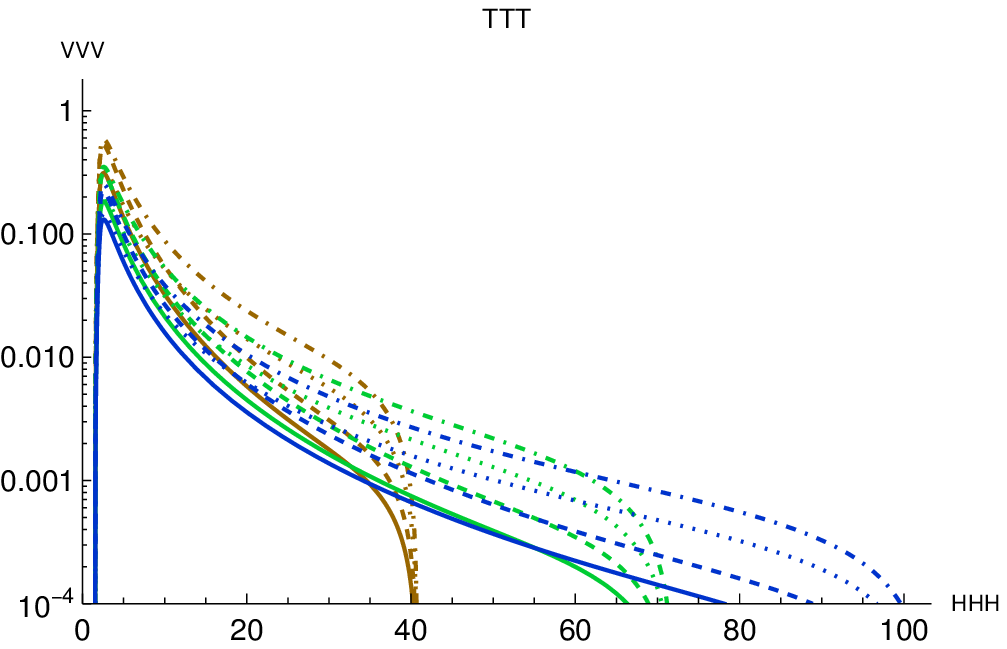}}}
	\vspace{0.2cm}
	\caption{\small The single differential cross-section for $ \pi^{+} $ ($ \pi^{-} $) is shown as a function of $  M_{\gamma \meson}^{2}  $ on the left (right) for different values of $ S_{\gamma N} $. The brown, green and blue curves correspond to $ S_{\gamma N} = 80,\,140,\,200\,\GeV^{2} $. The dashed (non-dashed) lines correspond to holographic (asymptotic) DA, while the dotted (non-dotted) lines correspond to the standard (valence) scenario. The holographic DA with the standard scenario has the largest contribution for every $ \SgN $.}
	\label{fig:compass-sing-diff}
\end{figure}

The variation of the single differential cross-section with $ M_{\gamma \meson}^2 $ for different values of $ S_{\gamma N} $ is shown in Figure \ref{fig:compass-sing-diff}. The values of $ S_{\gamma N} $ chosen are 80, 140 and 200 GeV$ ^2 $. Note that a log scale is used for the vertical axis, as variations in the cross-section over the full range of $ \Msq $ are quite large. From the plots, it is clear that the cross-section is dominated by the region of very small $ \Msq $.

\subsubsection{Integrated cross-section}

\begin{figure}[h!]
	\psfrag{HHH}{\hspace{-1.5cm}\raisebox{-.6cm}{\scalebox{.8}{$ S_{\gamma N} ({\rm 
					GeV}^{2})$}}}
	\psfrag{VVV}{\raisebox{.3cm}{\scalebox{.9}{$\hspace{-.4cm}\displaystyle
				\sigma_{\gamma\pi^+}({\rm pb})$}}}
	\psfrag{TTT}{}
	\vspace{0.2cm}
	\centerline{
		{\includegraphics[width=18pc]{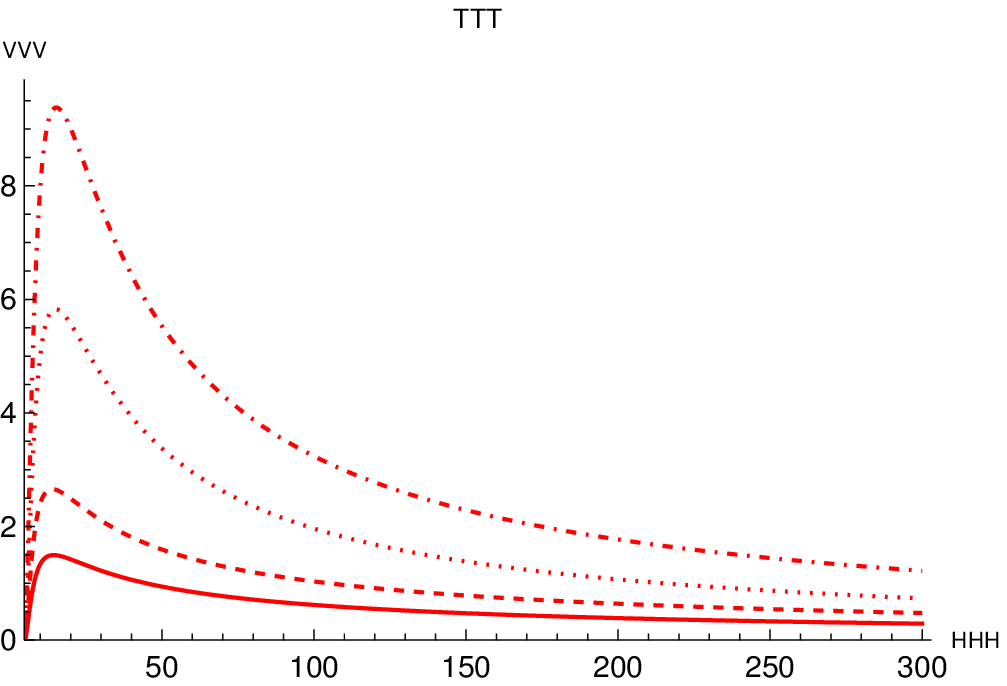}}
		\psfrag{VVV}{\raisebox{.3cm}{\scalebox{.9}{$\hspace{-.4cm}\displaystyle
					\sigma_{\gamma\pi^-}({\rm pb})$}}}
		{\includegraphics[width=18pc]{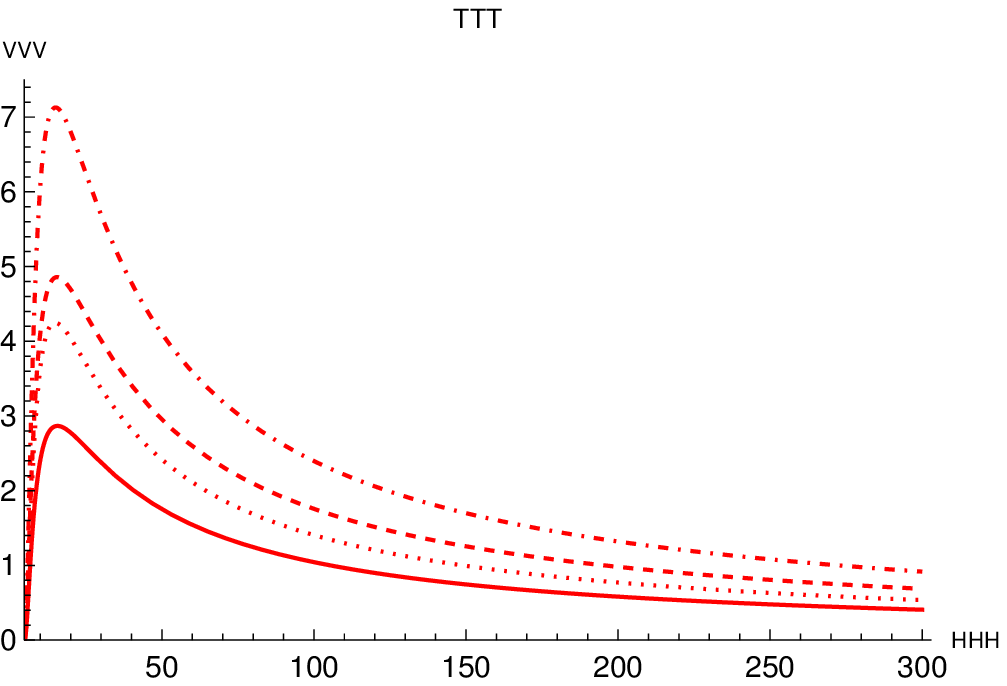}}}
	\vspace{0.2cm}
	\caption{\small The integrated cross-section for $ \pi^{+} $ ($ \pi^{-} $) is shown as a function of $   S_{\gamma N}  $ on the left (right). The dashed (non-dashed) lines correspond to holographic (asymptotic) DA, while the dotted (non-dotted) lines correspond to the standard (valence) scenario. We thus find that the maximum cross-section appears at around 20 GeV$ ^2 $, a feature which was not totally clear in Figure \ref{fig:jlab-int-sigma}.}
	\label{fig:compass-int-sigma}
\end{figure}

The variation of the integrated cross-section as a function of $ \SgN $ is shown in Figure \ref{fig:compass-int-sigma}. The full kinematical range of $ \SgN $ (up to 300 GeV$ ^{2} $) at COMPASS is covered in the plots. The peak of the cross-section occurs at around $ 20 \GeV^2 $. Similar comments as in \SEC\ref{sec:int-X-section-JLab} apply.

\subsubsection{Polarisation asymmetries}

\label{sec:pol-asym-COMPASS}

We now show the results for the linear polarisation asymmetries (LPAs) corresponding to COMPASS kinematics. As before, for the fully differential and single differential plots, we choose the reference value of 200 GeV$ ^2 $ for $ \SgN $.

We start by showing the LPA at the fully differential level as a function of $ (-u') $ for different values of $ \Msq $. As in \SEC\ref{sec:compass-fully-diff}, The values of $ \Msq $ used are 3, 4 and 5 GeV$ ^2 $, and $ \SgN=200\GeV^2 $. Similar comments as for the JLab kinematics case apply (see \SEC\ref{sec:pol-asym-jlab}), i.e. the LPAs distinguish between GPD models well (standard vs valence scenarios), but they are rather insensitive to the DA models (asymptotic vs holographic).

\begin{figure}[h!]
	\psfrag{HHH}{\hspace{-1.5cm}\raisebox{0.4cm}{\scalebox{.8}{$-u' ({\rm 
					GeV}^{2})$}}}
	\psfrag{VVV}{LPA}
	\psfrag{TTT}{}
	\vspace{0.2cm}
	\centerline{
		{\includegraphics[width=18pc]{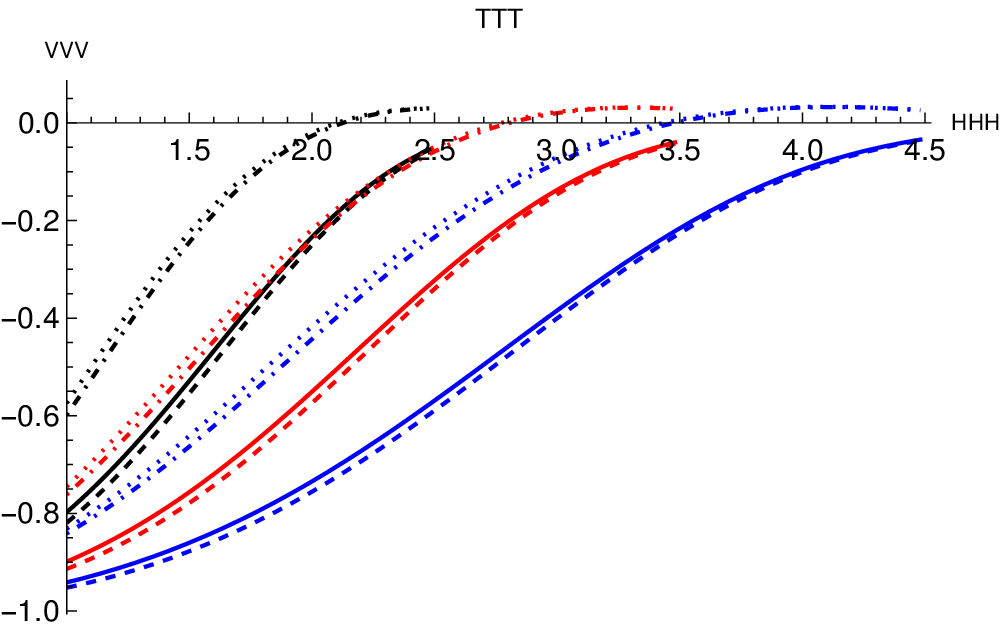}}
		\psfrag{VVV}{LPA}
		{\includegraphics[width=18pc]{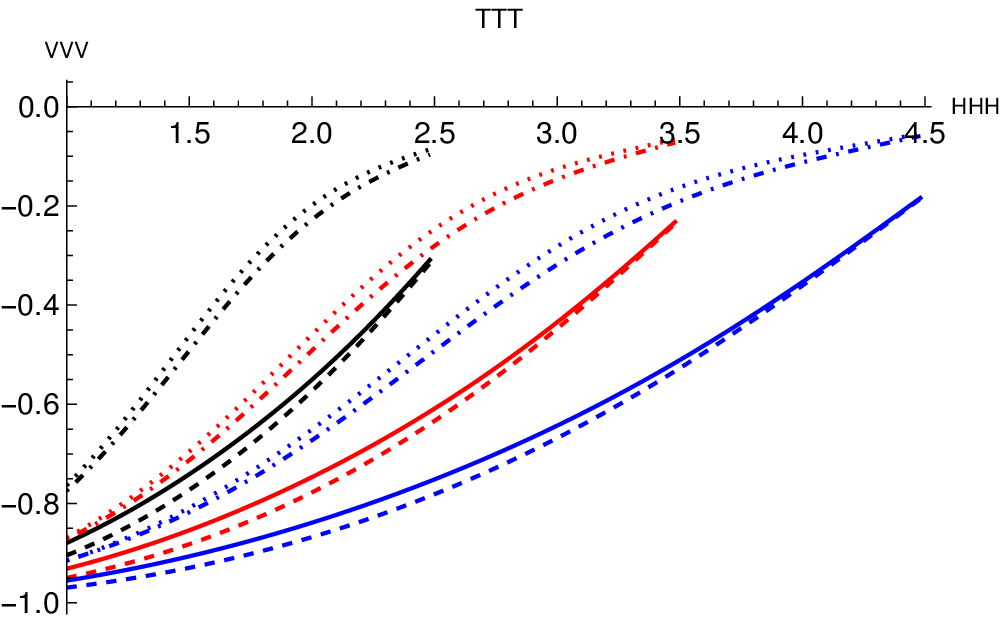}}}
	\vspace{0.2cm}
	\caption{\small The LPA at the fully-differential level for for $ \pi^{+} $ ($ \pi^{-} $) is shown as a function of $  \left( -u' \right)  $ on the left (right) for different values of $ M_{\gamma \meson}^2 $. The black, red and blue curves correspond to $ M_{\gamma \meson}^{2}=3,\,4,\,5\, $ GeV$ ^2 $ respectively, and $ \SgN = 200 \GeV^{2}$. The dashed (non-dashed) lines correspond to holographic (asymptotic) DA, while the dotted (non-dotted) lines correspond to the standard (valence) scenario.}
	\label{fig:compass-pol-asym-fully-diff-diff-M2}
\end{figure}

\begin{figure}[h!]
	\psfrag{HHH}{\hspace{-1.1cm}\raisebox{-.5cm}{\scalebox{.8}{$-u' ({\rm 
					GeV}^{2})$}}}
	\psfrag{VVV}{LPA}
	\psfrag{TTT}{}
	\vspace{0.2cm}
	\centerline{
		{\includegraphics[width=18pc]{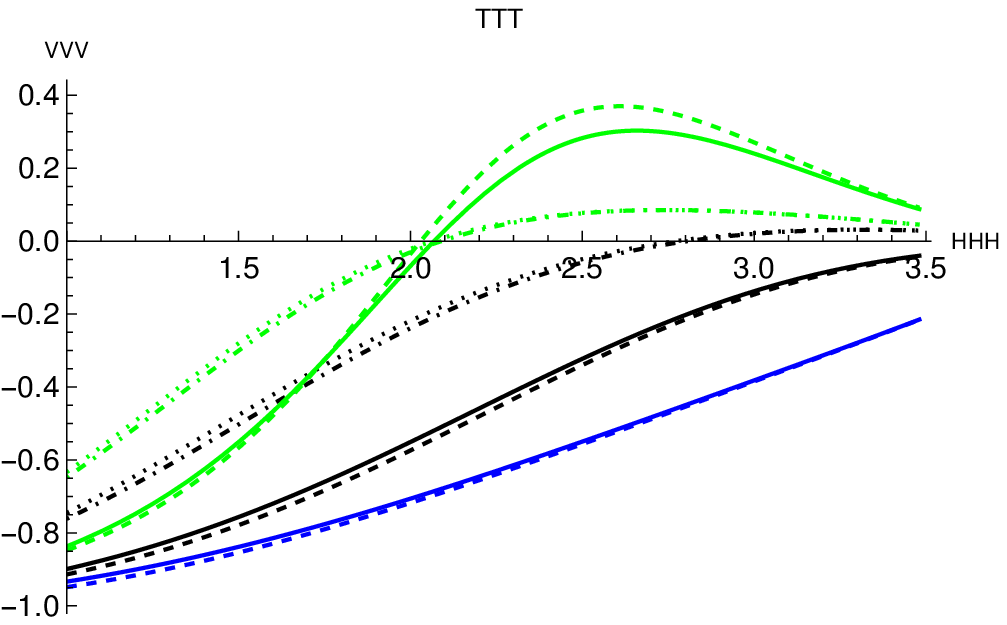}}
		\psfrag{VVV}{LPA}
		\psfrag{HHH}{\hspace{-1.1cm}\raisebox{.2cm}{\scalebox{.8}{$-u' ({\rm 
						GeV}^{2})$}}}
		{\includegraphics[width=18pc]{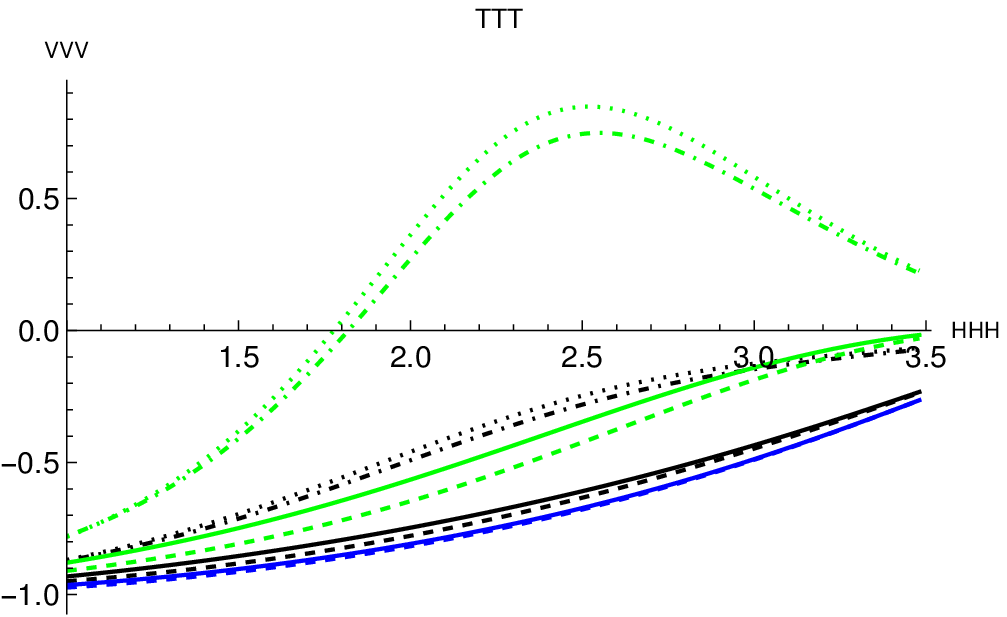}}}
	\vspace{0.2cm}
	\caption{\small The LPA at the fully-differential level for for $ \pi^{+} $ ($ \pi^{-} $) is shown as a function of $  \left( -u' \right)  $ on the left (right), using $ \Msq = 4 \GeV^2$ and $ \SgN = 200 \GeV^2 $. The black curves correspond to the total contribution, i.e. vector and axial GPD contributions combined. As before, the dashed (non-dashed) lines correspond to holographic (asymptotic) DA, while the dotted (non-dotted) lines correspond to the standard (valence) scenario. Note that the vector contributions consist of only two curves in each case, since they are insensitive to either valence or standard scenarios. The effect of using the valence or standard scenario is significant, while the difference between using asymptotic and holographic DA is minimal.}
	\label{fig:compass-pol-asym-fully-diff-VandA}
\end{figure}

In Figure \ref{fig:compass-pol-asym-fully-diff-VandA}, the relative contributions of the vector and axial GPDs to the LPA at the fully differential level are shown. $ \SgN =200 \GeV^2$ and $ \Msq=4 \GeV^2 $ were used to generate the plots. The sensitivity of the axial GPD model used (standard vs valence scenario) is yet again apparent from the plots (green curves).

\begin{figure}[h!]
	\psfrag{HHH}{\hspace{-1.2cm}\raisebox{0.4cm}{\scalebox{.8}{$-u' ({\rm 
					GeV}^{2})$}}}
	\psfrag{VVV}{LPA}
	\psfrag{TTT}{}
	\vspace{0.2cm}
	\centerline{
		{\includegraphics[width=18pc]{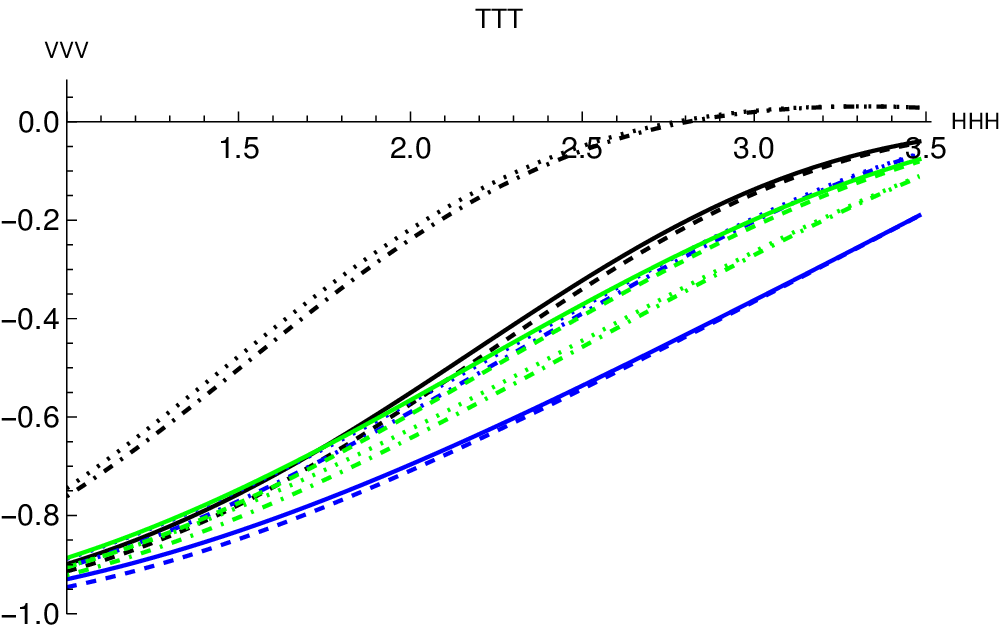}}
		\psfrag{VVV}{LPA}
		{\includegraphics[width=18pc]{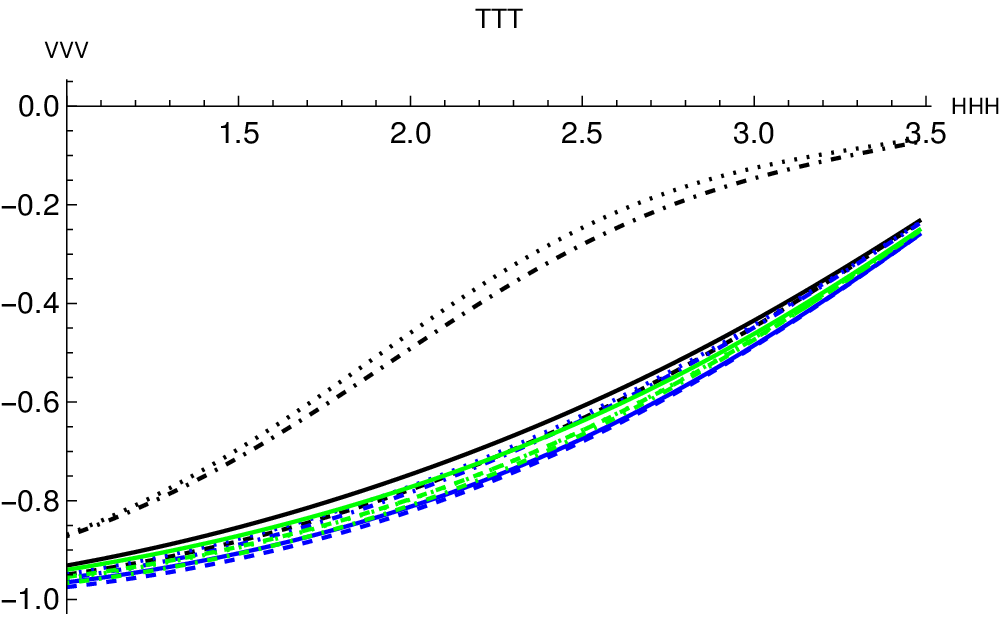}}}
	\vspace{0.2cm}
	\caption{\small The LPA at the fully-differential level for for $ \pi^{+} $ ($ \pi^{-} $) is shown as a function of $  \left( -u' \right)  $ on the left (right), using $ \Msq = 4 \GeV^2$ and $ \SgN = 200 \GeV^2 $. The blue and green curves correspond to contributions from the u quark ($ H_{u} $ and $  \tilde{H} _{u} $) and d quark ($ H_{d} $ and $  \tilde{H} _{d} $) GPDs respectively. The black curves correspond to the total contribution. The dashed (non-dashed) lines correspond to holographic (asymptotic) DA, while the dotted (non-dotted) lines correspond to the standard (valence) scenario.}
	\label{fig:compass-pol-asym-fully-diff-uandd}
\end{figure}

Next, we show the relative contributions of the u quark and d quark GPDs to the LPA at the fully differential level in Figure \ref{fig:compass-pol-asym-fully-diff-uandd}. The values $ \SgN =200 \GeV^2$ and $ \Msq=4 \GeV^2 $ were used to generate the plots.

\begin{figure}[h!]
	\psfrag{HHH}{\hspace{-1.5cm}\raisebox{-.6cm}{\scalebox{.8}{ $ M_{\gamma \meson}^{2}({\rm 
					GeV}^{2}) $}}}
	\psfrag{VVV}{LPA}
	\psfrag{TTT}{}
	\vspace{0.2cm}
	\centerline{
		{\includegraphics[width=18pc]{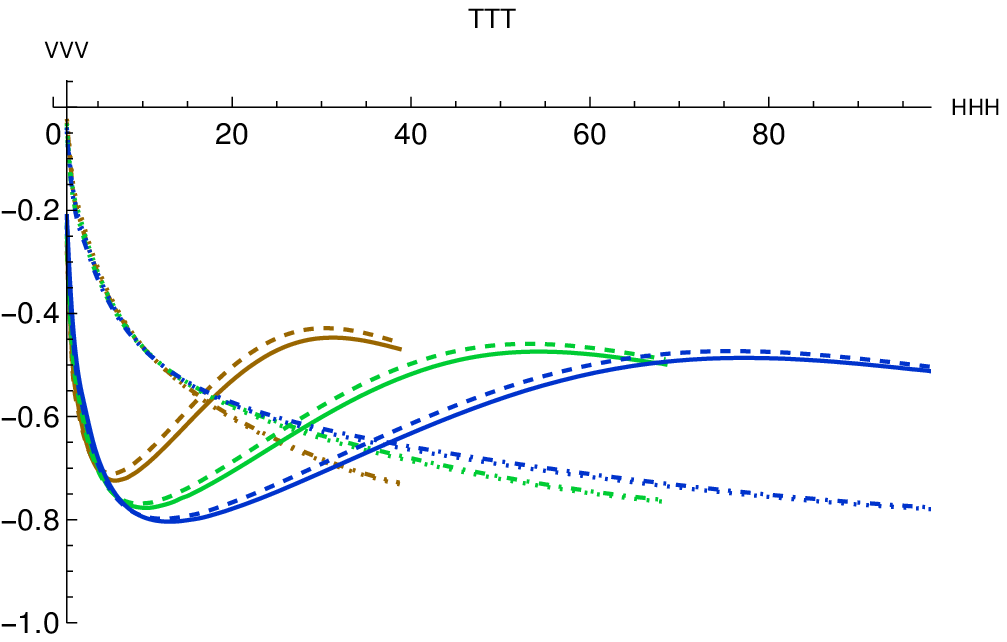}}
		\psfrag{VVV}{LPA}
		{\includegraphics[width=18pc]{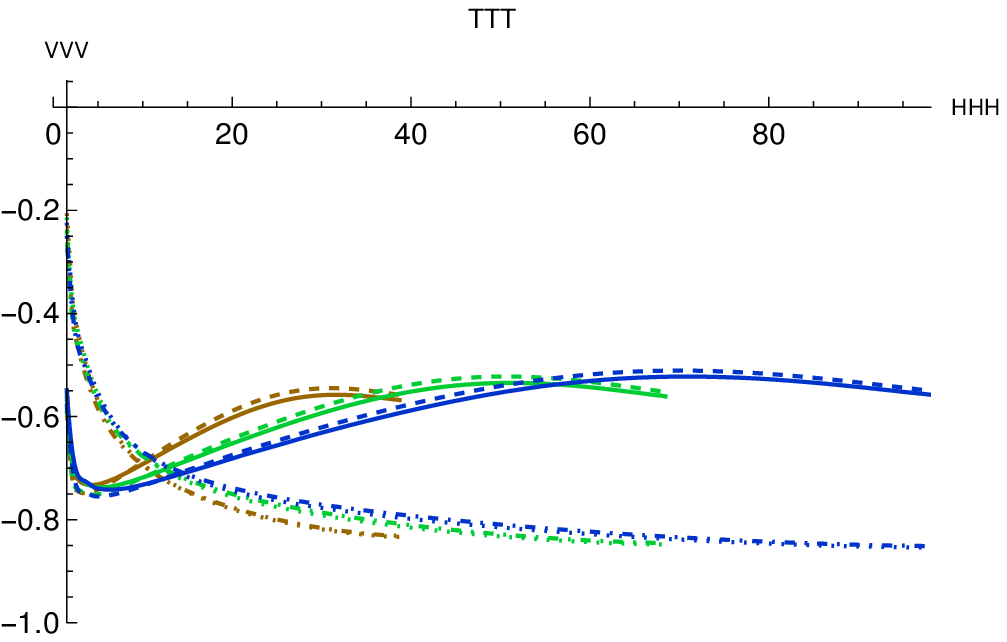}}}
	\vspace{0.2cm}
	\caption{\small The LPA at the single differential level for for $ \pi^{+} $ ($ \pi^{-} $) is shown as a function of $  M_{\gamma \meson}^{2}  $ on the left (right). The brown, green and blue curves correspond to $ S_{\gamma N} = 80,\,140,\,200\,\GeV^{2} $. The dashed (non-dashed) lines correspond to holographic (asymptotic) DA, while the dotted (non-dotted) lines correspond to the standard (valence) scenario. As was the case in Figure \ref{fig:jlab-pol-asym-sing-diff}, we note that the choice of the GPD model (valence or standard) gives a completely different shape for the LPA.}
	\label{fig:compass-pol-asym-sing-diff}
\end{figure}

Now, we show the variation of the LPA at the single differential level as a function of $ \Msq $ for different values of $ S_{\gamma N} $ in Figure \ref{fig:compass-pol-asym-sing-diff}. The three values of $ S_{\gamma N} $ used are 80, 140 and 200 GeV$ ^2 $, and correspond to the colours brown, green and blue respectively. As in the JLab kinematics plots in Figure \ref{fig:jlab-pol-asym-sing-diff}, we note that such LPAs can very easily distinguish between the standard and valence scenarios, as they have completely different shapes as a function of $ \Msq $.

\begin{figure}[h!]
	\psfrag{HHH}{\hspace{-1.5cm}\raisebox{-.6cm}{\scalebox{.8}{ $ S_{\gamma N}({\rm 
					GeV}^{2}) $}}}
	\psfrag{VVV}{LPA}
	\psfrag{TTT}{}
	\vspace{0.2cm}
	\centerline{
		{\includegraphics[width=18pc]{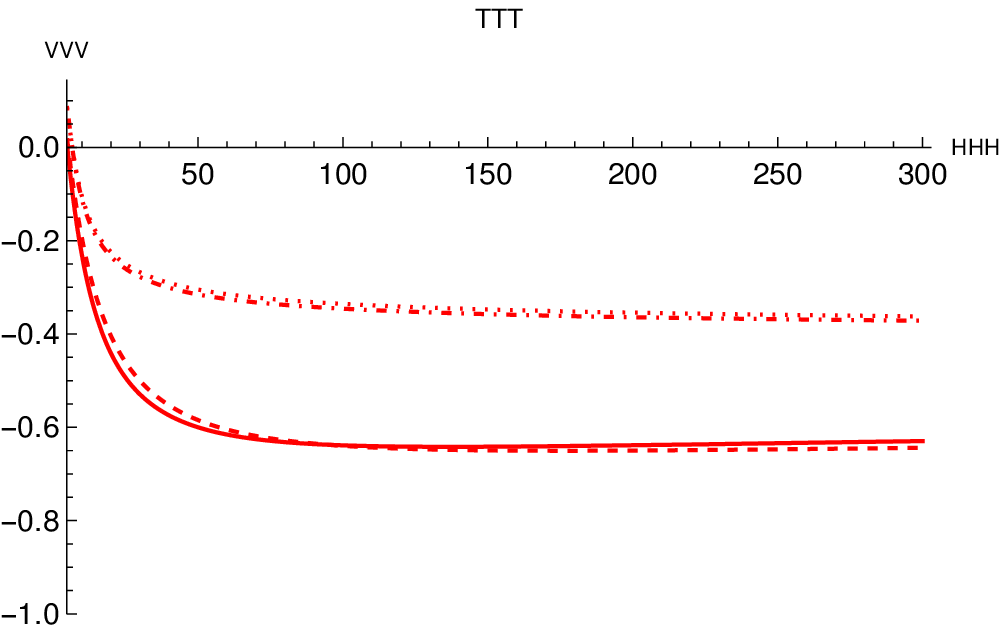}}
		\psfrag{VVV}{LPA}
		{\includegraphics[width=18pc]{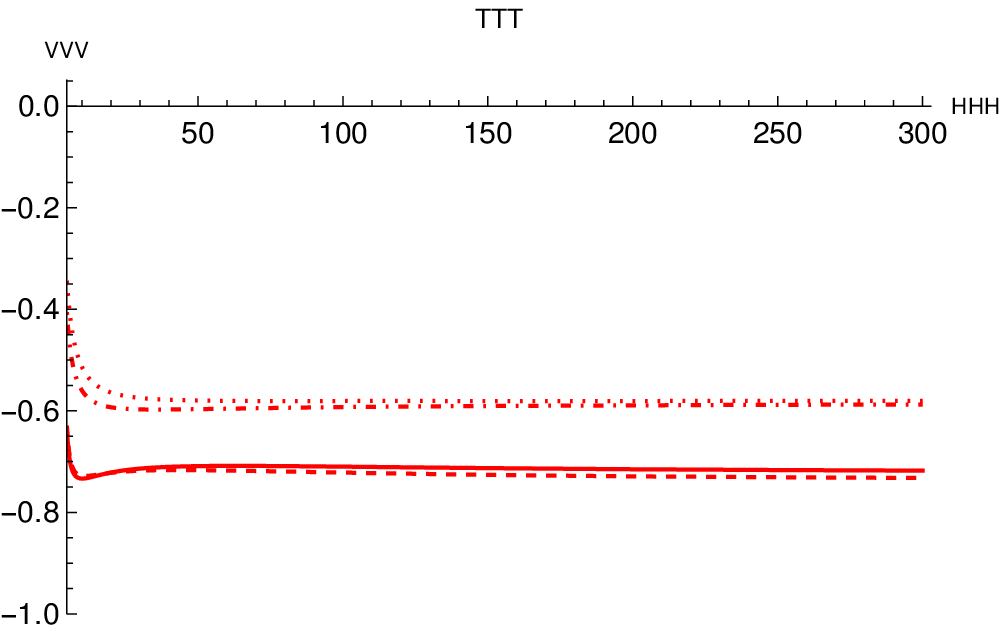}}}
	\vspace{0.2cm}
	\caption{\small The LPA integrated over all differential variables for $ \pi^{+} $ ($ \pi^{-} $) is shown on the left (right) as a function of $ \SgN $. The dashed (non-dashed) lines correspond to holographic (asymptotic) DA, while the dotted (non-dotted) lines correspond to the standard (valence) scenario.}
	\label{fig:compass-pol-asym-int-sigma}
\end{figure}

To conclude this section on COMPASS kinematics, we show the variation of the LPA, integrated over all differential variables, as a function of $ \SgN $ in Figure \ref{fig:compass-pol-asym-int-sigma}.

\FloatBarrier

\subsection{EIC and UPC at LHC kinematics}

In this section, we consider photon-nucleon centre-of-mass energies $ \SgN $ of up to 20000 GeV$ ^2 $. This covers the whole range of the expected EIC kinematics, and the most relevant part of UPCs at LHC kinematics. At EIC, the maximum centre of mass energy of the electron-proton system, $ S_{eN} $,  is expected to be roughly 19600 GeV$ ^2 $ \cite{AbdulKhalek:2021gbh}.

On the other hand, studying our process at LHC kinematics in UPCs in principle requires centre of mass energies of the order of the TeV scale. However, both the cross-section and the photon flux drop very rapidly as $ \SgN $ increases, such that only a tiny contribution is lost by neglecting contributions which are beyond the kinematics of EIC, i.e. above $ \SgN=20000 \GeV^2 $.

\subsubsection{Fully differential cross-section}

\begin{figure}[h!]
	\psfrag{HHH}{\hspace{-1.5cm}\raisebox{-.6cm}{\scalebox{.8}{$-u' ({\rm 
					GeV}^{2})$}}}
	\psfrag{VVV}{\raisebox{.3cm}{\scalebox{.9}{$\hspace{-.4cm}\displaystyle\left.\frac{d 
					\sigma_{\gamma\pi^+}}{d M^2_{\gamma \pi^+} d(-u') d(-t)}\right|_{(-t)_{\rm min}}({\rm pb} \cdot {\rm GeV}^{-6})$}}}
	\psfrag{TTT}{}
	\vspace{0.2cm}
	\centerline{
		{\includegraphics[width=18pc]{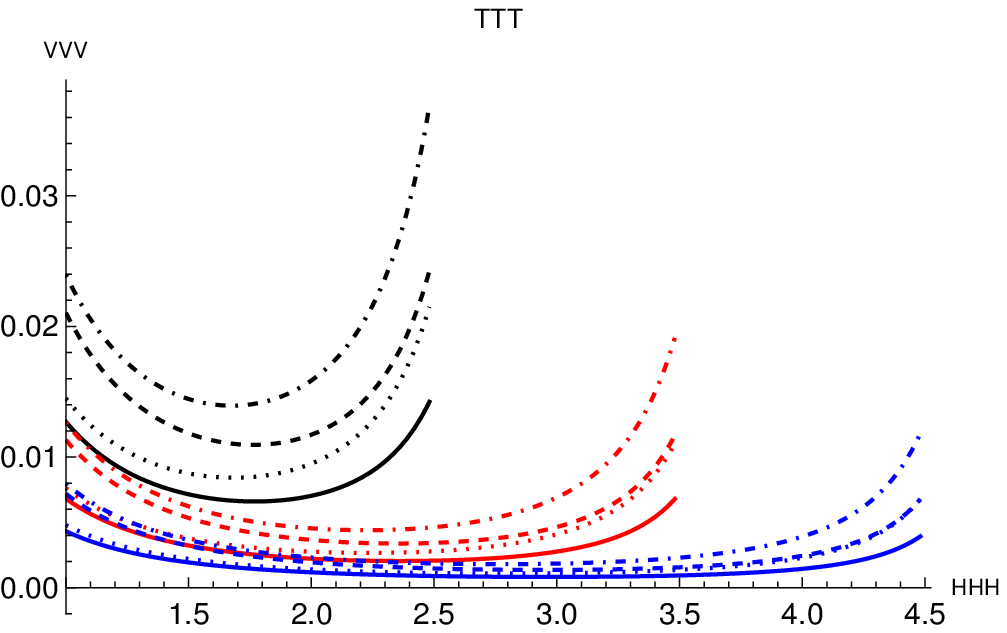}}
		\psfrag{VVV}{\raisebox{.3cm}{\scalebox{.9}{$\hspace{-.4cm}\displaystyle\left.\frac{d 
						\sigma_{\gamma\pi^-}}{d M^2_{\gamma \pi^-} d(-u') d(-t)}\right|_{(-t)_{\rm min}}({\rm pb} \cdot {\rm GeV}^{-6})$}}}
		{\includegraphics[width=18pc]{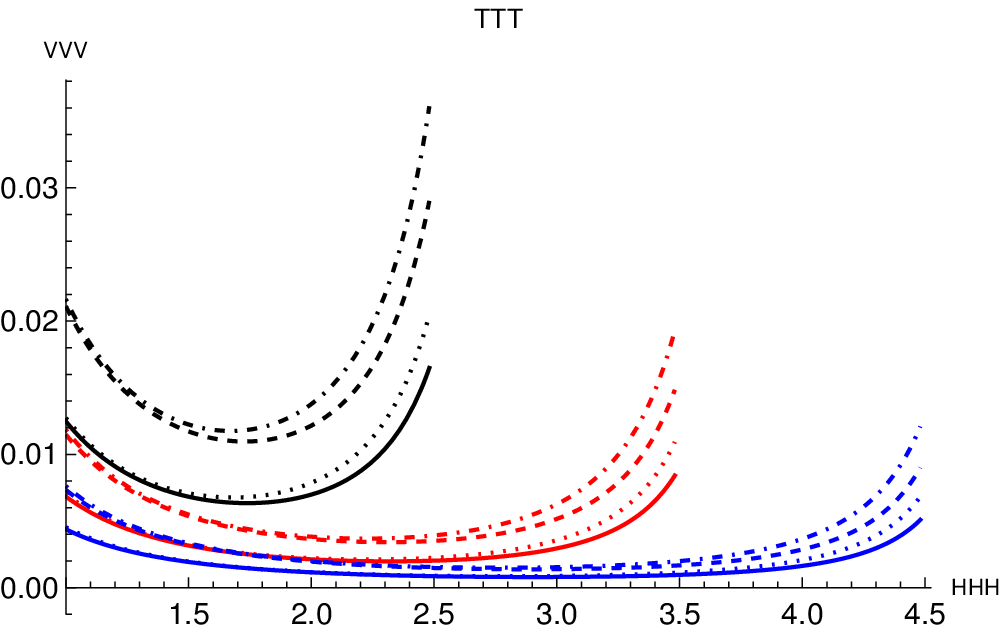}}}
	\vspace{0.2cm}
	\caption{\small The fully differential cross-section for $ \pi^{+} $ ($ \pi^{-} $) is shown as a function of $  \left( -u' \right)  $ on the left (right) for different values of $ M_{\gamma \meson}^2 $. The black, red and blue curves correspond to $ M_{\gamma \meson}^{2}=3,\,4,\,5\, $ GeV$ ^2 $ respectively. The dashed (non-dashed) lines correspond to holographic (asymptotic) DA, while the dotted (non-dotted) lines correspond to the standard (valence) scenario. As mentioned in the text, $ S_{\gamma N} $ is fixed at 20000 GeV$ ^2 $ here.}
	\label{fig:EIC-LHC-UPC-fully-diff-diff-M2}
\end{figure}

The fully differential cross-section as a function of $ (-u') $ for different values of $ \Msq  $ is shown in Figure \ref{fig:EIC-LHC-UPC-fully-diff-diff-M2}. $ \SgN $ is fixed at 20000 GeV$ ^2 $. The three values of $ \Msq $ that we used are 3, 4 and 5 GeV$ ^2 $. We do not pick larger values of $\Msq$ as the values of the cross-section become too small in that case. The values of the cross-section here are suppressed by roughly a factor of 100 compared to those for the COMPASS kinematics, c.f. \FIG\ref{fig:compass-fully-diff-diff-M2}.

\begin{figure}[h!]
	\psfrag{HHH}{\hspace{-1.5cm}\raisebox{-.6cm}{\scalebox{.8}{$-u' ({\rm 
					GeV}^{2})$}}}
	\psfrag{VVV}{\raisebox{.3cm}{\scalebox{.9}{$\hspace{-.4cm}\displaystyle\left.\frac{d 
					\sigma_{\gamma\pi^+}}{d M^2_{\gamma \pi^+} d(-u') d(-t)}\right|_{(-t)_{\rm min}}({\rm pb} \cdot {\rm GeV}^{-6})$}}}
	\psfrag{TTT}{}
	\vspace{0.2cm}
	\centerline{
		{\includegraphics[width=18pc]{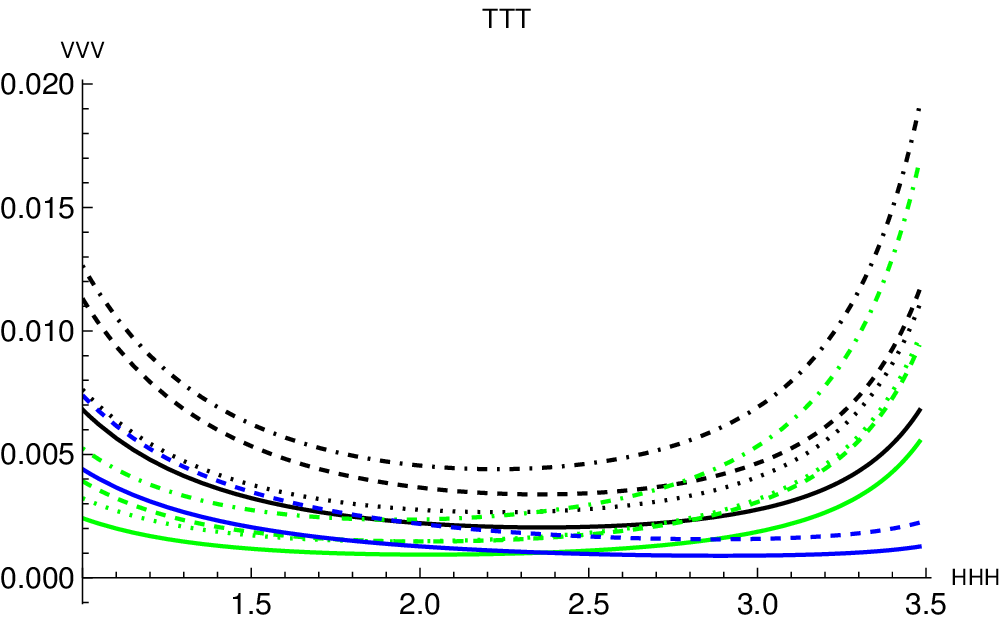}}
		\psfrag{VVV}{\raisebox{.3cm}{\scalebox{.9}{$\hspace{-.4cm}\displaystyle\left.\frac{d 
						\sigma_{\gamma\pi^-}}{d M^2_{\gamma \pi^-} d(-u') d(-t)}\right|_{(-t)_{\rm min}}({\rm pb} \cdot {\rm GeV}^{-6})$}}}
		{\includegraphics[width=18pc]{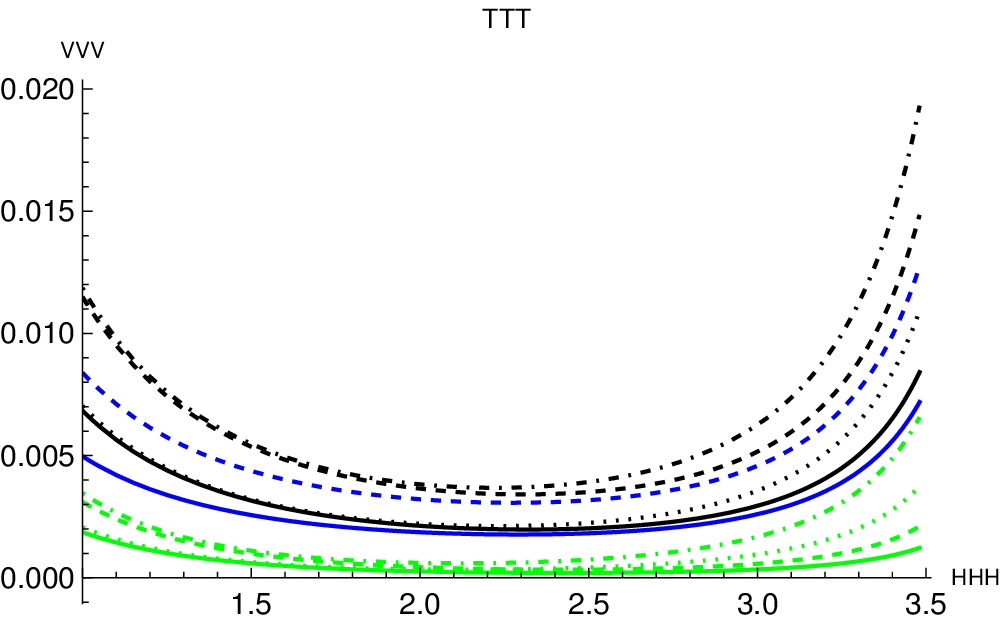}}}
	\vspace{0.2cm}
	\caption{\small The fully differential cross-section for $ \pi^{+} $ ($ \pi^{-} $) is shown as a function of $  \left( -u' \right)  $ on the left (right). The blue and green curves correspond to contributions from the vector and axial GPDs respectively. The black curves correspond to the total contribution, i.e. vector and axial GPD contributions combined. As before, the dashed (non-dashed) lines correspond to holographic (asymptotic) DA, while the dotted (non-dotted) lines correspond to the standard (valence) scenario. We fix $ S_{\gamma N}= 20000\,  \mathrm{GeV}^{2}  $ and $ M_{\gamma \meson}^{2}= 4\,  \mathrm{GeV}^{2}  $. Note that the vector contributions consist of only two curves in each case, since they are insensitive to either valence or standard scenarios.}
	\label{fig:EIC-LHC-UPC-fully-diff-VandA}
\end{figure}

Next, we show the relative contributions of the vector and axial GPDs to the fully differential cross-section in Figure \ref{fig:EIC-LHC-UPC-fully-diff-VandA}, as a function of $ (-u') $. To generate the plots, $ \Msq= 4\GeV^2 $ and $ \SgN = 20000 \GeV^2 $ were used.

\begin{figure}[h!]
	\psfrag{HHH}{\hspace{-1.5cm}\raisebox{-.6cm}{\scalebox{.8}{$-u' ({\rm 
					GeV}^{2})$}}}
	\psfrag{VVV}{\raisebox{.3cm}{\scalebox{.9}{$\hspace{-.4cm}\displaystyle\left.\frac{d 
					\sigma_{\gamma\pi^+}}{d M^2_{\gamma \pi^+} d(-u') d(-t)}\right|_{(-t)_{\rm min}}({\rm pb} \cdot {\rm GeV}^{-6})$}}}
	\psfrag{TTT}{}
	\vspace{0.2cm}
	\centerline{
		{\includegraphics[width=18pc]{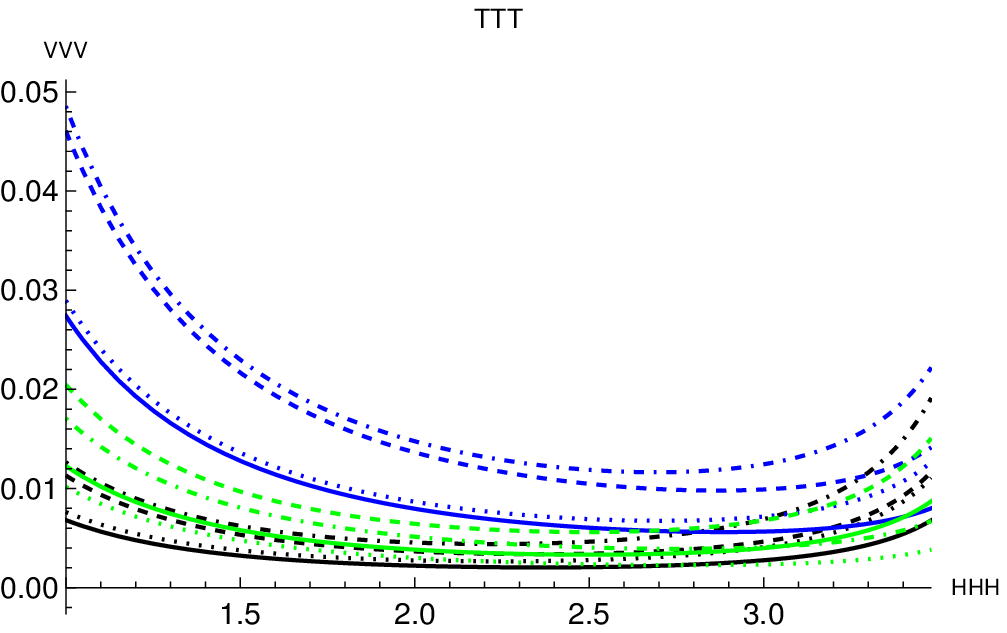}}
		\psfrag{VVV}{\raisebox{.3cm}{\scalebox{.9}{$\hspace{-.4cm}\displaystyle\left.\frac{d 
						\sigma_{\gamma\pi^-}}{d M^2_{\gamma \pi^-} d(-u') d(-t)}\right|_{(-t)_{\rm min}}({\rm pb} \cdot {\rm GeV}^{-6})$}}}
		{\includegraphics[width=18pc]{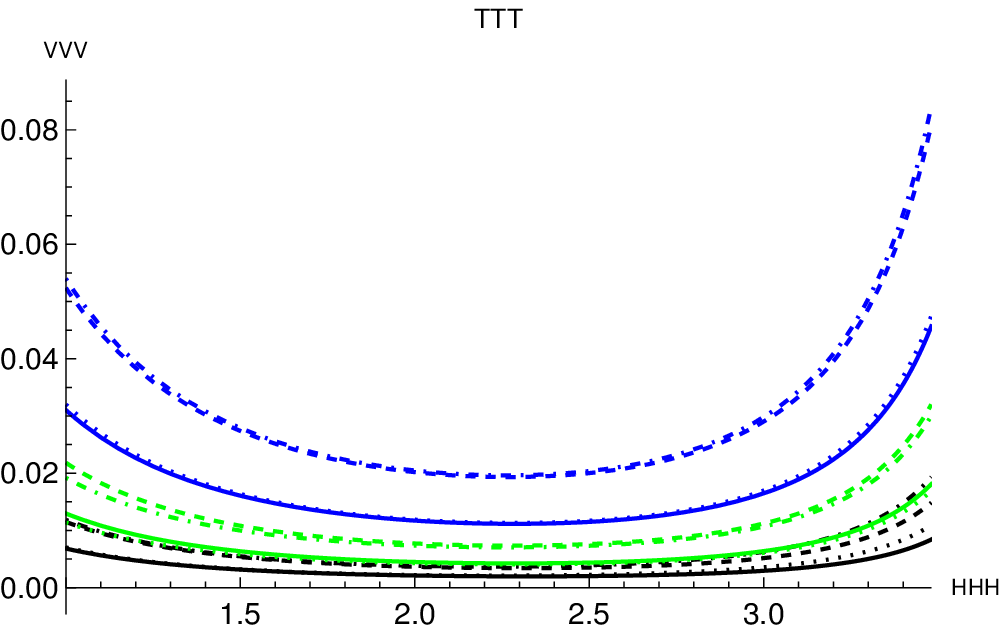}}}
	\vspace{0.2cm}
	\caption{\small The fully differential cross-section for $ \pi^{+} $ ($ \pi^{-} $) is shown as a function of $  \left( -u' \right)  $ on the left (right). The blue and green curves correspond to contributions from the u quark ($ H_{u} $ and $  \tilde{H} _{u} $) and d quark ($ H_{d} $ and $  \tilde{H} _{d} $) GPDs respectively. The black curves correspond to the total contribution. The dashed (non-dashed) lines correspond to holographic (asymptotic) DA, while the dotted (non-dotted) lines correspond to the standard (valence) scenario. We fix $ S_{\gamma N}= 20000\,  \mathrm{GeV}^{2}  $ and $ M_{\gamma \meson}^{2}= 4\,  \mathrm{GeV}^{2}  $. Similar comments as in Figure \ref{fig:compass-fully-diff-uandd} apply.}
	\label{fig:EIC-LHC-UPC-fully-diff-uandd}
\end{figure}

Finally, we show the relative contributions of the u quark and d quark GPDs to the fully differential cross-section in Figure \ref{fig:EIC-LHC-UPC-fully-diff-uandd}, as a function of $ (-u') $. We used $ \SgN = 20000 \GeV^2 $ and $ \Msq = 4 \GeV^2 $ to generate the plots.

\subsubsection{Single differential cross-section}

\label{sec:EIC-LHC-UPC-single-diff-X-section}

\begin{figure}[h!]
	\psfrag{HHH}{\hspace{-1.5cm}\raisebox{-.6cm}{\scalebox{.8}{$\Msq  ({\rm 
					GeV}^{2})$}}}
	\psfrag{VVV}{\raisebox{.3cm}{\scalebox{.9}{$\hspace{-.4cm}\displaystyle\frac{d 
					\sigma_{\gamma\pi^+}}{d M^2_{\gamma \pi^+}}({\rm pb} \cdot {\rm GeV}^{-2})$}}}
	\psfrag{TTT}{}
	\vspace{0.2cm}
	\centerline{
		{\includegraphics[width=18pc]{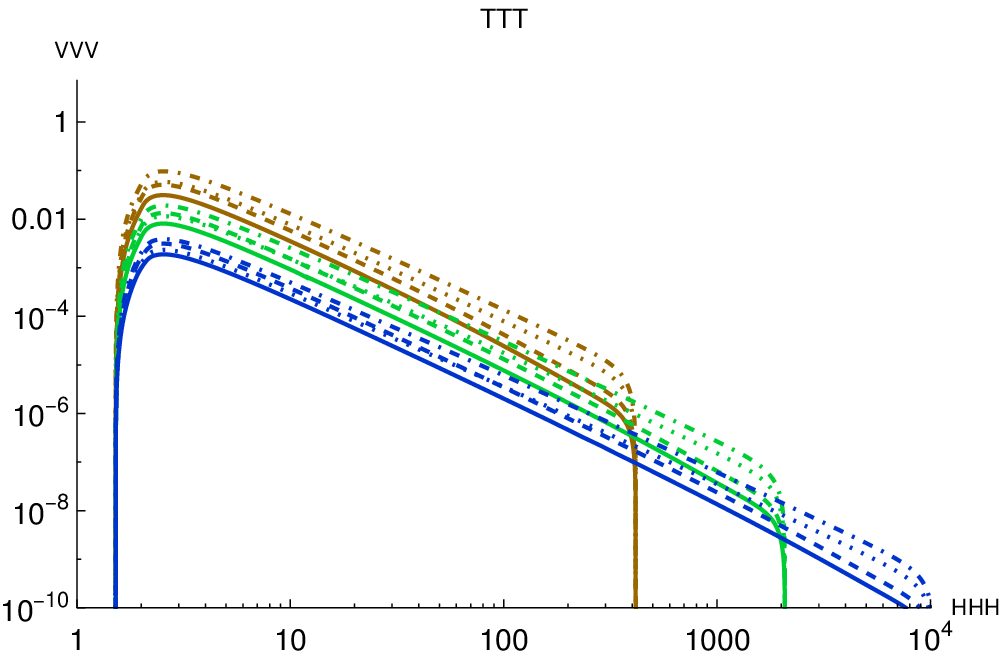}}
		\psfrag{VVV}{\raisebox{.3cm}{\scalebox{.9}{$\hspace{-.4cm}\displaystyle\frac{d 
						\sigma_{\gamma\pi^-}}{d M^2_{\gamma \pi^-}}({\rm pb} \cdot {\rm GeV}^{-2})$}}}
		{\includegraphics[width=18pc]{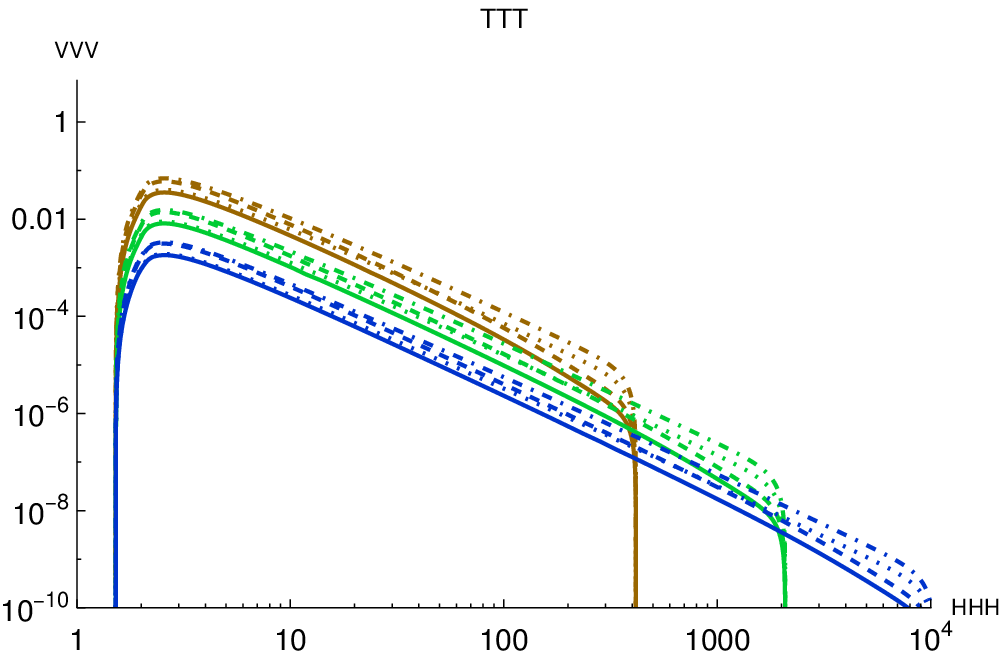}}}
	\vspace{0.2cm}
	\caption{\small The single differential cross-section for $ \pi^{+} $ ($ \pi^{-} $) is shown as a function of $  M_{\gamma \meson}^{2}  $ on the left (right) for different values of $ S_{\gamma N} $. The brown, green and blue curves correspond to $ S_{\gamma N} = 800,\,4000,\,20000\,\GeV^{2} $. The dashed (non-dashed) lines correspond to holographic (asymptotic) DA, while the dotted (non-dotted) lines correspond to the standard (valence) scenario. The holographic DA with the standard scenario has the largest contribution for every $ \SgN $. Note that both axes are log scales.}
	\label{fig:EIC-LHC-UPC-sing-diff}
\end{figure}

We now turn to the cross-section integrated over $ (-u') $ and $ (-t) $, i.e. single differential cross-section in $ \Msq $. The results are shown in Figure \ref{fig:EIC-LHC-UPC-sing-diff}, where 3 typical different values of $ \SgN $ are used, namely 800, 4000 and 20000 GeV$ ^{2} $. Note that we choose to show the plots using a log scale for the both axes, since the peak of the cross-section is very close to zero (roughly 3-4 $ \GeV^2 $) on such a wide range, and the variations in the values of the cross-section are huge.

\subsubsection{Integrated cross-section}

\begin{figure}[h!]
	\psfrag{HHH}{\hspace{-1.5cm}\raisebox{-.6cm}{\scalebox{.8}{$ S_{\gamma N} ({\rm 
					GeV}^{2})$}}}
	\psfrag{VVV}{\raisebox{.3cm}{\scalebox{.9}{$\hspace{-.4cm}\displaystyle
				\sigma_{\gamma\pi^+}({\rm pb})$}}}
	\psfrag{TTT}{}
	\vspace{0.2cm}
	\centerline{
		{\includegraphics[width=18pc]{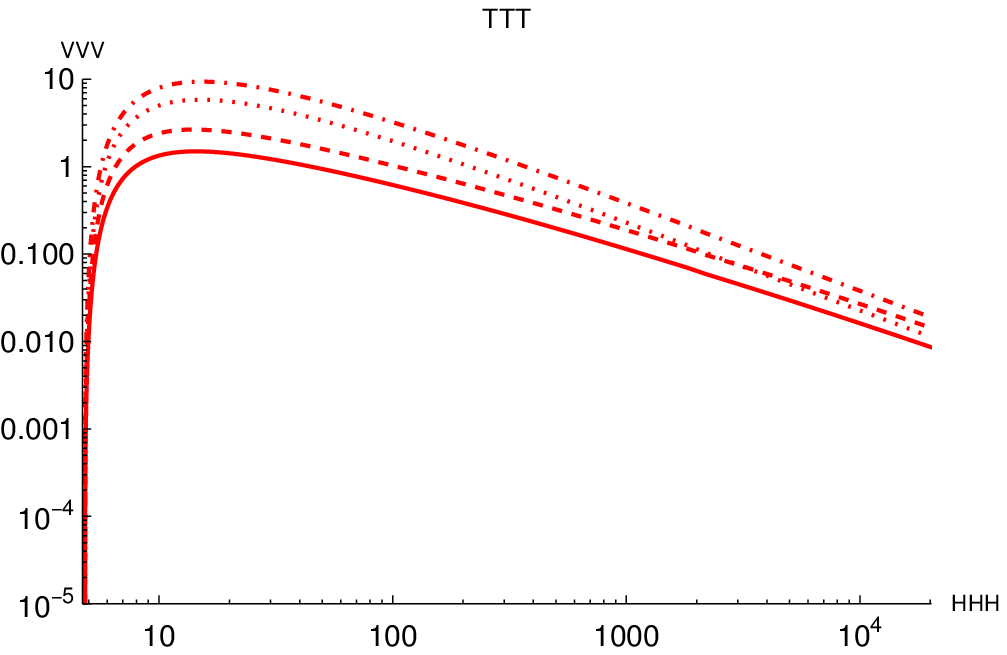}}
		\psfrag{VVV}{\raisebox{.3cm}{\scalebox{.9}{$\hspace{-.4cm}\displaystyle
					\sigma_{\gamma\pi^-}({\rm pb})$}}}
		{\includegraphics[width=18pc]{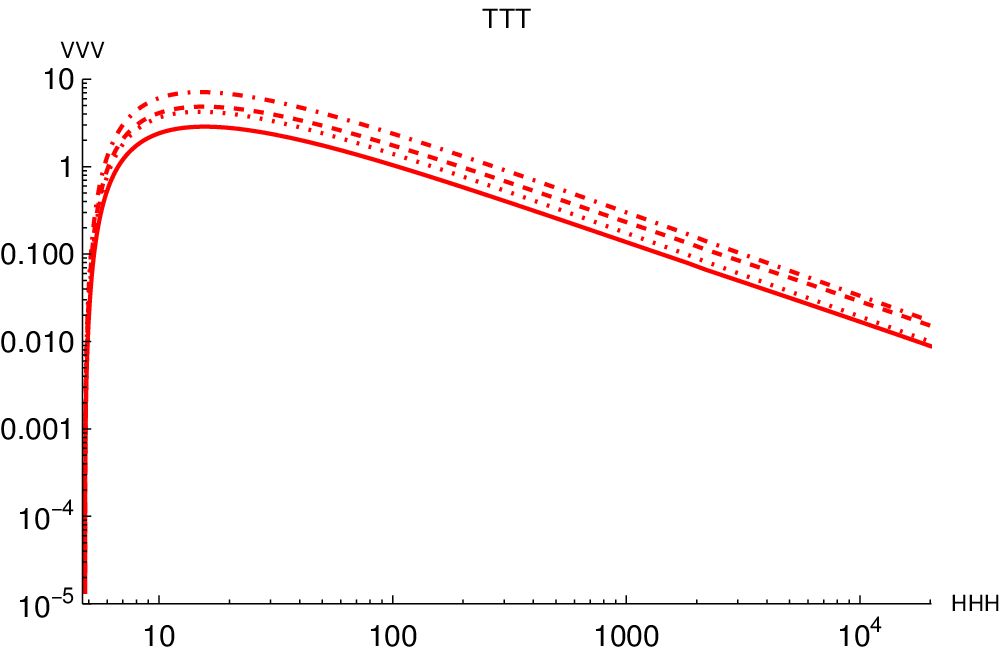}}}
	\vspace{0.2cm}
	\caption{\small The integrated cross-section for $ \pi^{+} $ ($ \pi^{-} $) is shown as a function of $   S_{\gamma N}  $ on the left (right). The dashed (non-dashed) lines correspond to holographic (asymptotic) DA, while the dotted (non-dotted) lines correspond to the standard (valence) scenario.}
	\label{fig:EIC-LHC-UPC-int-sigma}
\end{figure}

The fully integrated cross-section is shown as a function of $ \SgN $ in Figure \ref{fig:EIC-LHC-UPC-int-sigma}. As with the single differential cross-section plot in the previous section, we use a log scale for both the vertical and the horizontal axes, since the peak of the curve occurs at relatively small $ \SgN $ (roughly 20 GeV$ ^2 $), and the variations in the values of the cross-section are huge. As before, the holographic DA case with the standard scenario has the largest cross-section among the 4 possible cases.

We note that the cross-section falls to very low values at $ \SgN=20000 \GeV^2 $, about 200 times less than its value at the peak. Therefore, as far the cross-section as a function of $ \SgN $ is concerned, truncating at $ \SgN=20000 \GeV^2 $ for UPCs at LHC kinematics, which involves TeV energies, is a very good approximation. In addition to the decrease in the cross-sectional values themselves, the photon flux also decreases rapidly with $ \SgN $, making this approximation even stronger.

\subsubsection{Polarisation asymmetries}

\begin{figure}[h!]
	\psfrag{HHH}{\hspace{-1.5cm}\raisebox{0.4cm}{\scalebox{.8}{$-u' ({\rm 
					GeV}^{2})$}}}
	\psfrag{VVV}{LPA}
	\psfrag{TTT}{}
	\vspace{0.2cm}
	\centerline{
		{\includegraphics[width=18pc]{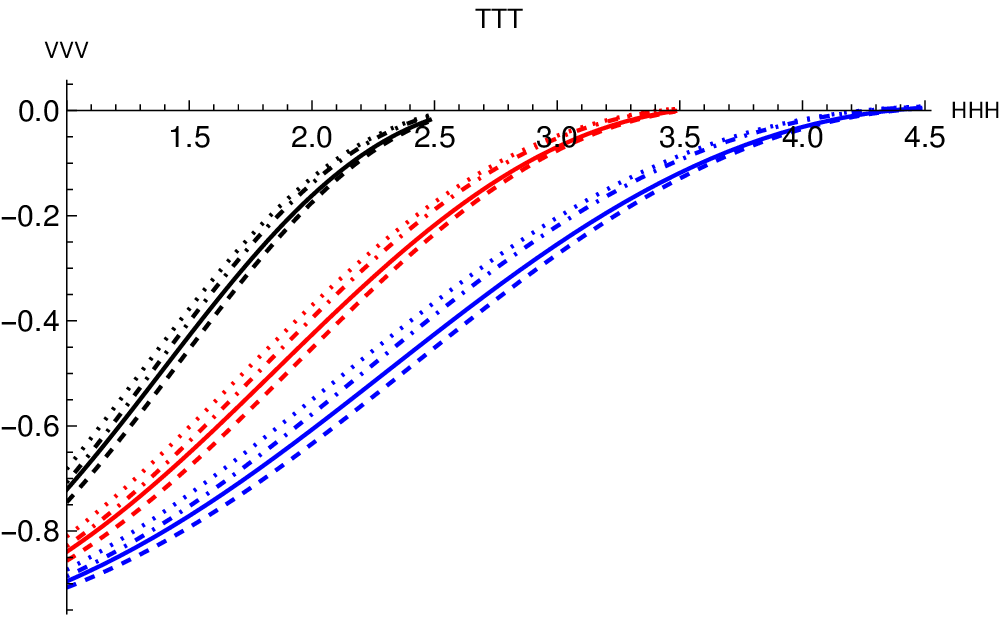}}
		\psfrag{VVV}{LPA}
		{\includegraphics[width=18pc]{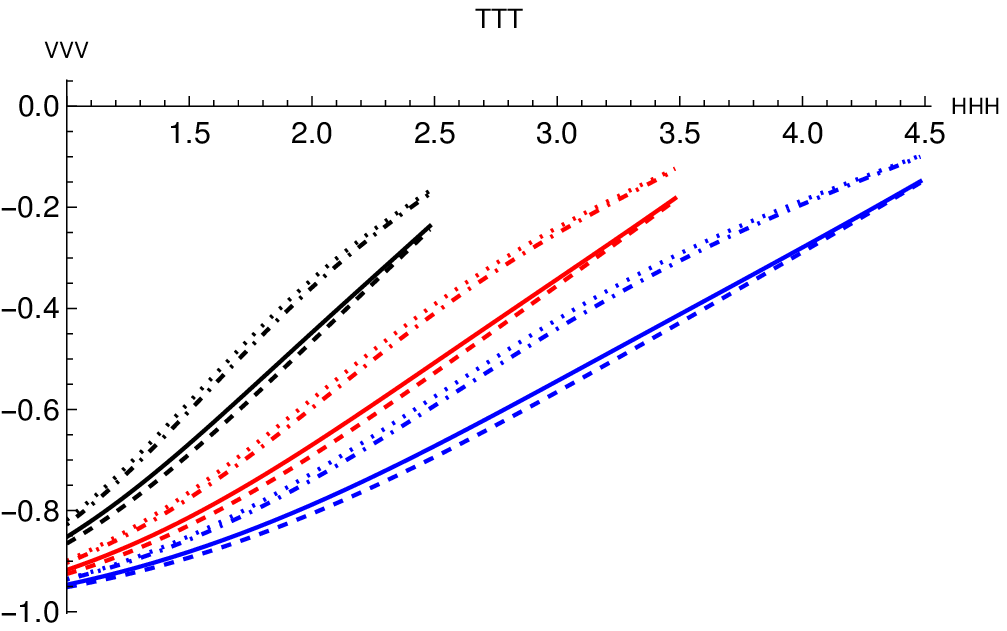}}}
	\vspace{0.2cm}
	\caption{\small The LPA at the fully-differential level for $ \pi^{+} $ ($ \pi^{-} $) is shown as a function of $  \left( -u' \right)  $ on the left (right) for different values of $ M_{\gamma \meson}^2 $. The black, red and blue curves correspond to $ M_{\gamma \meson}^{2}=3,\,4,\,5\, $ GeV$ ^2 $ respectively, and $ \SgN = 20000 \GeV^{2}$. The dashed (non-dashed) lines correspond to holographic (asymptotic) DA, while the dotted (non-dotted) lines correspond to the standard (valence) scenario.}
	\label{fig:EIC-LHC-UPC-pol-asym-fully-diff-diff-M2}
\end{figure}

The LPA at the fully differential level is shown in Figure \ref{fig:EIC-LHC-UPC-pol-asym-fully-diff-diff-M2} as a function of $ (-u') $. The values used are 3, 4 and 5 GeV$ ^2 $ for $ \Msq $, and $ 20000 \GeV^2 $ for $ \SgN $. Similar comments as in \SEC\ref{sec:pol-asym-COMPASS} hold.

\begin{figure}[h!]
	\psfrag{HHH}{\hspace{-1.1cm}\raisebox{0.4cm}{\scalebox{.8}{$-u' ({\rm 
					GeV}^{2})$}}}
	\psfrag{VVV}{LPA}
	\psfrag{TTT}{}
	\vspace{0.2cm}
	\centerline{
		{\includegraphics[width=18pc]{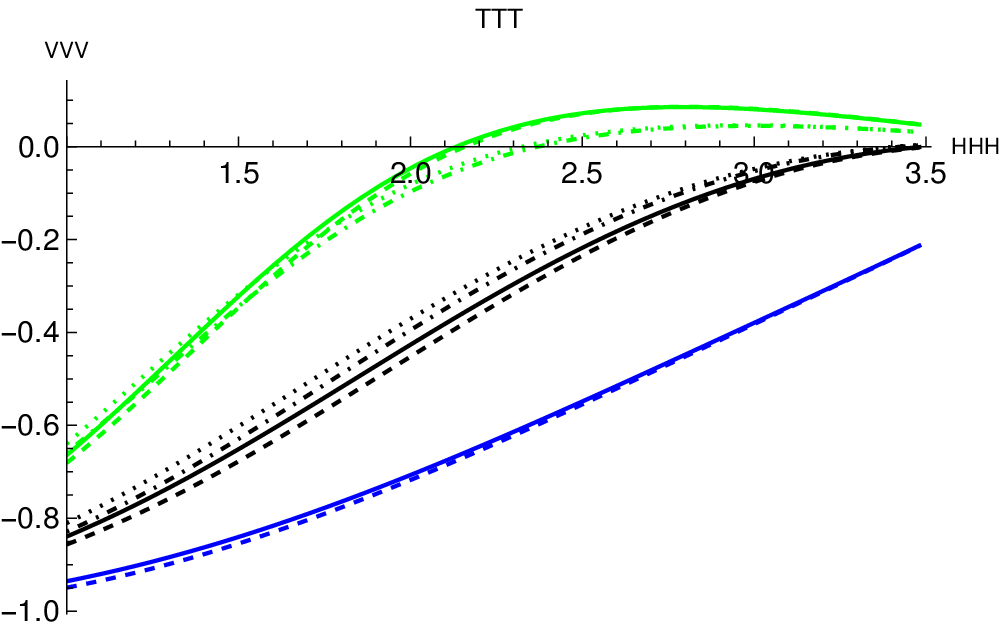}}
		\psfrag{VVV}{LPA}
		\psfrag{HHH}{\hspace{-1.1cm}\raisebox{0.15cm}{\scalebox{.8}{$-u' ({\rm 
					GeV}^{2})$}}}
		{\includegraphics[width=18pc]{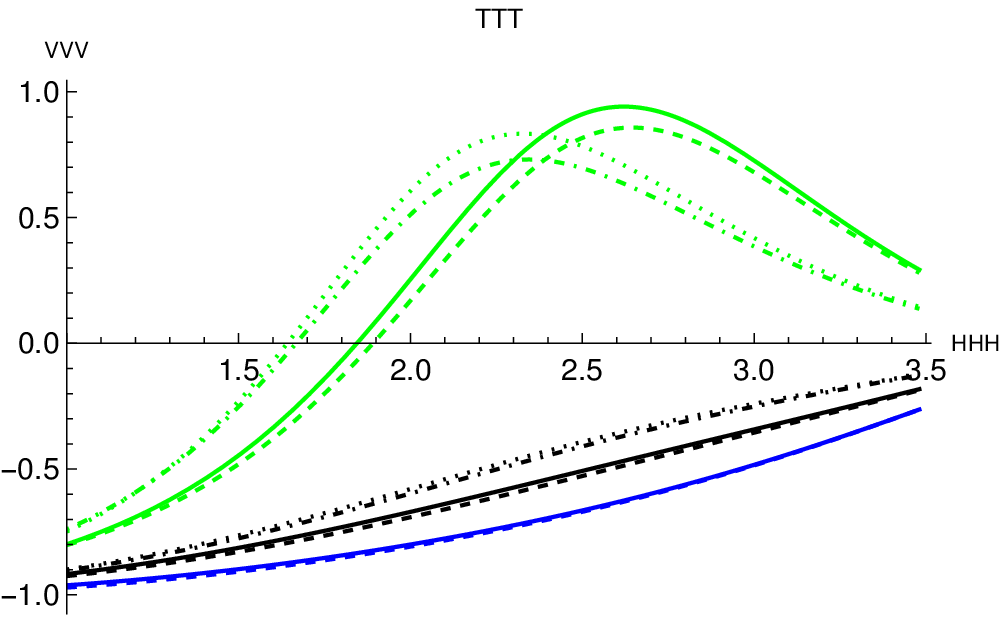}}}
	\vspace{0.2cm}
	\caption{\small The LPA at the fully-differential level for for $ \pi^{+} $ ($ \pi^{-} $) is shown as a function of $  \left( -u' \right)  $ on the left (right), using $ \Msq = 4 \GeV^2$ and $ \SgN = 20000 \GeV^2 $. The black curves correspond to the total contribution, i.e. vector and axial GPD contributions combined. As before, the dashed (non-dashed) lines correspond to holographic (asymptotic) DA, while the dotted (non-dotted) lines correspond to the standard (valence) scenario. Note that the vector contributions consist of only two curves in each case, since they are insensitive to either valence or standard scenarios. The effect of using the valence or standard scenario is significant, while the difference between using asymptotic and holographic DA is minimal.}
	\label{fig:EIC-LHC-UPC-pol-asym-fully-diff-VandA}
\end{figure}

Next, the relative contributions from the vector and axial GPDs to the LPA are shown in Figure \ref{fig:EIC-LHC-UPC-pol-asym-fully-diff-VandA}. $ \SgN =20000 \GeV^2$ and $ \Msq=4 \GeV^2 $ were used to generate the plots. As before, the axial GPD contributions using the standard and valence scenarios are significantly different, while the DA model has little effect on the LPA.

\begin{figure}[h!]
	\psfrag{HHH}{\hspace{-1.2cm}\raisebox{0.4cm}{\scalebox{.8}{$-u' ({\rm 
					GeV}^{2})$}}}
	\psfrag{VVV}{LPA}
	\psfrag{TTT}{}
	\vspace{0.2cm}
	\centerline{
		{\includegraphics[width=18pc]{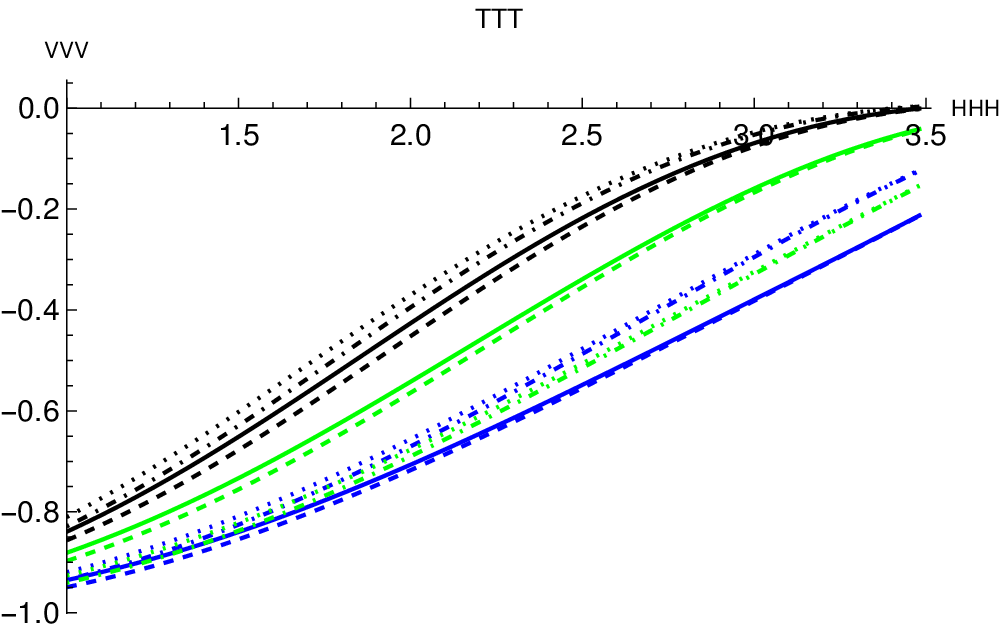}}
		\psfrag{VVV}{LPA}
		{\includegraphics[width=18pc]{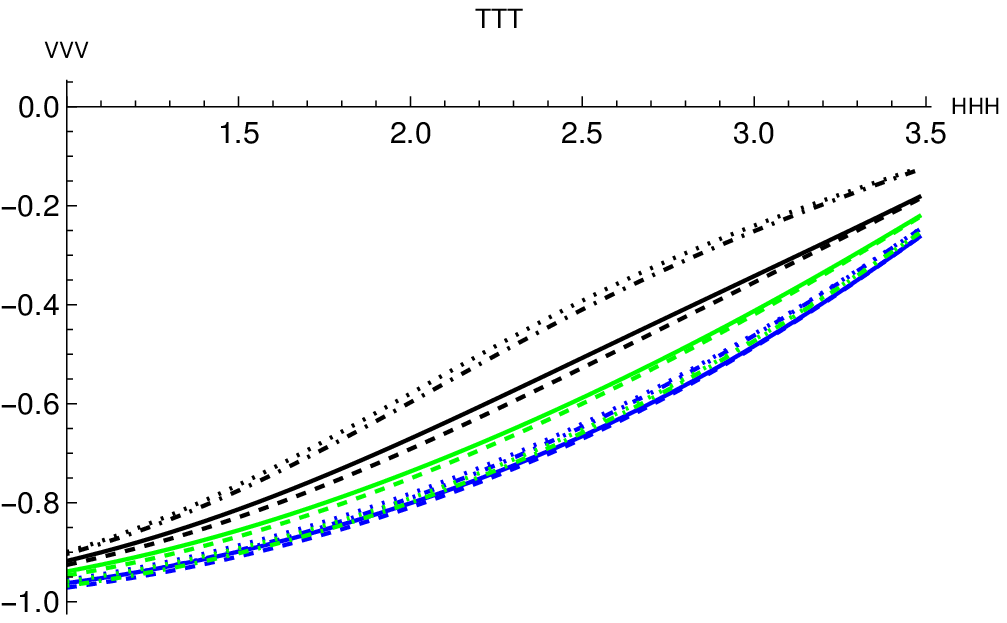}}}
	\vspace{0.2cm}
	\caption{\small The LPA at the fully-differential level for for $ \pi^{+} $ ($ \pi^{-} $) is shown as a function of $  \left( -u' \right)  $ on the left (right), using $ \Msq = 4 \GeV^2$ and $ \SgN = 20000 \GeV^2 $. The blue and green curves correspond to contributions from the u quark ($ H_{u} $ and $  \tilde{H} _{u} $) and d quark ($ H_{d} $ and $  \tilde{H} _{d} $) GPDs respectively. The black curves correspond to the total contribution. The dashed (non-dashed) lines correspond to holographic (asymptotic) DA, while the dotted (non-dotted) lines correspond to the standard (valence) scenario.}
	\label{fig:EIC-LHC-UPC-pol-asym-fully-diff-uandd}
\end{figure}

The last plot for the LPA at the fully differential level is shown in Figure \ref{fig:EIC-LHC-UPC-pol-asym-fully-diff-uandd}. It corresponds to the relative contributions of the u quark and d quark GPDs. The values used to generate the plots are $ \Msq = 4 \GeV^2 $ and $ \SgN = 20000 \GeV^2 $. 

\begin{figure}[h!]
	\psfrag{HHH}{\hspace{-1.5cm}\raisebox{-.6cm}{\scalebox{.8}{ $ M_{\gamma \meson}^{2}({\rm 
					GeV}^{2}) $}}}
	\psfrag{VVV}{LPA}
	\psfrag{TTT}{}
	\vspace{0.2cm}
	\centerline{
		{\includegraphics[width=18pc]{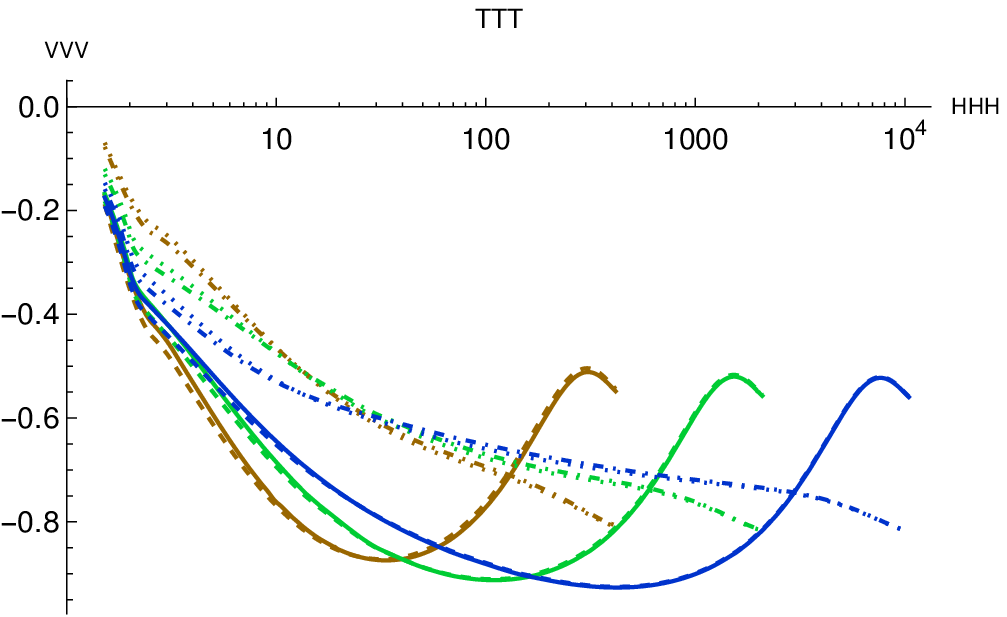}}
		\psfrag{VVV}{LPA}
		{\includegraphics[width=18pc]{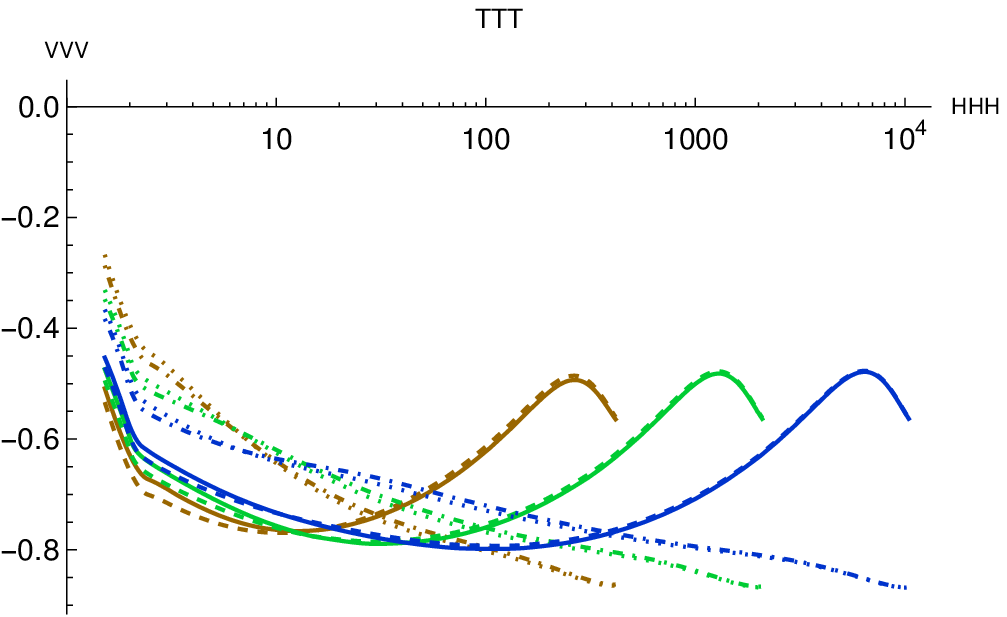}}}
	\vspace{0.2cm}
	\caption{\small The LPA at the single differential level for $ \pi^{+} $ ($ \pi^{-} $) is shown as a function of $  M_{\gamma \meson}^{2}  $ on the left (right). The brown, green and blue curves correspond to $ S_{\gamma N} = 800,\,4000$ and 20000 GeV$ ^2 $. The dashed (non-dashed) lines correspond to holographic (asymptotic) DA, while the dotted (non-dotted) lines correspond to the standard (valence) scenario. As was the case in Figure \ref{fig:compass-pol-asym-sing-diff}, we note that the choice of the GPD model (valence or standard) gives a completely different shape for the LPA, whereas the model for the DA has minimal effect. Note that a log scale is used for the horizontal axis.}
	\label{fig:EIC-LHC-UPC-pol-asym-sing-diff}
\end{figure}

Next, we turn to results for the LPA at the single differential level, as a function of $ \Msq $. This is shown in Figure \ref{fig:EIC-LHC-UPC-pol-asym-sing-diff} for $ \SgN= 800,\,4000$ and 20000 GeV$ ^2 $. Again, we note that the LPA can be used to distinguish between the GPD models (standard vs valence), but is not so efficient for separating DA models.

\begin{figure}[h!]
	\psfrag{HHH}{\hspace{-1.5cm}\raisebox{-.6cm}{\scalebox{.8}{ $ S_{\gamma N}({\rm 
					GeV}^{2}) $}}}
	\psfrag{VVV}{LPA}
	\psfrag{TTT}{}
	\vspace{0.2cm}
	\centerline{
		{\includegraphics[width=18pc]{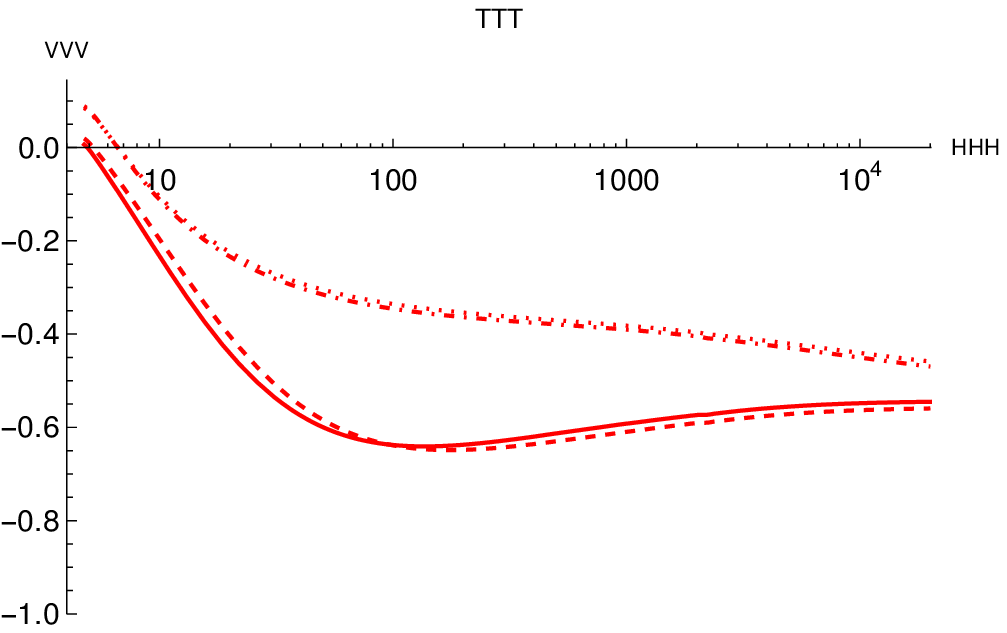}}
		\psfrag{VVV}{LPA}
		{\includegraphics[width=18pc]{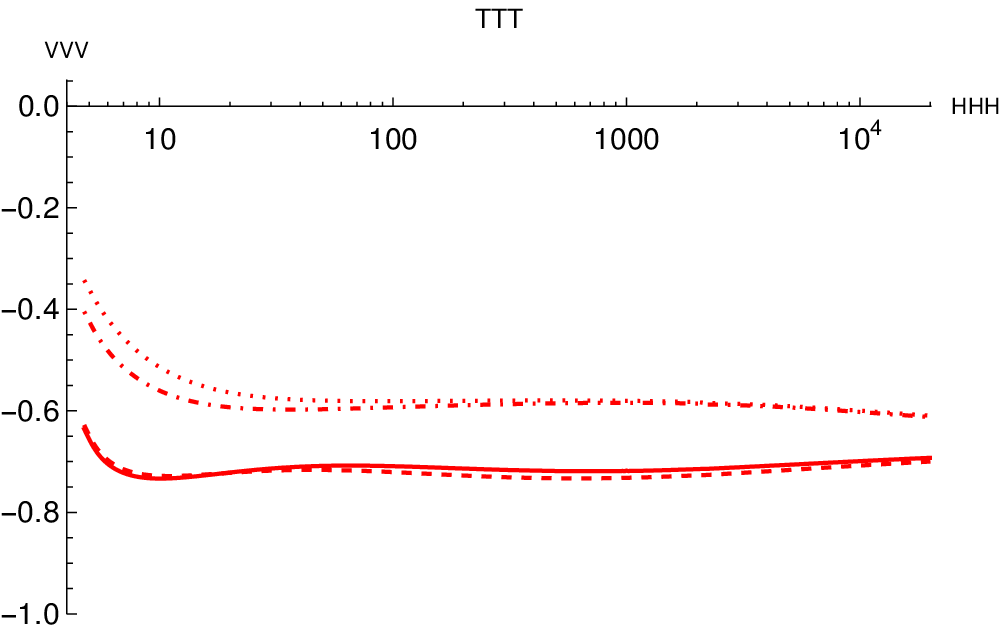}}}
	\vspace{0.2cm}
	\caption{\small The LPA integrated over all differential variables for $ \pi^{+} $ ($ \pi^{-} $) is shown on the left (right) as a function of $ \SgN $. The dashed (non-dashed) lines correspond to holographic (asymptotic) DA, while the dotted (non-dotted) lines correspond to the standard (valence) scenario.}
	\label{fig:EIC-LHC-UPC-pol-asym-int-sigma}
\end{figure}

Finally, we show how the LPA, integrated over the differential variables, behaves as a function of $ \SgN $ in Figure \ref{fig:EIC-LHC-UPC-pol-asym-int-sigma}. As in the plots at the single differential level in Figure \ref{fig:EIC-LHC-UPC-pol-asym-sing-diff}, the model used for the axial GPDs (valence or standard) has a significant effect, while the model used for the DAs (asymptotic or holographic) has a negligible effect.

\subsection{Counting rates}

\label{sec:counting-rates}

\subsubsection{JLab}

\label{sec:jlab-counting-rates}

In JLab, the source of photons is the electron beam. Hence, one uses the Weizs\"acker-Williams distribution to calculate the photon flux. The details of the formulae used are found in \APP\ref{app:WW-distribution}. The parameter $ Q^{2}_{ \mathrm{max} } $ appearing in the WW distribution is fixed at 0.1 GeV$ ^{2} $.

In the case of a lepton beam, one should also consider Bethe-Heitler-type 
processes, in which the final real photon is emitted by the lepton beam.
As discussed in ref.~\cite{Boussarie:2016qop}, such a Bethe-Heitler contribution is 
suppressed with respect to the production mechanism studied here.

The angular coverage of the final state particles is in principle a potential 
experimental issue. It can be shown that the angular distribution of the outgoing photon at JLab Hall B, which 
might evade detection, does not affect our predictions. A detailed discussion of this matter can be found in \cite{Boussarie:2016qop, Duplancic:2018bum}, and therefore, we do not repeat it here.

Table \ref{tab:jlab-counting-rates} shows the counting rates expected at JLab for our process for $ \pi^{\pm} $, using a luminosity of 100 nb$ ^{-1} $s$ ^{-1} $, and 100 days of run. The minimum and maximum values of the counting rates correspond to the boundaries obtained by considering all the 4 different possibilities, i.e. the 2 models for the GPDs (standard and valence scenarios) and the 2 models for the DAs (asymptotic and holographic DA). In general, the lowest value is obtained for an asymptotic DA with valence scenario, while the largest value is obtained for a holographic DA with standard scenario. We find that the values obtained for the JLab experiment are very promising.

\begin{table}[h!]
	\centering
\begin{tabular}{|c|c|}
\hline
Meson & Counting rates \\
\hline
\hline
$  \pi ^{+} $ & 0.29-1.76 $  \times 10^5  $\\
\hline
$  \pi ^{-} $ & 0.53-1.33 $  \times 10^5$ \\
\hline
\end{tabular}
\caption{Estimated counting rates at JLab for $ \pi^{\pm} \gamma  $ photoproduction.}
\label{tab:jlab-counting-rates}
\end{table}

\subsubsection{COMPASS}

At COMPASS, the source of photons is the muon beam. Thus, like in the previous section, one uses the Weizs\"acker-Williams distribution to obtain the photon flux, with the small modification that the electron mass is replaced by the muon mass. As in the previous subsection, $ Q^{2}_{ \mathrm{max} } $ is fixed at 0.1 GeV$ ^2 $.

Table \ref{tab:compass-counting-rates} shows the counting rates expected at COMPASS for our process for $ \pi^{\pm} $. Like before, the minimum and maximum values of the counting rates correspond to the boundaries obtained by considering all the different possibilities, i.e. the 2 models for the GPDs (standard and valence scenarios) and the 2 models for the DAs (asymptotic and holographic DA). In general, the lowest value is obtained for an asymptotic DA with valence scenario, while the largest value is obtained for a holographic DA with standard scenario. To obtain the values in the table, we assumed a luminosity of 0.1 nb$ ^{-1} $s$ ^{-1} $, and 300 days of run.

\begin{table}[h!]
	\centering
	\begin{tabular}{|c|c|}
		\hline
		Meson & Counting rates \\
		\hline
		\hline
		$  \pi ^{+} $ & 1.23-7.37 $ \times 10^2 $ \\
		\hline
		$  \pi ^{-} $ & 2.27-5.55 $ \times 10^2 $ \\
		\hline
	\end{tabular}
	\caption{Estimated counting rates at COMPASS for $ \pi^{\pm}\gamma  $ photoproduction.}
	\label{tab:compass-counting-rates}
\end{table}

\subsubsection{EIC}

\label{sec:EIC-counting-rates}

In the computation of the counting rates, the only difference between EIC and JLab is the centre of mass energy of the electron-nucleon system, $ S_{eN} $. 

Table \ref{tab:EIC-counting-rates} shows the expected counting rates for EIC, assuming a total integrated luminosity of $ 10^7 $ nb$ ^{-1} $. In particular, we use the highest expected electron-nucleon centre of mass energy, corresponding to  $ S_{eN}= 19600 \GeV^2 $ \cite{AbdulKhalek:2021gbh}. In addition, because of the high centre-of-mass energy available, one can study the kinematic region where $  \xi  $ is small. Therefore, we also show the values of the counting rates with the constraint that $ \SgN > 300 \GeV^2 $, which corresponds to $  \xi \lesssim 5  \cdot 10^{-3} $ \footnote{Note that the relation between $ \SgN $ and $  \xi  $ involves $ \Msq $, which is why a cut in $\SgN$ does not directly correspond to a cut in $\xi$. However, as can be seen in \SEC\ref{sec:EIC-LHC-UPC-single-diff-X-section}, the cross-section is dominated by small $ \Msq $, so the region of small $  \xi  $ is actually the one where most of the contribution comes from.}. In fact, values of $  \xi  $ as small as $ 7.5 \cdot 10^{-6} $ can be probed. With the cut in $ \SgN $, the counting rate decreases by roughly a factor of 20, and this is due to the fact that the peak of the cross-section is located at low $ \SgN $, roughly $ 20 \GeV^2 $, as can be seen in Figures \ref{fig:compass-int-sigma} and \ref{fig:EIC-LHC-UPC-int-sigma}. The minimum and maximum values for the counting rates in Table \ref{tab:EIC-counting-rates} are obtained as described in previous sections.

\begin{table}[h!]
	\centering
	\begin{tabular}{|c|c|c|}
		\hline
		Meson & Total Counting rates & Counting rates with $ \SgN > 300 \GeV^2 $\\
		\hline
		\hline
		$  \pi ^{+} $ & 0.23-1.3 $ \times 10^{4}$ & 1.40-5.00 $ \times 10^2 $ \\
		\hline
		$  \pi ^{-} $ & 0.42-1.0 $ \times 10^{4}$ & 1.76-3.87 $ \times 10^2 $\\
		\hline
	\end{tabular}
	\caption{Estimated counting rates at EIC kinematics for $ \pi^{\pm}\gamma  $ photoproduction.}
	\label{tab:EIC-counting-rates}
\end{table}

\subsubsection{Ultraperipheral collisions at LHC}

\label{sec:UPC-LHC}

In ultraperipheral collisions (UPCs), the beam and target are far enough apart such that there are no hadronic interactions between them. Thus, the nucleus/proton interacts by the exchange of photons. In particular, heavy nuclei, such as lead, can act as a good source of photons, since the photon flux scales as $ Z^2 $, where $ Z $ is the charge of the nucleus. The details on how the photon flux is obtained can be found in \APP\ref{app:photon-flux-UPC}. We work in the limit where the lead nucleus acts as the source of the quasi-real photon.

Table \ref{tab:LHC-UPC-counting-rates} shows the counting rates corresponding to p-Pb UPCs at LHC, assuming an integrated luminosity of 1200 nb$ ^{-1} $, which corresponds to the expected data taking for runs 3 and 4 \cite{Citron:2018lsq}. As in \SEC\ref{sec:EIC-counting-rates}, there is an order of magnitude drop in the counting rate when a cut of $ \SgN > 300 \GeV^2 $ is imposed. We note that preliminary results for UPCs at LHC have already been reported in \cite{Duplancic:2022yzi}.

\begin{table}[h!]
	\centering
	\begin{tabular}{|c|c|c|}
		\hline
		Meson & Total Counting rates & Counting rates with $ \SgN > 300 \GeV^2 $\\
		\hline
		\hline
		$  \pi ^{+} $ & 1.6-9.3 $ \times 10^{3}$ & 1.00-3.40 $ \times 10^2 $ \\
		\hline
	\end{tabular}
	\caption{Estimated counting rates at p-Pb UPCs at LHC for $ \pi^{+}\gamma  $ photoproduction.}
	\label{tab:LHC-UPC-counting-rates}
\end{table}

\section{Conclusion}

\label{sec:conclusion}


In this work, we extend the analysis 
of $\gamma N \to (\gamma \meson^{\pm}) N'$ process
introduced in \cite{Duplancic:2018bum} 
by including the linear polarisation asymmetries, 
extending the kinematics to selected future experiments (COMPASS, EIC and UPCs at LHC), 
and computing predictions for an alternative 'holographic' DA
(\ref{DA-hol}). 
Since we consider the  large angle scattering kinematics, 
which amounts to large $ (-u') $ and $ \Msq $, and small $ (-t) $, we are able to employ the collinear factorization.
In fact, QCD factorisation has been recently proven to hold for a family of $2 \to 3 $ exclusive processes \cite{Qiu:2022bpq, Qiu:2022pla}, which includes our process,
for large $   |\vec{p}_{t}|  $.
Therefore, it is expected to hold for
large $ \Msq $, which is a more strict kinematical limit.

Our results show that the exclusive photoproduction of a $ \gamma {\meson}^{\pm} $ pair with a large invariant mass provides another interesting channel to study GPDs, besides the extensively studied channels such as DVCS, DVMP and TCS. The counting rates at various experiments have been estimated in \SEC\ref{sec:counting-rates}, and the values look promising, especially at JLab where they were estimated to be of the order of $ 10^5 $. In fact, the GPD model corresponding to the standard scenario, which is favoured by lattice results \cite{Alexandrou:2017oeh}, as well as its recent update in \cite{ Alexandrou:2020sml}, gives larger cross-sections. Furthermore, we found that the linear polarisation asymmetries wrt the incoming photon are significant, and become even larger at higher centre of mass energies. In particular, this is almost maximal in the case of the $\pi^-$. Moreover, by exploiting the high energies available at EIC and UPCs at LHC, one is able to probe GPDs in the region of small skewness $ \xi $, a region where very little is known about GPDs. 
We found that 
imposing kinematical cuts 
to the region  of small $ \xi  \leq 5\times 10^{-3} $ 
the counting rates drop by roughly a factor of 10,
which still leaves sufficient statistics. 

Phenomenologically, it is known that in $\pi^0$ electroproduction, higher twist-3 pion DA contributions are important at moderate energies. Indeed, the chiral nature of the $\pi$ meson leads to an anomalously large twist 3 chiral-odd DA and this component has been advocated to be the source of the large transverse amplitude for $\pi$ meson electroproduction measured at JLab~\cite{CLAS:2012cna, JeffersonLabHallA:2016wye} (let us remind the reader that the leading twist-2 factorization proof of DVMP applies only to the amplitude with longitudinally polarised photon). However, when naively calculated, one is faced with end-point singularities \cite{Kroll:2016aop}. We refer to \cite{Goloskokov:2009ia,Goloskokov:2011rd} for a possible treatment of such issues.
We would like to emphasise that for our process, which involves photoproduction, it is unclear whether such contributions would be important. This could be resolved by comparing our results, which are based on leading twist-2 DA, with experimental data. 
Nevertheless, it would be interesting to estimate the contribution of twist-3 pion DAs in our process.

Like was performed for the case of diphoton photoproduction in \cite{Grocholski:2021man}, we intend to perform the computation at NLO in $\alpha_{s}$, in the spirit of \cite{Nizic:1987sw,Duplancic:2006nv}. While QCD collinear factorisation was proved for our process, it is nevertheless desirable to determine the finite radiative NLO corrections after explicitly performing the cancellation of UV/IR divergences. The knowledge of such corrections, which are often significant for phenomenology, will increase the precision of our predictions and will give us the opportunity to estimate the uncertainties related to our process based on the collinear factorisation approach.

We will also extend our analysis to a final state neutral pion.
While the structure of the code allows this extension to be
performed with minimal effort for quark GPDs,
the quantum numbers of the $  \pi ^{0} $ meson, 
in particular its positive $ C $-parity, 
allow contributions from gluon GPDs\footnote{These are of course absent for charged pions on the basis of charge conservation, but such contributions are also absent in the case of $ \rho^{0} $, because of its negative $ C $-parity.}. Such gluonic contributions, with the pomeron quantum numbers, would eventually dominate at small $x$.

\section*{Acknowledgements}

We thank Nicole D'Hose,  Aude Glaenzer, Ronan McNulty, Kenneth Osterberg, Marco Santimaria, Daria Sokhan, Pawel Sznajder, Daniel Tapia Takaki, Charlotte van Hulse and Michael Winn for useful discussions. SN is supported by the GLUODYNAMICS project funded by the ``P2IO LabEx (ANR-10-LABEX-0038)" in the framework ``Investissements d’Avenir" (ANR-11-IDEX-0003-01) managed by the Agence Nationale de la Recherche (ANR), France. SN also acknowledges the hospitality of NCBJ where part of this work was done. The work of L. S. is supported by the grant 2019/33/B/ST2/02588 of the National Science Center in Poland. L. S. thanks the P2IO Laboratory of Excellence (Programme Investissements d’Avenir ANR-10-LABEX-0038) and the P2I - Graduate School of Physics of Paris-Saclay University for support. 
This publication is supported by the Croatian Science Foundation project IP-2019-04-9709,
and by the EU Horizon 2020 research and innovation programme, STRONG-2020
project, under grant agreement No 824093.

\appendix

\section{Amplitudes with a holographic distribution amplitude}

In this appendix, we show the explicit results for the amplitudes in terms of building block integrals. In particular, we discuss only the holographic DA case, since all the relevant results for the asymptotic DA case can be found in appendix D of \cite{Duplancic:2018bum}.

\label{app:holographic-DA}

\subsection{Integration over $z$}
\label{app:z-integration}

We consider the holographic DA
\beqa
\label{Def:DA-non-asymptotic}
\phiLC(z) = \frac{8}\pi \sqrt{z(1-z)}\,,
\eqa
which normalises to 1
\beqa
\label{norm-DA-non-asymptotic}
\int\limits_0^1 \phiLC(z) = 1\,,
\eqa
as seen immediately after using the change of variable $z =\cos^2t.$
To compute the contribution of the different Feynman diagrams  convoluted to the DA, we will need to compute a set of integrals.

The simplest one is
\beqa
\label{Def:int1-LC}
\int\limits_0^1 \frac{\phiLC(z)}{z \bar{z}} dz=8\,,
\eqa
analogous to the asymptotic DA case
\beqa
\label{Def:int1-AS}
\int\limits_0^1 \frac{\phiAS(z)}{z \bar{z}} dz=6 \,,
\eqa
which allows us to compute diagrams $A_1$, $A_2$, $A_3$, $B_1$, $B_2$, $B_3$. In particular, this means that for these diagrams, the same building block integrals as the asymptotic DA case can be used, provided the overall prefactor is changed from 6 to 8. In practice, what is done is that the prefactor is added a posteriori - When generating the tables, there is no prefactor.

We now consider the diagrams $A_{4,5}$ and $B_{4,5}$ which are more involved. First, let us consider the vector structure.
Denoting
\beqa
\label{Def:A-B}
A &=& 2 \xi +(1-\alpha)(x-\xi+ i \varepsilon)\,,\\
B &=& \alpha(x-\xi+i \varepsilon)\,,
\eqa
we consider the integral
\beqa
\label{Def:intK-LC}
K=\int\limits_0^1 \frac{\phiLC(z)}{z \bar{z}} \frac{\zb}{A \zb + B}dz =\frac{8}A-\frac{B}{A^2} I\,,
\eqa
with
\beqa
\label{Def:intI-LC}
I=\frac{8}\pi \int\limits_0^1 \frac{dz}{\left(z +\frac{B}A \right)\sqrt{z(1-z)}}\,,
\eqa
which enters the diagram $A_4^V$ (structure $T_A$). 
Using again the change of variable $z =\cos^2t,$ and then  introducing $u=e^{2it}$, a careful analysis of the pole structure in the complex $u$-plane in the various configurations depending on the signs of $A$, $B$, and on the position of $A/B$ with respect to $0$ and $1$, which are related to the values of $\xi$ and $\alpha$, leads to the following result valid for any configuration:
\beqa
\label{Res:intI-LC}
I=8 A B \sqrt{\frac{1}{B^3(A+B)}}\,.
\eqa
We stress that this result is valid for \textit{any} values of $  \alpha  $ and $  \xi  $ allowed by the kinematics of our process, and therefore, its form should not be modified during implementation.

Next, the diagram $A_4^V$ (structure $T_B$) involves the integral
\beqa
\label{Def:intM-LC}
M=\int\limits_0^1 \frac{\phiLC(z)}{z \bar{z}} \frac{\zb^2}{A \zb + B}dz =\frac{8}{A^2} \left(\frac{A}{2}-B\right)+\frac{B^2}{A^3} I\,.
\eqa
The diagram $A_5^V$ involves, for the structure $T_A$, the integral $I$  and
for the structure $T_B$ the integral
\beqa
\label{Def:intL-LC}
L=\int\limits_0^1 \frac{\phiLC(z)}{z \bar{z}} \frac{(\zb-z)(1-\alb z)}{A \zb + B}dz = 2 \alb M +(3 \alpha -1) K -\alpha \frac{I}A\,.
\eqa
The diagram $B_4^V$ involves, for both  $T_A$ and $T_B$ structures, the integral
\beqa
\label{Def:intR-LC}
R=\int\limits_0^1 \frac{\phiLC(z)}{z \bar{z}} \frac{z \zb}{A \zb + B}dz =
K-M\,.
\eqa
The diagram $B_5^V$ involves, for the structure $T_A$,
\beqa
\label{Def:intS-LC}
S=\int\limits_0^1 \frac{\phiLC(z)}{z \bar{z}} \frac{z}{A \zb + B}dz =
\frac{I}A-K\,,
\eqa
and
\beqa
\label{Def:intT-LC}
T=\int\limits_0^1 \frac{\phiLC(z)}{z \bar{z}} \frac{z^2}{A \zb + B}dz =
M-2 K+\frac{I}A\,,
\eqa
for the structure $T_B$.

We now consider the axial structure.
For the diagram $A_4^A$, the structure $T_{A_5}$ only involves the integral $K$ while the structure $T_{B_5}$ involves the integral
\beqa
U=\int\limits_0^1 \frac{\phiLC(z)}{z \zb} \frac{\zb (\zb-z)}{A \zb + B}dz =
2M-K\,.
\eqa
For the diagram $A_4^A$, the structure $T_{A_5}$ involves the integral 
\beqa
\label{Def:intV-LC}
V=\int\limits_0^1 \frac{\phiLC(z)}{z \zb} \frac{2-\alpha -2 \alb z}{A \zb + B}dz =
\alpha \frac{I}A + 2 \alb K\,,
\eqa
while the structure $T_{B_5}$ involves the integral
\beqa
\label{Def:intW-LC}
W=\int\limits_0^1 \frac{\phiLC(z)}{z \zb} \frac{1 -2 \alb z}{A \zb + B}dz =
(2\alpha-1) \frac{I}A + 2 \alb K\,.
\eqa
For the diagram $B_4^A$, the integrals are the same as for $B_4^V.$

For the diagram $B_5^A$, for the structure $T_{A_5}$, the integral involved is
\beqa
Z=\int\limits_0^1 \frac{\phiLC(z)}{z \zb} \frac{z (1-2 z)}{A \zb + B}dz =
-2M+3K- \frac{I}A \,.
\eqa
In this way, all the integrals corresponding to any Lorentz structure are fully defined.

\subsection{Integration over $x$ and explicit expressions}

\label{app:hol-DA-explicit}

We now pass to the convolution with GPDs. The integration over $x$ can only be done numerically, and involves a set of basic integrals which we now discuss.

The new following set of basic integrals wrt the case of asymptotic DA is relevant for the holographic DA:
\beqa
\label{Def:chi1}
\chi_{1}&=&\int \frac{1}{\sqrt{(x-\xi+i \epsilon)(x+\xi+i \epsilon)}} \frac{1}{2\xi+\alb(x-\xi+i \epsilon)} \sgn (x+\xi) f(x) d x\,, \ \\
\label{Def:chi2}
\chi_{2}&=&\int \frac{1}{\sqrt{(x-\xi+i \epsilon)(x+\xi+i \epsilon)}} \frac{1}{\left[2 \xi+\alb(x-\xi+i \epsilon)\right]^{2}} \sgn (x+\xi) f(x) d x \,, \ \ \\
\label{Def:chi3}
\chi_{3}&=&\int \frac{1}{\sqrt{(x-\xi+i \epsilon)(x+\xi+i \epsilon)}}  \frac{1}{x+\xi+i \epsilon}\sgn (x+\xi) f(x) dx\,. 
\eqa
In practice,  in order to deal with convergent integrals around the pole located at
\beq
\label{Def:xp}
x_p = -\xi \frac{1+\alpha}{1-\alpha},
\eq
one should rather consider the three following integrals:
\beqa
\chi_{a}&=&\int \left[ \frac{\sgn (x+\xi)}{\sqrt{x^2-\xi^2}}  + \frac{\alb}{2 \sqrt{\alpha} \xi}\right] \frac{1}{2\xi+\alb(x-\xi)} f(x) d x \,,\ \\
\chi_{b}&=&\int \left\{
\left[ \frac{\sgn (x+\xi)}{\sqrt{x^2-\xi^2}}  + \frac{\alb}{2 \sqrt{\alpha} \xi}\right] \frac{1}{\left[2\xi+\alb(x-\xi)\right]^2}+
\frac{(1+\alpha)\alb}{8 \xi^2 \alpha \sqrt{\alpha}}\frac{1}{2\xi+\alb(x-\xi)}\right\}f(x) d x \,,\nonumber  \\
\\
\chi_{c}&=& \int \frac{\sgn (x+\xi)}{\sqrt{x^2-\xi^2}}f(x) dx\,. 
\eqa

At this point, we stress again that the following sums of diagrams with the holographic DA
\beqa
&& N^q_{A_5}[(AB)_{123}]^s, \ N^q_{A_5}[(AB)_{123}]^a,
\nonumber \\
&& N^q_{B_5}[(AB)_{123}]^s, \ N^q_{B_5}[(AB)_{123}]^a,
\nonumber \\ 
&& \tilde{N}^q_A[(AB)_{123}]^s, \ \tilde{N}^q_A[(AB)_{123}]^a,
 \nonumber \\
&& \tilde{N}^q_B[(AB)_{123}]^s, \ \tilde{N}^q_B[(AB)_{123}]^a,
\eqa
can all be obtained through the expressions from Appendix D in \cite{Duplancic:2018bum} - Only a change in the overall prefactor from 6 to 8 is needed.

The remaining sums of diagrams are given by
\begin{align}
N^q_{A_5}[(AB)_{45}]&=8 i\left[-\frac{2-\alpha}{\alpha^2 \bar{\alpha}\xi^2}  I_{e}-\frac{1}{\alpha^{2} \bar{\alpha} \xi^{2}} I_{f}+\frac{2}{\alpha \bar{\alpha} \sqrt{\alpha} \xi^{2}} \chi_{c}-\frac{8}{\bar{\alpha} \sqrt{\alpha}} \chi_{b}\right]\,,\\[5pt]
N^q_{B_5}[(AB)_{45}]&= 8 i\left[-\frac{1}{\alpha \bar{\alpha} \xi^2} I_{e}+\frac{1-2\alpha}{\alpha^{2}\bar{\alpha}\xi^2} I_{f} +\frac{2}{\alpha \bar{\alpha} \sqrt{\alpha}\xi^{2}} \chi_{c}-\frac{8}{\bar{\alpha} \sqrt{\alpha}} \chi_b \right]\,,
\end{align}
for the $ SP $ case, and 
\begin{align}
 \tilde{N} ^q_{A}[(AB)_{45}]&=8 
 \left[\frac{2-\alpha}{\alpha \xi} I_{e} + \frac{-1+2 \alpha}{\alpha \xi} I_{f} - \frac{2}{\xi \sqrt{\alpha}} \chi_c -8 \xi \sqrt{\alpha} \chi_b  \right]\,,\\[5pt]
 \tilde{N} ^q_{B}[(AB)_{45}]&= 
\frac{8}{\alpha^2 \xi^2}  \left( I_{e} - I_{f} \right) \,,
\end{align}
for the $ PP $ case. The building block integrals $ I_{e} $ and $ I_{f} $ are defined in Appendix D of \cite{Duplancic:2018bum}.

Specifying the symmetry of the GPD implies that
\begin{align}
    I_f&=-\bar{I}_e\,, &&\textrm{(symmetric GPD)}\,,\\
    I_f&=\bar{I}_e\,, &&\textrm{(anti-symmetric GPD)}\,,
\end{align}
where $\bar{I}_e \equiv I_e^*$. This then leads to
\begin{align}
	N^q_{A_5}[(AB)_{45}]^a&=8 i\left[-\frac{2-\alpha}{\alpha^2 \bar{\alpha}\xi^2}  I_{e}-\frac{1}{\alpha^{2} \bar{\alpha} \xi^{2}} \bar{I}_{e}+\frac{2}{\alpha \bar{\alpha} \sqrt{\alpha} \xi^{2}} \chi_{c}-\frac{8}{\bar{\alpha} \sqrt{\alpha}} \chi_{b}\right]\,,\\[5pt]
	N^q_{B_5}[(AB)_{45}]^a&= 8 i\left[-\frac{1}{\alpha \bar{\alpha} \xi^2} I_{e}+\frac{1-2\alpha}{\alpha^{2}\bar{\alpha}\xi^2} \bar{I}_{e} +\frac{2}{\alpha \bar{\alpha} \sqrt{\alpha}\xi^{2}} \chi_{c}-\frac{8}{\bar{\alpha} \sqrt{\alpha}} \chi_b \right]\,,
\end{align}
for the $ SP $ case (and assuming an antisymmetric GPD), and 
\begin{align}
	\tilde{N} ^q_{A}[(AB)_{45}]^s&=8 
	\left[\frac{2-\alpha}{\alpha \xi} I_{e} - \frac{-1+2 \alpha}{\alpha \xi} \bar{I}_{e} - \frac{2}{\xi \sqrt{\alpha}} \chi_c -8 \xi \sqrt{\alpha} \chi_b  \right]\,,\\[5pt]
	\tilde{N} ^q_{B}[(AB)_{45}]^s&=  
 \frac{8}{\alpha^2 \xi^2}  \left( I_{e} +\bar{I}_{e} \right) \,,
\end{align}
for the $ PP $ case (and assuming a symmetric GPD). Note that $ 	N^q_{A_5}[(AB)_{45}]^s$, $N^q_{B_5}[(AB)_{45}]^s$, $\tilde{N} ^q_{A}[(AB)_{45}]^a $ and $\tilde{N} ^q_{B}[(AB)_{45}]^a $ never appear in computations, see \SEC\ref{sec:organising-amplitude}. Recall that the sums of diagrams are normalised by $ s $, so that they are dimensionless, see Appendix D in \cite{Duplancic:2018bum}. Note also that the building block integral $  \chi _{a} $ cancels in the sum of diagrams.

\subsection{Analytical results in the case of a constant GPD}

As a benchmark, we now give the analytical results for the particular case, though unrealistic, of a constant GPD (taken to be 1) for which the $x$ integration can be performed analytically.

The building block integrals from the asymptotic DA case (see Appendix D in \cite{Duplancic:2018bum} for their definitions) then read 
\begin{eqnarray}
	\label{I2-f-1}
I_b&=& -\frac{2}{(1-\alpha)^2} \frac{1}{1-x_p^2}\,,\\[5pt]
\label{I4-f-1}
I_c&=&\frac{1}{2\alpha \alb \xi 
	\left(1-x_p^2\right)}+\frac{2 x_p \ln \alpha
}{\alb^3\left(1-x_p^2\right)^2}
-\frac{1+x_p^2}{\alb^3\left(1-x_p^2\right)^2 } \ln \frac{1+\xi }{1-\xi
}  \\
&&
+\frac{1}{8 \alb \alpha ^2
	\xi ^2}
\left(\ln \frac{1+\xi}{1-\xi }-\ln
\frac{1-x_p}{1+x_p}\right)
-\frac{1}{8 \alb \xi ^2} \left(-\ln \frac{1+\xi}{1-\xi
}-\ln
\frac{1-x_p}{1+x_p}\right)\,,\nonumber\\[5pt]
\label{I6-f-1}
I_{e}&=&\ln \frac{1 - \xi}{1 + \xi} - i \pi\,,\\[5pt]
\label{I8-f-1}
I_h&=&\frac{1}{1-\alpha}\!
\left\{\ln (1-x_p)
\ln \frac{1+\xi}{\alpha  (1-\xi )}
-\ln (-1-x_p) \ln
\frac{1-\xi }{\alpha  (1+\xi)}
-\ln \alpha \left(\ln \frac{1+\xi
}{1-\xi }-i \pi \right)\right. \!\!  \\
&&\left.
+ 2 \ln \frac{1+\xi}{1-\xi } \ln \frac{1-\alpha }{2
	\xi }
-\text{Li}_2\left(-\frac{(1-\alpha) (1-\xi )}{2 \xi }\right)-\text{Li}_2\left(\frac{(1-\alpha )
	(1-\xi )}{2 \alpha  \xi }\right) 
\right. \nonumber \\
&&\left.
+\text{Li}_2\left(\frac{(1-\alpha
	) (1+\xi)}{2 \xi }\right)+\text{Li}_2\left(-\frac{(1-\alpha)
	(1+\xi)}{2 \alpha  \xi }\right)\right\}\,,\nonumber\\[5pt]
\label{I9-f-1}
I_i&=&\frac{1}{1 - \alpha} (\ln (1 - x_p) - \ln (-1 - x_p))\,.
\end{eqnarray}

Straightforward, though lengthy, calculations show that the building block integrals for the holographic DA case are
%
%
\beqa
\label{chia-f-1}
\chi_a&=&-\frac{1}{2 \xi \sqrt{\alpha}}\left[\ln \left(1+\alpha
-(1-\alpha) \xi+ 2 \sqrt{\alpha} \sqrt{1-\xi^{2}}\right)
\right.\\
&&\left.
+\ln \left(1+\alpha+(1-\alpha)\xi-2 \sqrt{\alpha}\sqrt{1-\xi^{2}}\right)-2 \ln \left( 1-\alpha+(1+\alpha)\xi\right)+i \pi \right], \nonumber\\[5pt]
\label{chib-f-1}
\chi_b&=&\frac{1+\alpha}{4 \alpha \xi} \chi_{a}+\frac{(1-\alpha)(1+\alpha)}{2 \xi \alpha} \frac{\sqrt{1-\xi^{2}}}{(1-\alpha)^{2}-\xi^{2}(1+\alpha)^{2}} \\
&&-\frac{1-\alpha}{\xi \sqrt{\alpha}} \frac{1}{(1-\alpha)^{2}-\xi^{2}(1+\alpha)^{2}}\,,\nonumber\\[5pt]
\label{chic-f-1}
\chi_c &=& - i \pi\,.
\eqa

\section{Photon flux}

\label{app:photon-flux}

\subsection{Weizs\"acker-Williams distribution}

\label{app:WW-distribution}

To obtain the photon flux from an incoming lepton, the Weizs\"acker-Williams distribution is used. This is given by~\cite{Kessler:1975hh,Frixione:1993yw}
\beqa
\label{WW}
f(x)=\frac{\alpha_{\rm em}}{2 \pi}
\left\{2 m_e^2 x
\left(\frac{1}{Q^2_{\rm max}} -\frac{1-x}{m_e^2 x^2}  \right)
+ \frac{\left((1-x)^2+1\right) 
	\ln \frac{Q^2_{\rm max}(1-x)}{m_e^2 x^2}}x
\right\},
\eqa
where $x$ is the fraction of energy lost by the incoming electron, $m_e$ is the 
electron mass and $Q^2_{\rm max}$ is the typical maximal value of the virtuality 
of the exchanged photon, which we take to be $0.1$~GeV$^{2}$ here.

Using the expression for $x$ as a function of the incoming electron energy 
$E_e$ in the \textit{target rest frame},
\beqa
\label{xSgammaN_Ee}
x[S_{\gamma N}] = \frac{S_{\gamma N} - M^2}{2 E_e M},
\eqa
it is easy to obtain integrated cross-sections at the level
of the $e N$ process, using the relation
\beqa
\label{sigma-WW}
\sigma_{e N} = \int \sigma_{\gamma N}(x)\, f(x)\, dx = \int_{S_{\gamma N \, {\rm 
			crit}}}^{S_{\gamma N \, {\rm max}}} \frac{1}{2 E_e M} 
\, \sigma_{\gamma N}(x[S_{\gamma N}]) \, f(x[S_{\gamma N}])
\,dS_{\gamma N} \,.
\eqa

The limits of integration can be found as follows: The minimum value $S_{\gamma N {\rm crit}} \simeq 4.75~{\rm GeV}^2$ corresponds to the imposed necessary kinematical cuts, see appendix~\ref{app:phase-space}. On the other hand, $ S_{\gamma N {\rm max}} $ can be found by finding the value of $ \SgN $ for which $ f $ in \eqref{WW} vanishes, which corresponds to $x[S_{\gamma N}] \approx 1$ in \eqref{xSgammaN_Ee}. For example, at JLab Hall B (CLAS12), with $E_e=12~{\rm GeV}$, one finds that $S_{\gamma N {\rm max}} \simeq 23 \GeV^2$.

\subsection{Photon Flux in UPC}

\label{app:photon-flux-UPC}

\subsubsection{Derivation}

To facilitate the reader, we provide here the main steps of the derivation of the photon flux from a heavy nucleus in ultra-peripheral collisions (UPCs) \cite{Klein:1999qj,Baur:2001jj,Baltz:2007kq}.

UPCs are dominated by photon interactions. To extract the photon flux from heavy ions in UPCs, we follow largely the procedure implemented in STARlight \cite{Klein:2016yzr}, which is a C++ code used in many simulations. 

The photon flux (number of photons per unit area per unit energy) from a relativistic heavy nucleus is given by
\begin{align}
\label{eq:photonflux}
\frac{d^3 N_{ \gamma }}{dk d^{2} \vec{b} }=\frac{Z^2  \alpha  x^2}{ \pi ^2 k  |\vec{b}| ^2}K_{1}^{2}(x)\,,
\end{align}
where $ b $ is the impact parameter, $ k $ is the energy of the photon, $ Z $ is the charge of the heavy nucleus, $  \alpha  $ is the coupling constant of QED, $ K_{1} $ is the modified Bessel function, $  \gamma  $ is the Lorentz factor of the particle emitting the photon, and $ x=\frac{k  |\vec{b}| }{  \gamma \hbar c } $. We note that this equation can be used in any frame, provided both the Lorentz factor $  \gamma  $ and the photon energy $ k $ are measured in the same frame.


To obtain the photon spectrum, one integrates eq.~\eqref{eq:photonflux} over the impact parameter space, assuming circular symmetry,
\begin{align}
\label{eq:photonspectrum}
\frac{dN_{ \gamma }(k)}{dk}=\int^{b_{ \mathrm{max} }}_{b_{ \mathrm{min} }} db\;2 \pi b\,\frac{d^3 N_{ \gamma }}{dk d^{2} \vec{b} }\,P_{ \mathrm{NOHAD} }(b)\,,
\end{align}
where $ P_{ \mathrm{NOHAD} }(b) $ is the probability for no hadronic interactions. The exact way $  P_{ \mathrm{NOHAD} }(b) $ is modelled depends on the nuclei in question. For our purposes, we use the Glauber model and hence
\begin{align}
\label{eq:PNOHAD}
P_{ \mathrm{NOHAD} }(b)=\exp \left[ -  \sigma _{NN}T_{A}(b) \right] \,,
\end{align}
where $ T_{A}(b) $ is the \textit{nuclear thickness function}. The latter is calculated from the nuclear density function $  \rho _{A}(r) $ via
\begin{align}
\label{eq:thicknessfn}
T_{A}(b)&=\int_{-z^{*}}^{z^{*}} dz\, \rho _{A} \left( \sqrt{b^2+z^2} \right) \,,
\end{align}
where $ z^{*} $ is an input to the calculation. The nuclear density is assumed to follow a Woods-Saxon distribution \cite{Woods:1954zz}, given by
\begin{align}
 \rho _{A}(r)&=\frac{ \rho^{A} _{0}}{1+\exp  \left[ \frac{r-r_{A}}{d} \right] }\,,
\end{align}
where $ r_{A} $ is the heavy ion radius, $ d $ is the \textit{Woods-Saxon skin depth}, and $  \rho _{0}^{A} $ is a parameter that is fixed by normalisation, see \eqref{eq:rhoAnormalisation}.

Finally, in \eqref{eq:PNOHAD}, $  \sigma _{NN} $, the nucleon-nucleon interaction cross-section, is based on the parameterisation of pp collisions with centre of mass energy $ \sqrt{s} $ above 7 GeV \cite{ParticleDataGroup:2014cgo},
\begin{align}
 \sigma _{NN}&= \left( 33.73+0.2838 \ln^2 (r)+13.67 r^{-0.412}-7.77r^{-0.5626} \right)  \mathrm{mb} \,,
\end{align}
where $ r\equiv \frac{s}{1  \mathrm{GeV}^2 } $.

In this way, applying the above results to our case, the photon flux can now be fully determined as a function of the photon energy, through \eqref{eq:photonspectrum}. To obtain the cross-section, one simply needs to perform the integration
\begin{eqnarray}
	 \sigma _{ \mathrm{UPC} }&=&\int \frac{dN_{ \gamma }(k)}{dk}  \sigma \left[ \SgN  \left( k \right)  \right]  dk \\[5pt]
		 &=&\frac{1}{2M}\int _{S_{\gamma N \, {\rm 
	 			crit}}}^{S_{\gamma N \, {\rm max}}} \frac{dN_{ \gamma }(\frac{\SgN-M^2}{2 M})}{dk}  \sigma \left[ \SgN   \right] d \SgN \,,
\end{eqnarray}
using
\begin{align}
	\SgN \left( k \right) &=2 M k+M^2\,,
\end{align}
where $ M $ is the target (proton) mass.

A description of the parameters used in the computation of the photon flux is given in the next subsection.

\subsubsection{Parameters used for the calculation of the photon flux}

Since we focus on p-Pb collisions at UPC, the case where the source of the photon flux is the heavy ion dominates. The parameters involving the Pb nucleus which we choose are shown in Table \ref{tab:upcparam}.
\begin{table}[h]
	\centering
\begin{tabular}{|c|c|}
	\hline
	Parameter & Value\\
	\hline
	\hline
	Pb charge, $ Z $ & 82\\
	\hline
	Pb radius, $ r_{A} $ & 6.624 fm\\
	\hline
	Woods-Saxon skin depth, $ d $ & 0.53 fm\\
	\hline
	$  \rho _{0}^{A} $ & 0.160696 $  \mathrm{fm}^{-3}  $\\
	\hline
		p radius, $ r_{p} $ & 0.7 fm\\
	\hline
		$ \hbar c $  & 0.1973 GeV fm\\
	\hline
\end{tabular}
\caption{Values for the parameters used for the calculation of the photon flux from the Pb nucleus.}
	\label{tab:upcparam}
\end{table}

In fact, $  \rho _{0}^{A} $ is fixed by normalisation, by requiring
\begin{align}
\label{eq:rhoAnormalisation}
\int d^{3} \mathbf{r} \,\frac{ \rho _{0}^{A}}{1+\exp  \left[ \frac{r-r_{A}}{d} \right] }&=A\,.
\end{align}

For the limits of the $z$-integration in eq.~\eqref{eq:thicknessfn} for the thickness function $ T_{A}(b) $, we choose $ z^{*}=15\,  \mathrm{fm}   $. 

Finally, for the impact parameter integration in eq.~\eqref{eq:photonspectrum}, we choose the limits of integration to be
\begin{align}
b_{ \mathrm{min} }&=0.7 \,R_{ \mathrm{sum} }\,,\\[5pt]
b_{ \mathrm{max} }&=2\,R_{ \mathrm{sum} }+8\frac{ \gamma \hbar c}{k}\,,
\end{align}
where $ R_{ \mathrm{sum} }$ is the sum of the nuclear radii, equal to $ r_{A}+r_{p} $.

Finally, we compute the Lorentz factor $ \gamma _{A} $ of the Pb nucleus in the p rest frame to be 
\begin{align}
	\gamma _{A}&=3.804 \times 10^7\,,
\end{align}
using a proton energy of $ 6.5 $ TeV and a Pb nucleus energy of $ 82 \times 6.5 = 533 $ TeV.

\section{$t$-dependence of GPDs and phase space integration}

\label{app:phase-space}

In this appendix, we briefly describe various aspects related to the phase space integration over $ (-u') $ and $(-t)$ to obtain the single differential cross-section. The discussion here is short in order to avoid repetition, since this has already been treated in Appendix E of \cite{Duplancic:2018bum}.

First, we note that the $ t $-dependence of the GPD is modelled by a simplistic ansatz, namely a factorised dipole form,	
\beq
\label{dipole}
F_H(t)= \frac{ \left( t_{ \mathrm{min} }-C \right)^{2} }{(t-C)^2}\,,
\eq
with $C=0.71~{\rm GeV}^2.$ We note that in the previous publications \cite{Boussarie:2016qop, Duplancic:2018bum}, the numerator in $ F_{H}(t) $ was simply taken to be $ C^2 $. We improve this here by replacing it with $  \left( t_{ \mathrm{min} }-C \right) ^2 $, which cancels the corresponding contribution at $ -t =  \left( -t \right)_{ \mathrm{min} }  $ in the fully differential cross-section, which is evaluated at $ -t =  \left( -t \right)_{ \mathrm{min} }  $.
The single differential cross-section then reads
\begin{equation}
	\label{difcrosec2}
	\frac{d\sigma}{dM^2_{\gamma\pi}} = \int_{(-t)_{ \mathrm{min} }}^{(-t)_{ \mathrm{max} }} \ d(-t)\ 
	\int_{(-u')_{ \mathrm{min} }}^{(-u')_{ \mathrm{max} }} \ d(-u') \ 
	F^2_H(t)\times\left.\frac{d\sigma}{d(-t) \, d(-u') d M^2_{\gamma\pi}}\right|_{\ 
		-t=(-t)_{ \mathrm{min} }} \,.
\end{equation}

The phase space integration in the $( -t,-u')$ plane should take care of several 
cuts. First, since we rely on factorisation at large angle, we enforce the two 
constraints $-u' > (-u')_{\rm min}\,,$ and
$-t' > (-t')_{\rm min}\,,$
and take $(-u')_{\rm min}=(-t')_{\rm min}=1~{\rm GeV}^2\,.$  Next, the variable $(-t)$ varies from $(-t)_{\rm min}$, determined by kinematics, up 
to a maximal value $(-t)_{\rm max}$ which we fix to be $(-t)_{\rm max}=0.5~{\rm 
	GeV}^2$. For the details of how $ (-t)_{ \mathrm{min} } $ is obtained, we refer the reader to Appendix E of \cite{Duplancic:2018bum}.

Due to the applied kinematical cuts, there are two values of $ \Msq $ for which the volume of the phase space vanishes, $ M^2_{\gamma \pi \, {\rm crit}} \approx 1.52 \GeV^2 $, which defines the lower boundary for $ \Msq $ and is independent of $ \SgN $, and $ M^2_{\gamma \pi \, {\rm Max}}  $, which defines the upper boundary for $ \Msq $ and is a function of $ \SgN $. The value of $ M^2_{\gamma \pi \, {\rm Max}}  $ decreases as $\SgN$ decreases. Therefore, the minimum value of $S_{\gamma N}$ is obtained from the constraint
$M^2_{\gamma \pi \, {\rm crit}}=M^2_{\gamma \pi \, {\rm Max}}$, leading to
$S_{\gamma N {\rm crit}} \simeq 4.75~{\rm GeV}^2.$

\section{Polarisation Asymmetries}

\label{app:polarisation-asymmetries}

\subsection{Kleiss-Sterling Spinor Techniques}

In this appendix, we present the details of the derivation of the formulae for the polarisation asymmetries using the Kleiss-Sterling spinor techniques \cite{Kleiss:1985yh}. This involves the introduction of two (arbitrary) four-vectors $ k_0 $ and $ k_1 $, satisfying
\begin{align}
	k_{0}^{2}=0\,,\qquad k_1^2 = -1\,,\qquad k_0   \cdot k_1 = 0\,.
\end{align}
One then defines the negative helicity spinor product at $ k_0 $  through
\begin{align}
	\label{eq:ref-KS-spinor}
	u_{-}(k_0) \bar{u}_{-}(k_0)= \frac{1}{2} \left( 1-\gamma ^{5} \right)  \slashed{k}_{0}\,.
\end{align}
The corresponding positive helicity state is obtained from the above by
\begin{align}
	u_{+}(k_0)= \slashed{k}_{1}u_{-}(k_0) \,.
\end{align}
A general spinor for \textit{lightlike} momentum $ l $ can be obtained from the above ones through
\begin{align}
	\label{eq:general-KS-spinor}
	u_{ \lambda }(l)=  \frac{1}{\sqrt{2l  \cdot k_{0}}}\slashed{l} u_{- \lambda }(k_0)\,,
\end{align}
where $  \lambda  $ denotes the helicity of the spinor. The only requirement for \eqref{eq:general-KS-spinor} to hold is that $ l \cdot k_{0}  \neq 0$.

For the polarisation vectors, we have
\begin{align}
	\varepsilon_{ \lambda }^{ \mu }(l) = \frac{1}{\sqrt{4 r  \cdot l }} \bar{u}_{ \lambda }(l)\gamma ^{ \mu }u_{ \lambda }(r)\,,
\end{align}
where $ r $ is any lightlike vector not collinear to $ l $ or $ k_0 $.

In this way, all spinor products, including the polarisation tensors, can be expressed as a Dirac trace at the \textit{amplitude} level. For our process, we find it convenient to choose the reference vectors such that
\begin{align}
	k_0^{ \mu } &\equiv p_\meson^\mu = \eqref{eq:p-pi}\,,\\[5pt]
		k_{1}^{ \mu }&\equiv - \frac{2}{s | \vec{p} _{t}|} \varepsilon^{p n p_{\perp} \mu }\\[5pt]
		r &\equiv p\,.
\end{align}
One can easily verify that the above choice satisfies all the required constraints.

Since we are interested in polarisation asymmetries wrt the incoming photon, we first compute the polarisation vectors corresponding to the circularly polarised states. We find that
\begin{align}
	\varepsilon_{\pm}(q)&= \frac{1}{\sqrt{2}} \left[ \frac{p_{\perp}^{ \mu }}{| \vec{p}_{t}| } \pm i k_{1}^{ \mu }\right] \,.
\end{align}
As we will later show, the circular asymmetry wrt the incoming photon vanishes for an unpolarised target, which is the case we consider here. Therefore, what we really need are the polarisation vectors corresponding to \textit{linearly} polarised states. These can be computed using\footnote{Note that the sign conventions used here are opposite to those of \cite{Leader:2011vwq}.}
\begin{align}
    \varepsilon_{x}&=\frac{1}{\sqrt{2}}\left[ \varepsilon_{+}+\varepsilon_{-} \right]\,,\\
    \varepsilon_{y}&=\frac{i}{\sqrt{2}}\left[ \varepsilon_{-}-\varepsilon_{+} \right]\,.
\end{align}

Thus, we have
\begin{align}
\varepsilon_{x}(q)&=\frac{p_{\perp}^{ \mu }}{| \vec{p}_{t}| }\,,\\[5pt]
\varepsilon_{y}(q)&=k_{1}^{ \mu }\,.
\end{align}
Thus, one readily obtains
\begin{align}
	\varepsilon_{x}^{*\, \mu }(q)  \varepsilon_{x}^{ \nu }(q)=\frac{p_{\perp}^{ \mu }p_{\perp}^{ \nu }}{| \vec{p}_{t}|^2 }\,,\\[5pt]
	\varepsilon_{y}^{*\, \mu }(q)  \varepsilon_{y}^{ \nu }(q)=k_{1}^{ \mu }k_{1}^{ \nu }\,.
\end{align}
As a consistency check, we note that the sum of the above two equations is exactly $ -g^{ \mu  \nu }_{\perp} $.

\subsection{Expressions for Polarisation Asymmetries}

Amplitudes corresponding to specific linear polarisation states can be defined as
\begin{align}
	 {\cal M } _{x} = \varepsilon_{x}^{ \mu }(q) {\cal M } _{ \mu }\,,\qquad 	 {\cal M } _{y} = \varepsilon_{y}^{ \mu }(q) {\cal M } _{ \mu }\,.
\end{align}
The linear polarisation asymmetry (LPA) can then be defined as
\begin{align}
 \mathrm{LPA} &\equiv \frac{	d \sigma _{x}-d \sigma _{y}}{d \sigma _{x}+d \sigma _{y}}\,,
\end{align}
where $ d \sigma _{i} $ ($i=x,\,y$) denotes the cross-section obtained by squaring the appropriate polarised amplitude $  {\cal M }_{i}  $. Note that our notation for $ d \sigma _{i} $ is loose and is used to represent either fully differential, single differential or fully integrated cross-sections, depending on the context.

For convenience, let us decompose the amplitude as (c.f.~equations \eqref{def:TA-TB} to \eqref{dec-tensors-quarks})
\begin{align}
	\label{eq:decomposition-tensors}
 {\cal M } =C_{A}T_{A}+C_{B}T_{B}+C_{A_{5}}T_{A_{5}}+C_{B_{5}}T_{B_{5}}\,,
\end{align}
i.e. in terms of the tensor structures defined in \eqref{def:TA-TB} and \eqref{def:TA5-TB5}. We note that the coefficients of the tensor structures include the spinors of the nucleons, as well as Dirac matrices associated with the definition of the GPDs. More explicitly, using \eqref{CEGPD} and \eqref{dec-tensors-quarks}, 
\begin{align}
\label{eq:CA}
    C_A&\equiv\frac{1}{n\cdot p}\bar{u}(p_2,\lambda_2) \slashed{n} \gamma^5 u(p_1,\lambda_1)\mathcal{\tilde{H}}_{\pi A} (\xi , t)\,,\\
    C_B&\equiv\frac{1}{n\cdot p}\bar{u}(p_2,\lambda_2) \slashed{n} \gamma^5 u(p_1,\lambda_1)\mathcal{\tilde{H}}_{\pi B} (\xi , t)\,,\\
    C_{A_5}&\equiv\frac{1}{n\cdot p}\bar{u}(p_2,\lambda_2) \slashed{n}  u(p_1,\lambda_1)\mathcal{{H}}_{\pi A_5} (\xi , t)\,,\\
    \label{eq:CB5}
    C_{B_5}&\equiv\frac{1}{n\cdot p}\bar{u}(p_2,\lambda_2) \slashed{n}  u(p_1,\lambda_1)\mathcal{{H}}_{\pi B_5} (\xi , t)\,.
\end{align}
By squaring the amplitude, and summing over the polarisation $  \lambda _{k} $ of the outgoing photon, we obtain
\begin{align}
\label{eq:pol-amp-sq-x}
\sum_{ \lambda _{k}}| {\cal M }_{x}|^2 &=|C_{A}|^2 + | \pt|^{4} |C_{B}|^2+\frac{s^2}{4}| \pt|^4 |C_{B_{5}}|^2-2 |\pt|^2  \mathrm{Re}(C_{A}^{*}C_{B}) \,,\\[5pt]
\label{eq:pol-amp-sq-y}
\sum_{ \lambda _{k}}| {\cal M }_{y}|^2 &=|C_{A}|^2 +\frac{s^2}{4}| \pt|^4 |C_{A_{5}}|^2 \,.
\end{align}
From the above polarised amplitude squared, one can easily compute the LPAs at various levels (from fully differential to integrated). 

\subsection{Vanishing of the circular asymmetry}

For the circular polarised amplitudes, the analogues of \eqref{eq:pol-amp-sq-x} and \eqref{eq:pol-amp-sq-y} are given by
\begin{align}
\sum_{ \lambda _{k}}| {\cal M }_{+}|^ 2&=\frac{1}{2} \bigg[ 2|C_{A}|^2 + | \pt|^{4} |C_{B}|^2+\frac{s^2}{4}| \pt|^4 |C_{A_{5}}|^2+\frac{s^2}{4}| \pt|^4 |C_{B_{5}}|^2-2 |\pt|^2  \mathrm{Re}(C_{A}^{*}C_{B}) \;\nonumber\\[5pt]
&\phantom{aaaaa} + s |\pt|^2    \mathrm{Im} \left( C_{A} \left( C_{A_{5}}^{*}+C_{B_{5}}^{*} \right)+|\pt|^2 C_{A_{5}} C_{B}^{*}  \right) \bigg]\,,\\[5pt]
\sum_{ \lambda _{k}}| {\cal M }_{-}|^ 2&=\frac{1}{2} \bigg[ 2|C_{A}|^2 + | \pt|^{4} |C_{B}|^2+\frac{s^2}{4}| \pt|^4 |C_{A_{5}}|^2+\frac{s^2}{4}| \pt|^4 |C_{B_{5}}|^2-2 |\pt|^2  \mathrm{Re}(C_{A}^{*}C_{B}) \;\nonumber\\[5pt]
&\phantom{aaaaa} - s |\pt|^2    \mathrm{Im} \left( C_{A} \left( C_{A_{5}}^{*}+C_{B_{5}}^{*} \right)+|\pt|^2 C_{A_{5}} C_{B}^{*}  \right) \bigg]\,.
\end{align}
So,
\begin{align}
\label{eq:circular-asymmetry}
\sum_{ \lambda _{k}}| {\cal M }_{+}|^ 2 - \sum_{ \lambda _{k}}| {\cal M }_{-}|^ 2=s |\pt|^2    \mathrm{Im} \left( C_{A} \left( C_{A_{5}}^{*}+C_{B_{5}}^{*} \right)+|\pt|^2 C_{A_{5}} C_{B}^{*}  \right) \,.
\end{align}
An interesting feature of the circular asymmetry is that it only contains terms that mix vector GPD and axial GPD contributions ($ A $ and $ B $, with $ A_{5} $ and $ B_{5} $). Thus, when averaging over the target helicity, it can be shown that all terms on the RHS of \eqref{eq:circular-asymmetry} vanish. Indeed, using equations \eqref{eq:CA} to \eqref{eq:CB5}, we obtain, after averaging and summing over the target helicities,
\begin{align}
\frac{1}{2} \sum_{\lambda_1,\lambda_2}\left(\sum_{ \lambda _{k}}| {\cal M }_{+}|^ 2 - \sum_{ \lambda _{k}}| {\cal M }_{-}|^ 2 \right)&= \frac{|\pt|^2}{  \left( n  \cdot p \right) } \left[  \tilde{{\cal 
		H}} _{\pi A}\left({\cal 
		H}_{\pi A_5}^{*}+{\cal 
		H}_{\pi B_5}^{*} \right) + |\pt|^2{{\cal 
		H}} _{\pi A_5} \tilde{{\cal 
		H}}_{\pi B}^{*}   \right]  \mathrm{tr} \left[  \slashed{p}_{2} \slashed{n}\gamma ^{5} \slashed{p}_{1} \slashed{n}     \right] \nonumber \\[5pt]
		&= 0\,.
\end{align}
This shows that for an unpolarised target, the circular asymmetry is identically zero. From a more physical point of view, the vanishing of the circular asymmetry is a consequence of parity invariance of QED and QCD. In particular, from \cite{Bourrely:1980mr}, one deduces that the amplitude for our process, ${\cal M } _{ \lambda _2  \lambda _k \,;\, \lambda _1  \lambda _q}$, has to obey the relation
\begin{equation}
	{\cal M } _{ \lambda _2  \lambda _k \,;\, \lambda _1  \lambda _q}= \eta\, (-1)^{ \lambda_1- \lambda_q- \left(  \lambda _2 -  \lambda _k \right) } {\cal M } _{- \lambda _2  -\lambda _k \,;\, -\lambda _1  -\lambda _q}\,,
\end{equation}
where $  \eta  $ represents a phase factor related to intrinsic spin. From this, we can deduce that
\begin{align}
	\sum_{ \lambda _i ,\,i\neq q}| {\cal M } _{ \lambda _2  \lambda _k \,;\, \lambda _1  +}|^2 = \sum_{ \lambda _i ,\,i\neq q}| {\cal M } _{ \lambda _2  \lambda _k \,;\, \lambda _1  -}|^2\,,
\end{align}
which implies that the circular asymmetry vanishes identically for an unpolarised target.

\bibliographystyle{utphys}

\bibliography{masterrefs.bib}

\end{document}